\documentclass{CUP-JNL-DCE}

\usepackage{latexsym}
\usepackage{graphicx}
\usepackage{subfig}
\graphicspath{{./}{./figures/}}
\usepackage{multicol,multirow}
\usepackage{amsmath,amssymb,amsfonts}
\usepackage{mathrsfs}
\usepackage{amsthm}
\usepackage{array} 
\usepackage{apacite}
\usepackage{rotating}
\usepackage{appendix}
\usepackage[authoryear]{natbib}
\usepackage{ifpdf}
\usepackage[T1]{fontenc}
\usepackage{times}
\usepackage{sourcesanspro}
\usepackage{newtxmath}
\usepackage{textcomp}%
\usepackage{xcolor}%
\usepackage{hyperref}
\usepackage{todonotes}
\usepackage{lineno}
\usepackage[final]{changes}
\definechangesauthor[name=Heng, color=purple]{hx}    
\definechangesauthor[name=Meelan, color=blue]{mc}
\definechangesauthor[name=Irfan, color=red]{iz}

\pdfoutput=1

\articletype{RESEARCH ARTICLE}
\jyear{2021}

\begin{document}
\begin{Frontmatter}

\title[]{Recurrent Neural Network for End-to-End Modeling of Laminar-Turbulent Transition }

\author[1]{Muhammad I. Zafar}\orcid{0000-0003-0664-2285} 
\author[2]{Meelan M. Choudhari}\email{m.m.choudhari@nasa.gov}\orcid{0000-0001-9120-7362}
\author[3]{Pedro Paredes}\orcid{0000-0003-1890-1811} 
\author*[1]{Heng Xiao}\email{hengxiao@vt.edu}\orcid{0000-0002-3323-4028}

\authormark{Muhammad I. Zafar \textit{et al.}}

\address[1]{\orgdiv{Kevin T. Crofton Department of Aerospace and Ocean Engineering}, \orgname{Virginia Tech}, \orgaddress{\street{Blacksburg}, \state{Virginia}, \country{USA}}}

\address*[2]{\orgdiv{Computational AeroSciences Branch}, \orgname{NASA Langley Research Center}, \orgaddress{\street{Hampton}, \state{Virginia}, \country{USA}}}

\address*[3]{\orgname{National Institute of Aerospace}, \orgaddress{\street{Hampton}, \state{VA}, \postcode{23666}, \country{USA}}}



\abstract{Accurate prediction of laminar-turbulent transition is a critical element of computational fluid dynamics simulations for aerodynamic design across multiple flow regimes. Traditional methods of transition prediction cannot be easily extended to flow configurations where the transition process depends on a large set of parameters. In comparison, neural network methods allow higher dimensional input features to be considered without compromising the efficiency and accuracy of the traditional data driven models. Neural network methods proposed earlier follow a cumbersome methodology of predicting instability growth rates over a broad range of frequencies, which are then processed to obtain the N-factor envelope, and then, the transition location based on the correlating N-factor. This paper presents an end-to-end transition model based on a recurrent neural network, which sequentially processes the mean boundary-layer profiles along the surface of the aerodynamic body to directly predict the N-factor envelope and the transition locations over a two-dimensional airfoil. The proposed transition model has been developed and assessed using a large database of 53 airfoils over a wide range of chord Reynolds numbers and angles of attack. The sequence-to-sequence transduction model proposed herein provides a more direct approach for accurate predictions of the transition location than the earlier neural network methods, which predict the local amplification rate of a single instability mode at a fixed location along the airfoil. The large universe of airfoils encountered in various applications causes additional difficulties. As such, we provide further insights on selecting training datasets from large amounts of available data. Although the proposed model has been analyzed for two dimensional boundary layers in this paper, it can be easily generalized to other flows due to embedded feature extraction capability of convolutional neural network in the model.
}

\begin{policy}[Impact Statement]
The recurrent neural network (RNN) proposed here represents a significant step toward an end-to-end prediction of laminar-turbulent transition in boundary-layer flows.
The general yet greatly simplified workflow should allow even nonexperts to the apply the proposed model for predicting transition due to a variety of instability mechanisms,  which is a significant advantage over traditional direct computations of the stability theory. The encoding of boundary layer profiles by the convolutional neural network (CNN) and sequence-to-sequence mapping enabled by the RNN faithfully represent the amplification of flow instability along the surface, exemplifying the direct correlation of the proposed model with the underlying physics.
Finally, we use a very large dataset  and provide insights and best-practice guidance toward the practical deployment of neural-network-based transition models in engineering environments.
\end{policy}
\end{Frontmatter}


\section{Introduction \label{introduction}}
Laminar-turbulent transition of boundary-layer flows has a strong impact on the performance of flight vehicles across multiple flow regimes due to its effect on surface skin friction and aerodynamic heating. Predicting the transition location in computational fluid dynamics (CFD) simulations of viscous flows remains a challenging area~\citep{slotnick2014}. Transition to turbulence in a benign disturbance environment is typically initiated by the amplification of modal instabilities of the laminar boundary layer. 
Depending on the flow configuration, these instabilities can be of several types, e.g., Tollmien-Schlichting (TS) waves, oblique first-mode instabilities, and planar waves of second-mode (or Mack-mode) type.

\begin{figure}
    \centering
    \subfloat[Linear Stability Theory (LST): Growth rates are computed for each instability wave characterized by frequency ($\omega$) and wave number ($\alpha$). Growth rates are integrated ($\int$) along the airfoil contour to obtain corresponding N-factor values, from which the N-factor envelope and transition location ($x/c_\text{tr}$) are determined for a given value of correlating N-factor ($\text{N}=\text{N}_\text{tr}$). The most amplified Tollmien-Schlichting (TS) waves in two-dimensional boundary layers correspond to two-dimensional instability waves ($\beta=0$) \citep{rajnarayan2013}. ]{\includegraphics[width=0.57\textwidth]{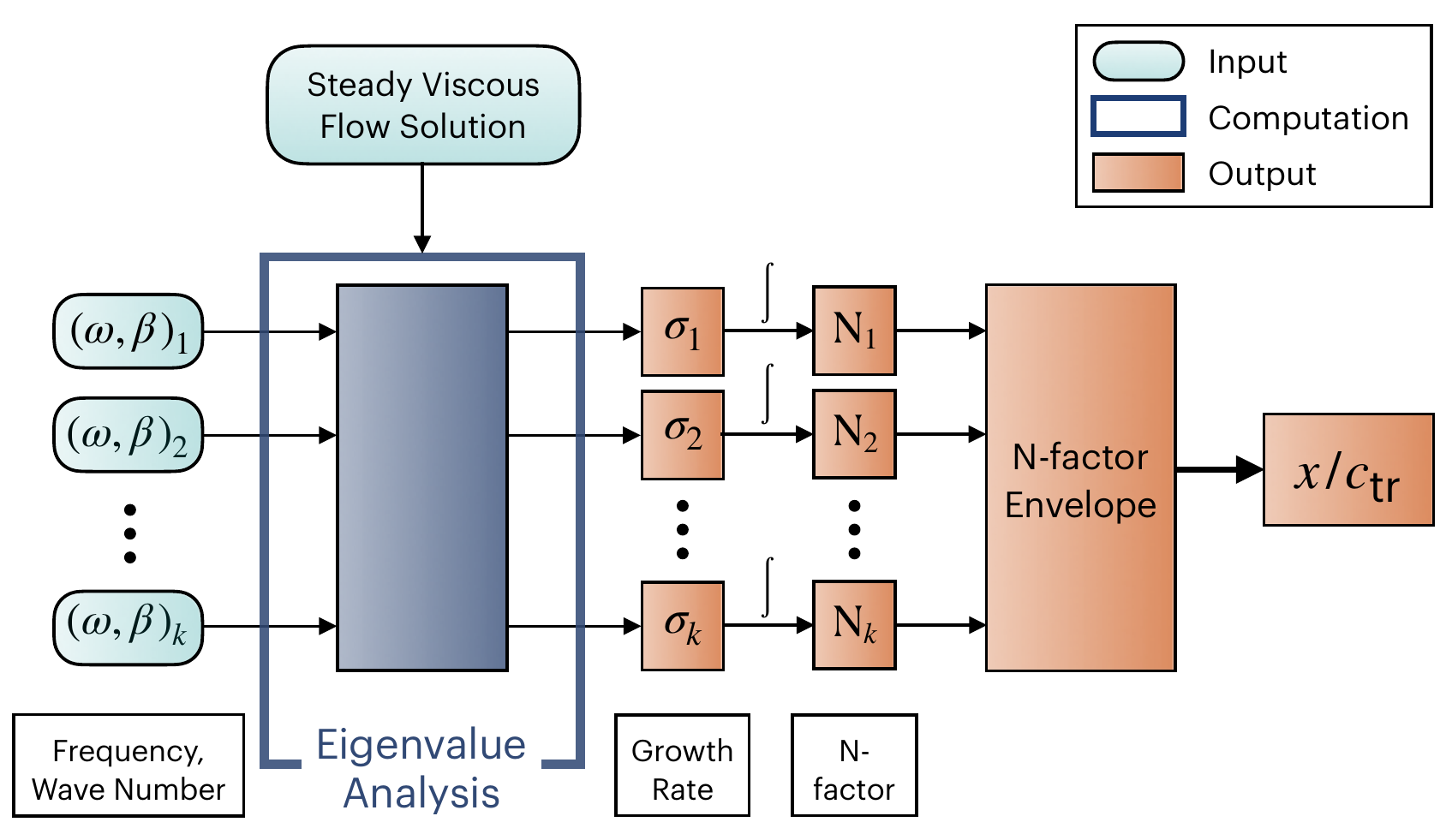}} \\
    \subfloat[Convolutional Neural Network-based model (CNN): Instead of using eigenvalue analysis, local growth rates corresponding to each frequency ($\omega$) are predicted using a neural network
    \citep{zafar2020}. ]{\includegraphics[width=0.57\textwidth]{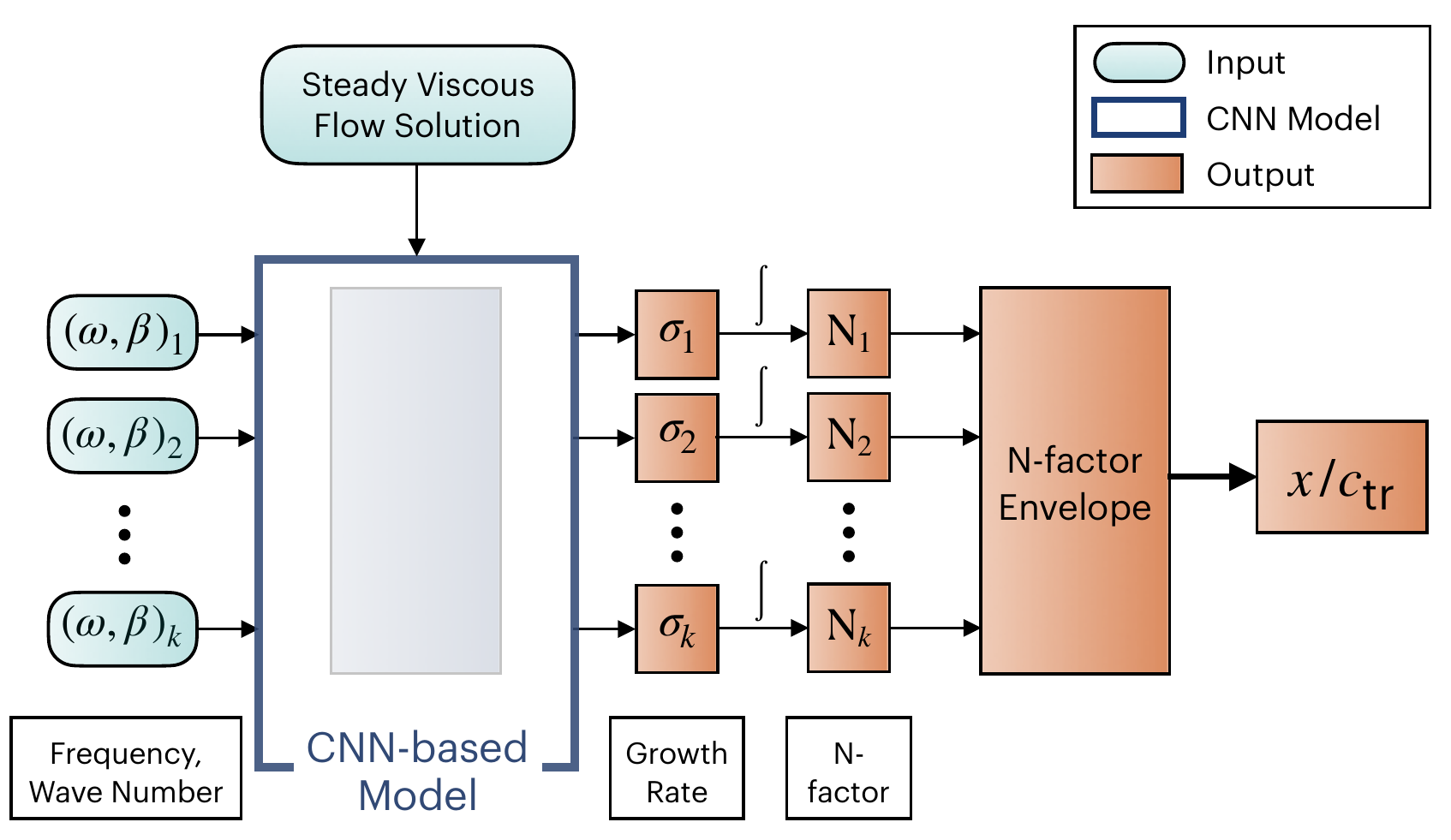}} \\
    \subfloat[Recurrent Neural Network (RNN) model for N-factor envelope modelling: Growth rate of the N-factor envelope ($dN/ds$) is directly predicted, which is then integrated ($\int$) along the airfoil contour to obtain the N-factor envelope and estimated transition location ($x/c_\text{tr}$) for a given value of correlating N-factor ($\text{N}=\text{N}_\text{tr}$).  ]{\includegraphics[width=0.57\textwidth]{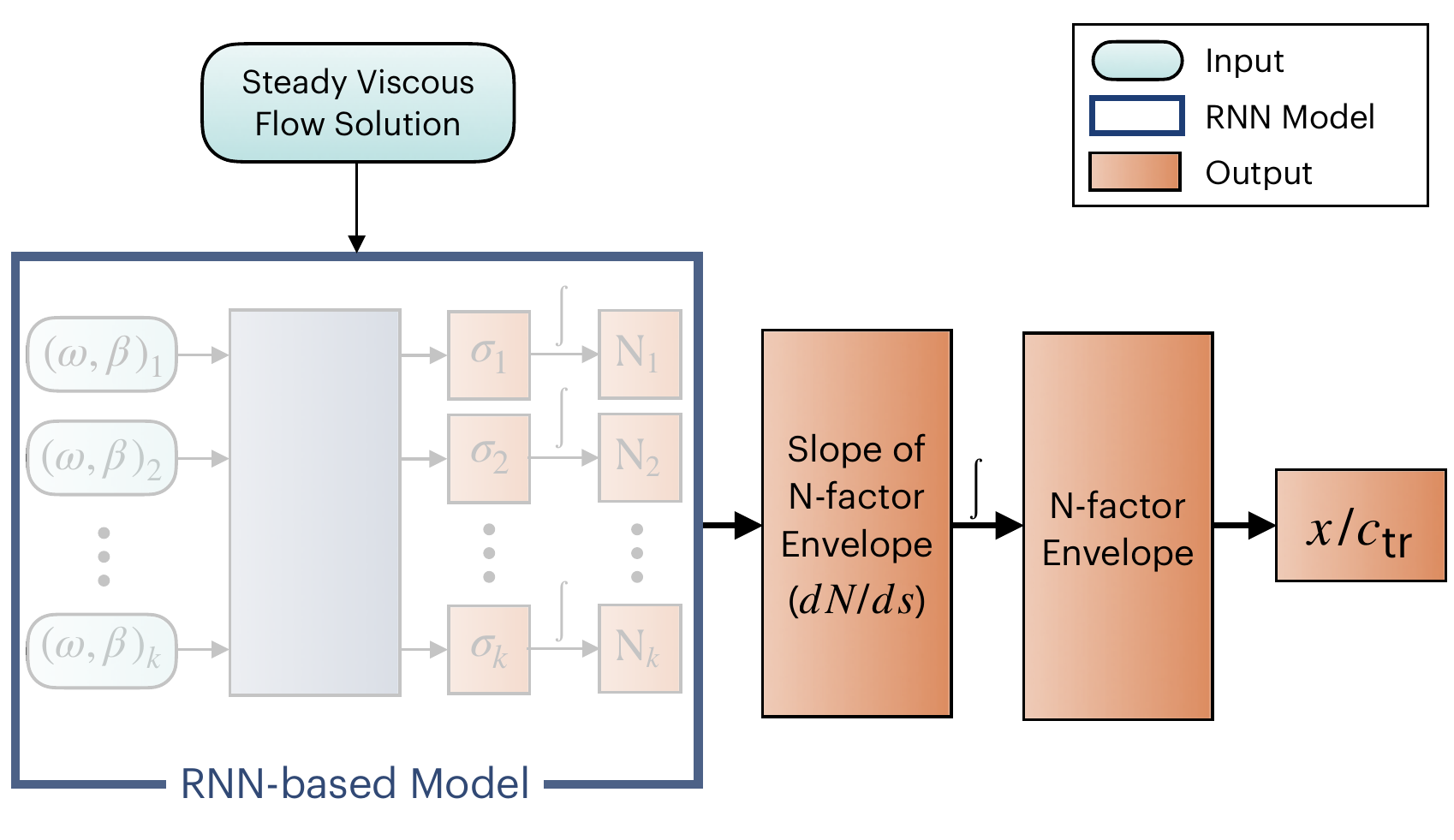}}
    \caption{Comparison of transition prediction methodologies}
    \label{fig:methodology_schematic}
\end{figure}

A description of transition prediction methods based on stability correlations can be found in a variety of references~\citep{ingen1956, ingen2008, smith1956}, but we provide a brief description here to make the paper self-contained. A somewhat expanded description may also be found in \cite{zafar2020}. The transition process begins with the excitation of linear instability waves that undergo a slow amplification in the region preceding the onset of transition.  The linear amplification phase is  followed by nonlinear processes that ultimately lead to turbulence. Since these nonlinear processes are relatively rapid, it becomes possible to predict the transition location based on the evolution of the most amplified instability mode. The linear amplification ratio, $e^N$, is generally computed using the classical linear stability theory~\citep{mack1987, reed1996, juniper2014, taira17, reshotko1976}. The local streamwise amplification rates ($\sigma$) along the aerodynamic surface can be determined by solving an eigenvalue problem for the wall-normal velocity perturbation, which is governed by the Orr-Sommerfeld (OS) equation \citep{drazin1981}. These local streamwise amplification rates corresponding to each frequency ($\omega$) are then integrated along the body curvature to obtain the logarithmic amplification of the disturbance amplitude (N-factor) for each disturbance frequency ($\omega$). The $e^N$ method assumes that the occurrence of transition correlates very well with the N-factor of the most amplified instability wave reaching a critical value $N_\text{tr}$. For subsonic and supersonic flows, the critical N-factor (denoted herein as $N_\text{tr}$) has been empirically found to lie in the range between 9 and 11 \citep{ingen2008,bushnell1989}. Such a prediction process may be schematically illustrated as shown in Fig.~\ref{fig:methodology_schematic}(a).

Linear stability computations rely on highly accurate computations of the mean boundary-layer flow. Solution of the linear stability equations is also computationally expensive and often leads to the contamination of the unstable part of the spectrum by spurious eigenvalues. The nonrobust nature of the stability computations requires a significant degree of expertise in stability theory on the user's part, making such computations inapt for nonexpert users. For these reasons, transition prediction based on stability computations has been difficult to automate and renders its direct integration in CFD solvers rather impractical. Several aerodynamic applications involving flow separation also entail viscous-inviscid interactions. Such interactions lead to a strong coupling between transition and the overall flow field, which requires an iterative prediction approach. Hence, the integration of the transition prediction method in the overall aerodynamic prediction method remains an important area of research \citep{slotnick2014}. 

Several methods have been proposed as simplifications or surrogate models of the $e^N$ methods, including database query techniques \citep{ingen2008, drela1987, perraud2016} and data fitting techniques \citep{dagenhart1981, stock1989, gaster1995, langlois2012, krumbein2008, rajnarayan2013, begou2017, pinna2018}. These methods are generally based on a small set of scalar input parameters representing the mean flow parameters and relevant disturbance characteristics. However, these methods do not scale well with larger sets of parameters, which tends to limit the expressive power of the transition model based on these traditional methods \citep{crouch2002}. In particular, the shape factor, is a commonly used scalar parameter to correlate the disturbance amplification rates to the mean flow of the boundary layer. However, the shape factor cannot be easily computed for many practical flows such as high speed flows over blunt leading edges, which results in a poor predictive performance of the database methods \citep{paredes2020_journal}. 

Neural networks provide a more generalized way of predicting the instability characteristics, while also accounting for their dependency on high-dimensional input features in a computationally efficient and robust manner.  \cite{fuller1997} applied neural network methods to instability problems in predicting the instability growth rates for a jet flow. \cite{crouch2002} used scalar parameters and the wall-normal gradient of the laminar velocity profile as an input of neural networks to predict the maximum instability growth rates. They demonstrated the generalizability of the neural network method for both Tollmien–Schlichting waves and stationary crossflow instabilities. The data for the gradient of the laminar velocity profile were coarsely defined at six equidistant points across the boundary layer. A fully connected neural network was used, which assumes no spatial structure on the input data. Such treatment of boundary-layer profiles may not be well suited for other instability mechanisms involving, for instance, Mack-mode instabilities in high-speed boundary layers that require input profiles of thermodynamics quantities along with the velocity profiles~\citep{paredes2020_journal} or the secondary instabilities of boundary-layer flows with finite-amplitude stationary crossflow vortices that include rapid variations along both wall-normal and spanwise coordinates. 

By utilizing recent developments in machine learning, \cite{zafar2020} proposed a transition model based on convolutional neural networks (CNNs), which has the ability to generalize across multiple instability mechanisms in an efficient and robust manner. CNNs were used to extract a set of latent features from the boundary-layer profiles, and the extracted features were used along with other scalar quantities as input to a fully connected network. The hybrid architecture was used to predict the instability growth rates for Tollmien-Schlichting instabilities in two-dimensional incompressible boundary layers. The extracted latent feature showed a strong, nearly linear correlation with the analytically defined shape factor ($H$) of the boundary-layer velocity profile. The model was trained using a database of Falkner--Skan family of selfsimilar boundary-layer profiles.
This CNN-based method is applicable to various instability mechanisms with higher-dimensional input features, since the boundary-layer profiles are treated in a physically consistent manner (i.e., as discrete representation of the profiles accounting for their spatial structures). It has been applied to predict the instability growth rates of Mack-mode instabilities in hypersonic flows over a moderately blunt-nosed body~\citep{paredes2020_journal}. This particular application requires additional input features in the form of boundary-layer profiles of thermodynamic quantities such as temperature and/or density, where the CNN-based model demonstrated highly accurate predictions of the instability growth rates. 

Despite clear advantages over earlier neural network-based models, the CNN-based transition model shares a few significant shortcomings with the direct integration of linear stability theory toward transition prediction.
Similar to the stability theory, the CNN-based transition model is based on the predictions of the instability growth rates corresponding to each selected disturbance frequency $\omega$ (and/or the spanwise wavenumber $\beta$ in the case of three-dimensional instabilities) and every station from the input set of boundary-layer profiles. The instability growth rates for each individual disturbance are integrated along the aerodynamic surface to predict the growth in disturbance amplitude, or equivalently, the N-factor curve for each combination of frequency and spanwise wavenumber. Finally, one must determine the envelope of the N-factor curves to predict the logarithmic amplification ratio for the most amplified disturbance at each station (denoted as $N_{env}$ herein), which can then be used in conjunction with the critical N-factor ($N_{cr}$) based on previous experimental measurements to predict the transition location (see Fig.~\ref{fig:methodology_schematic}b for a summary). This overall workflow not only extends over several steps, but also requires the user to estimate the range of frequencies (and/or spanwise wavenumbers) that would include the most amplified instability waves corresponding to the envelope of the N-factor curves. The selection of disturbance parameters for a given flow configuration can be somewhat challenging for nonexpert users. More important, however, the above workflow requires several redundant growth rate computations involving subdominant disturbances that do not contribute to the N-factor envelope used to apply the transition criterion, namely, $N = N_{cr}$. Finally, and similar to the earlier neural network models~\citep{crouch2002}, the growth rate prediction during the all important first step of the above workflow only uses the local boundary-layer profiles, and hence, does not utilize any information about the prior history of a given disturbance, e.g., any previously estimated instability growth rates at the upstream locations.  Since the boundary-layer profiles evolve in a continuous manner, the spatial variation in the disturbance growth rate represents an analytic continuation along the aerodynamic surface. Thus, embedding the upstream history of boundary-layer profiles and/or the disturbance growth rates should lead to more accurate, robust, and computationally efficient models for the onset of transition. 

A recurrent neural network (RNN) is a promising approach for modeling the history effects. The RNN is a general-purpose architecture for modeling sequence transduction by using an internal state (memory) that selectively keeps track of the information at the preceding steps during the sequence  \citep{graves2012}. RNNs provide a combination of multivariate internal state as well as nonlinear state-to-state dynamics, which make it particularly well-suited for dynamic system modeling. \cite{faller1997} exploited these attributes of RNNs to predict unsteady boundary-layer development and separation over a wing surface. The RNN architectures have also been used for the modeling of several other complex dynamic systems ranging from near-wall turbulence \citep{guastoni2019}, the detection of extreme weather events \citep{wan2018}, and the spatiotemporal dynamics of chaotic systems \citep{vlachas2020}, among others. 

The feature extraction capability of CNN and the sequence-to-sequence mapping enabled by RNN provide a direct correlation with the underlying physics of transition, exemplifying the machine learning models motivated by the physics of the problem, e.g. in the modeling of turbulence~\citep{wang2017-physics, wu2018-physics, duraisamy2019turbulence}. Transport equation based models, such as the well-known Langtry-Menter 4-equation model~\cite{menter2006}, are based on empirical transition correlations that are based on local mean flow parameters, therefore the connection with the underlying physics of the transition process is significantly weaker as compared to stability based correlation, whether it involves direct computations of linear stability characteristics or a proxy thereto as represented within the proposed RNN model. This paper is aimed at exploiting the sequential dependency of mean boundary-layer flow properties to directly predict maximum growth rates among all unstable modes at a sequence of stations along the airfoil surface. Such sequential growth rates can then be integrated along the airfoil surface to determine the N-factor envelope and corresponding transition location, as has been schematically illustrated in Fig.~\ref{fig:methodology_schematic}(c). To this end, an extensive airfoil database has been used that documents mean flow features and linear stability characteristics for a large set of airfoils at a range of flow conditions (Reynolds numbers and angles of attack). Furthermore, we provide insight on the similarity of stability characteristics among different families of airfoils and how a neural network trained on one set of airfoils can generalize to other ones, possibly at different flow conditions.

The rest of the manuscript is organized as follows. The proposed RNN model is introduced in \S\ref{rnn} along with the input and output features. Section~\ref{database} presents the airfoil database used to develop and evaluate the proposed transition model. Section~\ref{results} presents the results and discussion for different training and testing cases, which provide insight toward subsampling of training datasets for achieving a reasonable predictive performance from the RNN model. Section~\ref{conclusion} concludes the paper.

\section{Recurrent Neural Network \label{rnn}}
A neural network consists of successive composition of linear mapping and nonlinear squashing functions, which aims to learn the underlying relationship between an input vector ($\mathbf{q}$) and an output vector ($\mathbf{y}$) from a given set of training data. The series of functions are organized as a sequence of layers, each containing several neurons that represent specific mathematical functions. The mathematical functions in each layer are parameterized by the weight ($\mathbf{W}$) and bias ($\mathbf{b}$). Intermediate layers between the input layer ($\mathbf{q}$) and output layer ($\mathbf{y}$) are known as hidden layers. The functional mapping of a neural network with a single hidden layer can be expressed as:
\begin{equation}
    \boldsymbol{y}=\boldsymbol{W}^{(2)}\left(f\left[\boldsymbol{W}^{(1)} \boldsymbol{q}+\boldsymbol{b}^{(1)}\right]\right)+\boldsymbol{b}^{(2)}
\end{equation}
where $\mathbf{W}^{(l)}$ and $\mathbf{b}^{(l)}$ represent the weight matrix and bias vector for the $l^{th}$ layer, respectively, and $f$ is an activation function. Activation functions enable the representation of complex functional mapping by introducing nonlinearity in the composite functions. The training of a neural network is a process of learning the weights and biases with the objective of fitting the training data.

\begin{figure}
    \centering
    \includegraphics[width=0.67\textwidth]{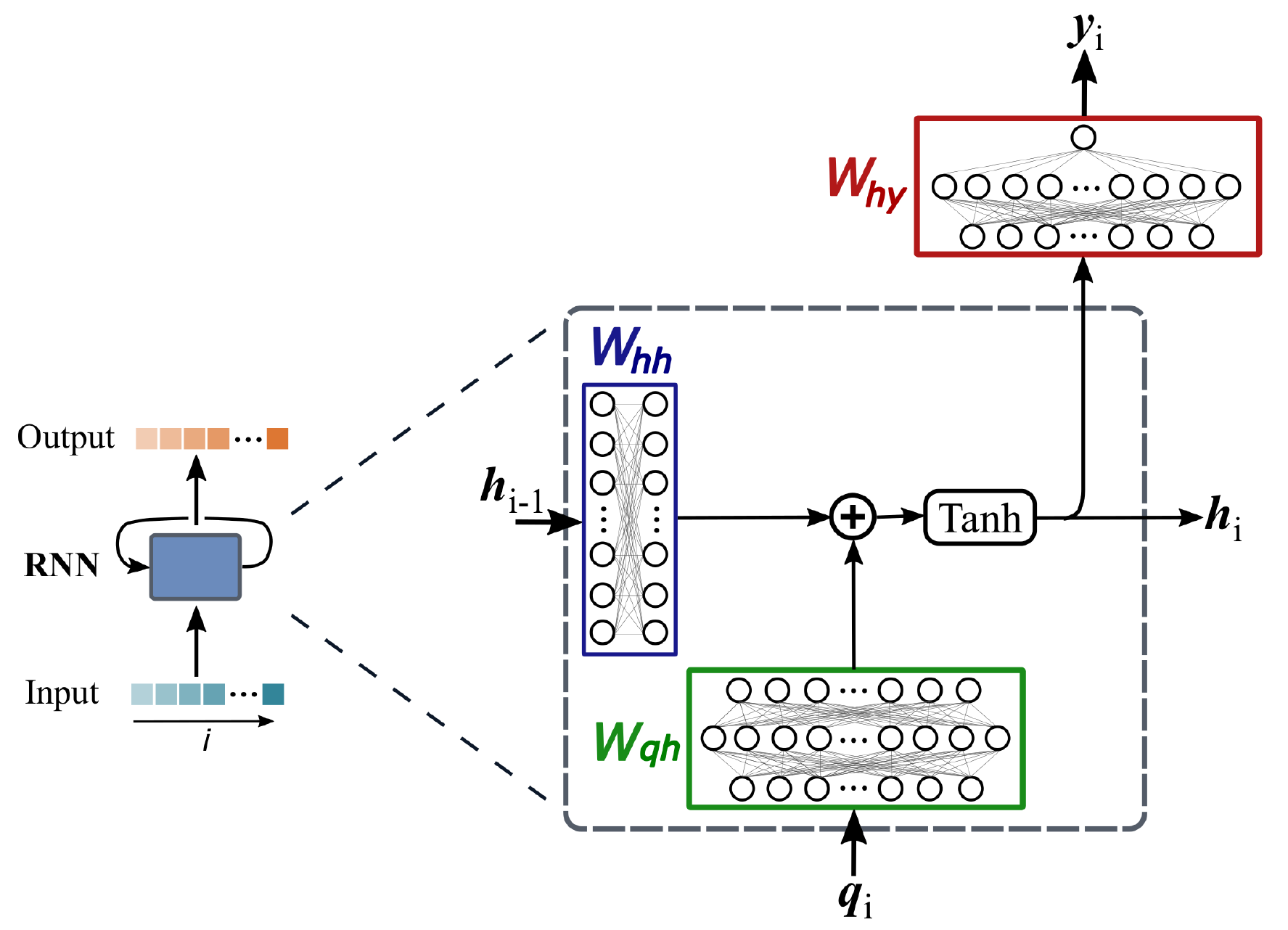}
    \caption{Schematic of the RNN cell shown as a blue box on the left. Within each RNN cell, the arrangement of the weight matrices is shown on the right. At any step $i$ of the sequence, the RNN cell takes input $q_i$ and previous hidden state $h_{i-1}$ and provides updated hidden state $h_i$ and output $y_i$}
    \label{fig:rnn_cell}
\end{figure}

\begin{figure}
    \centering
    \includegraphics[width=0.63\textwidth]{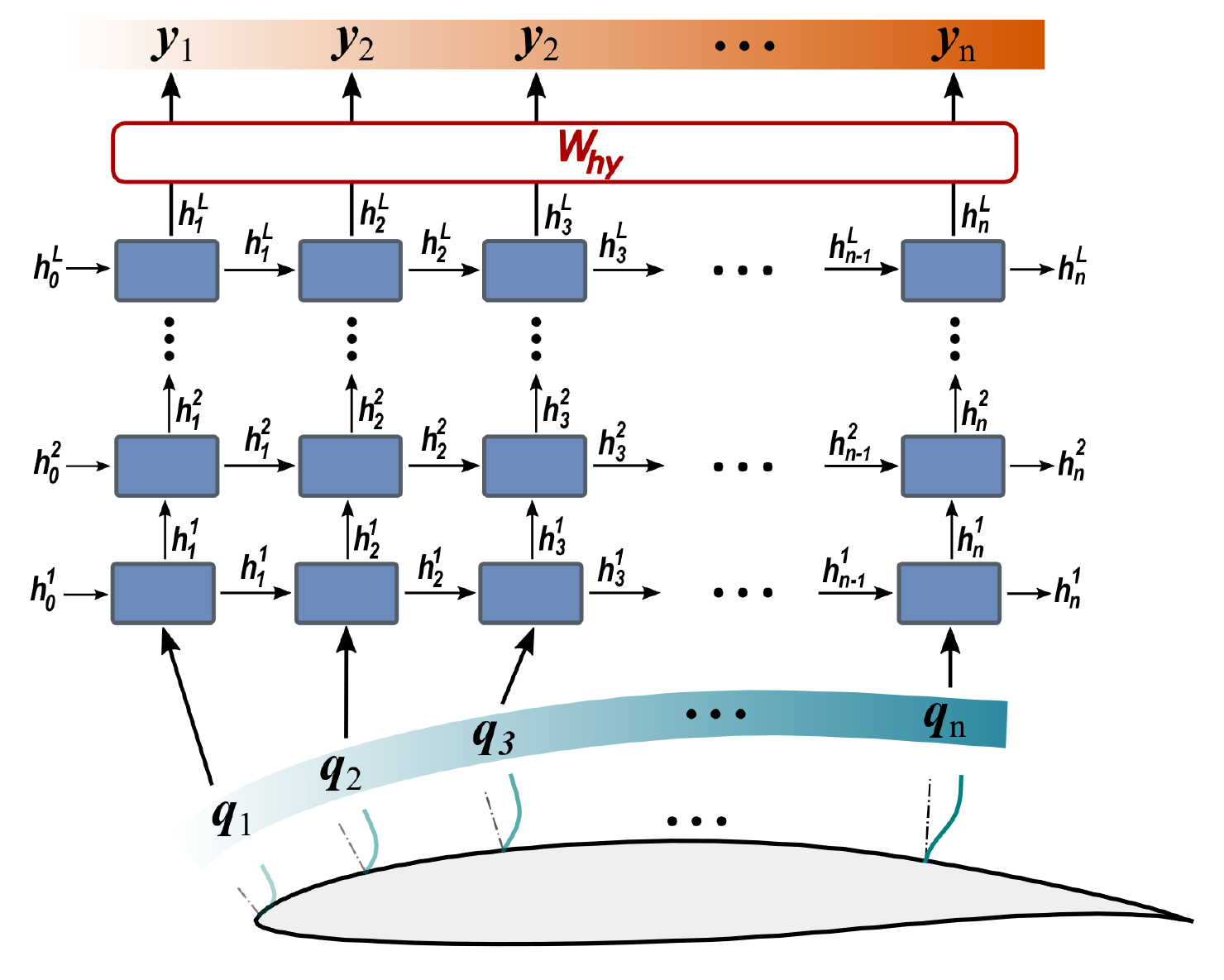}
    \caption{Sequences of input features and output for deep RNN architecture have been illustrated with respect to stations along the airfoil surface}
    \label{fig:rnn_architecture}
\end{figure}

Recurrent neural networks (RNNs) are architectures with internal memory (known as the \emph{hidden states}), which make them particularly suitable for sequential data such as time series, spatial sequences, and words in a text. The RNN processes the sequence of inputs in a step-by-step manner while selectively passing the information across a sequence of steps encoded in a hidden state. At any given step $i$, the RNN operates on the current input vector ($\mathbf{q}_i$) in the sequence and the hidden state $\mathbf{h}_{i-1}$ passed on from the previous step, to produce an updated hidden state $\mathbf{h}_{i}$ and an output $\mathbf{y}_i$. Figure~\ref{fig:rnn_cell} shows the schematic of a recurrent neural network.
Multiple RNNs can be stacked over each other, as shown in Fig.~\ref{fig:rnn_architecture}, to provide a deep RNN. The functional mapping for an architecture with $L$ layers of RNN stacked over each other can be expressed as:
\begin{subequations}
\label{eq:rnn}
\begin{align}
        \boldsymbol{h}_{i}^{l}&=f\left[\boldsymbol{W}_{hh}^{l} \cdot \boldsymbol{h}_{i-1}^{l}+\boldsymbol{W}_{qh}^{l} \cdot \boldsymbol{h}_{i}^{l-1}\right] \label{eq:rnn_A} \\ 
        \boldsymbol{y}_{i}&=\boldsymbol{W}_{hy} \cdot \boldsymbol{h}_{i}^{L} \label{eq:rnn_B}
\end{align}
\end{subequations}
where $\boldsymbol{W}_{\text{hh}}^{l}$, $\boldsymbol{W}_{\text{qh}}^{l}$, and $\boldsymbol{W}_{\text{hy}}$ are the model parameters corresponding to the mapping from a previous hidden state to subsequent hidden state, from an input vector to a hidden state, and from a hidden state to an output vector, respectively. The model parameters ($\boldsymbol{W}_{\text{hh}}^{l}$ and $\boldsymbol{W}_{\text{qh}}^{l}$) have sequential invariance across each layer, i.e., the input vector and hidden state at each step along the sequence are processed by the same parameters within a given layer of the RNN architecture. For the first layer, $\boldsymbol{h}_{i}^{l-1}$ is equivalent to the input vector $\boldsymbol{q}_{i}$, while for subsequent layers, $\boldsymbol{h}_{i}^{l-1}$ denotes the hidden state from the previous layer at the current step. In this manner, a multilayer RNN transmits the information encoded in the hidden state to the next step in the current layer and to the current step of the next layer by implementing Eq.~\ref{eq:rnn_A}. For the sake of brevity, the bias terms have not been mentioned in these equations.

Like deep feed-forward neural networks, deep RNNs can lead to more expressive models. However, the depth of an RNN can have multiple interpretations. In general, RNN architectures with multiple RNN layers stacked over each other are considered \emph{deep RNNs}, as shown in Fig.~\ref{fig:rnn_architecture}. Such deep RNN architectures have multiple internal memories (hidden states), one in each RNN layer. RNN architectures with multiple recurrent hidden states, stacked over each other, can model varying levels of dependencies (i.e., from short-term to long-term) in each hidden state~\citep{hermans2013}. These stacked-RNN architectures can still be considered shallow networks with limited expressivity, as all model parameters ($\boldsymbol{W}_{\text{hh}}$, $\boldsymbol{W}_{\text{qh}}$, and $\boldsymbol{W}_{\text{hy}}$) are generally represented by single linear layers. To allow for more complex functional representation, these single linear layers can be replaced by multiple nonlinear layers. \cite{pascanu2014} has shown that introducing depth via multiple nonlinear layers to represent $\boldsymbol{W}_{\text{hh}}$, $\boldsymbol{W}_{\text{qh}}$ and $\boldsymbol{W}_{\text{hy}}$ can lead to better expressivity of the RNN model. For the transition modeling problem addressed in this paper, multiple nonlinear layers are used within each RNN cell to express the complex physical mapping between the input features and output, as shown in Fig.~\ref{fig:rnn_cell}. Such architecture resulted in better learning and predictive capability of the RNN model.
\begin{table}[tbp]
\centering
\caption{Input features and output for the RNN model. \label{features}}
\begin{tabular}{>{\centering\arraybackslash}m{2cm}  >{\centering\arraybackslash}m{8cm}  >{\centering\arraybackslash}p{3.2cm}} \hline
\textbf{Feature/Output} & \textbf{Description} & \textbf{Definition} \\ \hline \hline
$q_1$ & Reynolds number based on edge velocity and momentum thickness & $Re_{\theta}$ \\
$q_2$ & Velocity profile as a function of wall normal coordinate y  & $ u_j$, $j=1, 2, \dots, 41$ \\
$q_3$ & First-order derivative of velocity profile & $\left. \frac{du}{dy}\right\vert_j$, $j=1, 2, \dots, 41$ \\
$q_4$ & Second-order derivative of velocity profile & $\left. \frac{d^2u}{dy^2}\right\vert_j$, $j=1, 2, \dots, 41$ \\ \hline
$y_1$ & Slope of N-factor envelope, corresponding to local growth rate of the most amplified disturbance at that location & $dN_\text{env}/ds$ \\
\hline
\end{tabular}
\end{table}

The underlying physics of transition doesn't require long term memory, unlike Natural language processing (NLP) for which more involved models like long short-term memory (LSTM) and transformers have proved to be very effective. For the transition problem, keeping track of last one or two stations have proved to be sufficient. In a study at the start of this research work, an informal investigation showed no advantage of LSTMs over RNNs, despite their added complexity and higher training cost.

With this perspective, the RNN model being proposed in this paper maps the sequential dependency of mean boundary-layer flow properties as input features to instability growth rates corresponding to the N-factor envelope as output features. Such input and output features, summarized in Table~\ref{features}, have been taken at a sequence of stations along the airfoil surface as shown in Fig.~\ref{fig:rnn_architecture}. Mean boundary-layer flow properties have been introduced in terms of the Reynolds number ($Re_{\theta}$) based on the local momentum thickness of the boundary layer, the velocity boundary-layer profile ($u$), and its derivatives ($du/dy$ and $d^2u/dy^2$). \cite{zafar2020} proposed a convolutional neural network model that encodes the information from boundary-layer profiles to a vector of latent features while accounting for the spatial patterns in the input profiles~\citep{wu2018seeing,carlos2018}. Such a treatment of boundary-layer profiles allows the trained neural network models to generalize to all practical flows with different instability mechanisms~\citep{paredes2020_journal}. 

\begin{figure}
    \centering
    \includegraphics[width=0.99\textwidth]{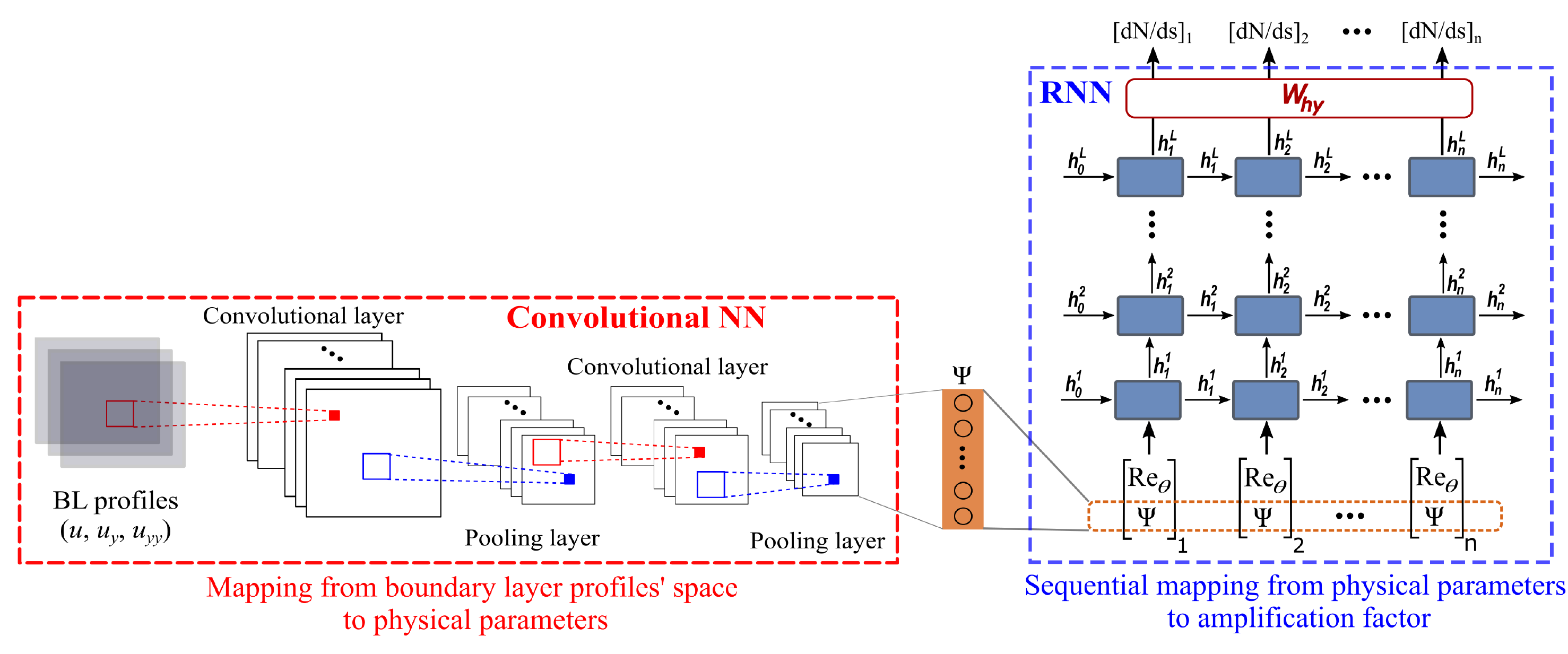}
    \caption{Proposed neural network for transition modeling. Convolutional neural network encodes the information from boundary-layer profiles ($u, u_y, u_{yy}$) into latent features ($\Psi$) at each station. RNN processes the input features ($Re_{\theta}$ and $\Psi$) in sequential manner to predict the growth rate ($dN/ds$) of the N-factor envelope}
    \label{fig:rnn_model}
\end{figure}

The RNN model presented in this paper builds upon this idea and further combines CNN with RNN to account for nonlocal physics in both streamwise and wall-normal directions. This is shown in Fig.~\ref{fig:rnn_model}. 
The hyperparameters of the proposed neural network have been empirically tuned to yield adequate complexity for learning all the required information, without causing an overfitting of the training data. After such hyperparameters tuning of the neural network model, early stopping was not required. With the boundary-layer profiles defined by using 41 equidistant points in the wall-normal direction, the CNN architecture contains 3 convolutional layers with 6, 8, and 4 channels, respectively, in those layers. Kernel size of 3$\times$1 has been used in each convolutional layer. Rectified Linear Unit (ReLU) is used as the activation function.
The CNN encodes the spatial information in the boundary-layer profiles along each station to a vector of latent features, $\Psi$. The results are not significantly sensitive to the number of latent features in the vector $\Psi$.  However, following sensitivity study on the size of latent features, the number of elements in $\Psi$ has been set to 8. The latent features $\Psi$ extracted from the boundary-layer profiles are then concatenated with $Re_{\theta}$ at each station, which provides the sequential input features for the RNN architecture to predict the local growth rate of the most amplified instability mode, or equivalently, the slope ($dN/ds$) of the N-factor envelope at a sequence of stations along the airfoil surface. Each RNN cell (Fig.~\ref{fig:rnn_cell}) consists of nonlinear mappings from the input to the hidden state ($\boldsymbol{W}_{\text{qh}}$) and from the hidden state to the output ($\boldsymbol{W}_{\text{hy}}$), with each mapping involving two hidden layers with 72 neurons each. The rectified linear activation function (ReLU) is used to introduce nonlinearity in these layers. The hidden state is represented by a vector of length 9 and a linear layer is used for the mapping $\boldsymbol{W}_{\text{hh}}$ between the hidden states. The RNN architecture consists of three RNN layers stacked over each other, with corresponding three internal memories (hidden states), each representing varying level of dependency (short to long term) between the output at the current station and the input at the current as well as the preceding stations.

As the CNN architecture is intrinsically dependent on the shape of the training data, the architecture can be tuned to different shapes of training data and the proposed model is expected to maintain its efficiency and accuracy. Future work will explore the vector-cloud neural network~\cite{zhou2021frame} which can deal with any number of arbitrarily arranged grid points across the boundary layer profiles. Since empirical tuning of hyperparemeters provided good results, we did not undertake an extensive optimization of the whole model. In a related unpublished work, more extensive hyperparameter optimization resulted in minor adjustments of the CNN architecture with comparable results.

\section{Database of Linear Amplification Characteristics for Airfoil Boundary Layers \label{database}}
A large database of the linear stability characteristics of two-dimensional incompressible boundary-layer flows over a broad set of airfoils was generated for the training and evaluation of the proposed model.
These boundary-layer flows can support the amplification of Tollmien-Schlichting (TS) waves and the most amplified TS waves at any location along the airfoil correspond to two-dimensional disturbances (i.e., spanwise wavenumber $\beta = 0$). This database documents the amplification characteristics of unstable TS waves under the quasiparallel, no-curvature approximation. A value of $N_\text{tr}=9$ has been empirically found to correlate with the onset of laminar-turbulent transition in benign freestream disturbance environments characteristic of external flows at flight altitudes. The airfoil contours were obtained from public domain sources, such as the UIUC Airfoil Coordinates Database \citep{uiuc}.
Linear stability characteristics for laminar boundary-layer flows were computed using a combination of potential flow solutions~\citep{drela1989} and a boundary-layer solver~\citep{wie1992}. The computational codes used are industry standard and have been used in number of research works over the years. Inviscid computations using panel method have been performed with 721 points around the airfoil. For boundary layer solver, 300-400 grid points have been considered in wall normal direction using a second order finite difference scheme. We note that the focus of this database is on transition due to TS waves in attached boundary layers, and therefore, flows involving a separation bubble (which cannot be computed with the viscous-inviscid-interactive procedure adopted herein) are not considered in the present work.

\begin{figure}
    \centering
    \includegraphics[width=0.97\textwidth]{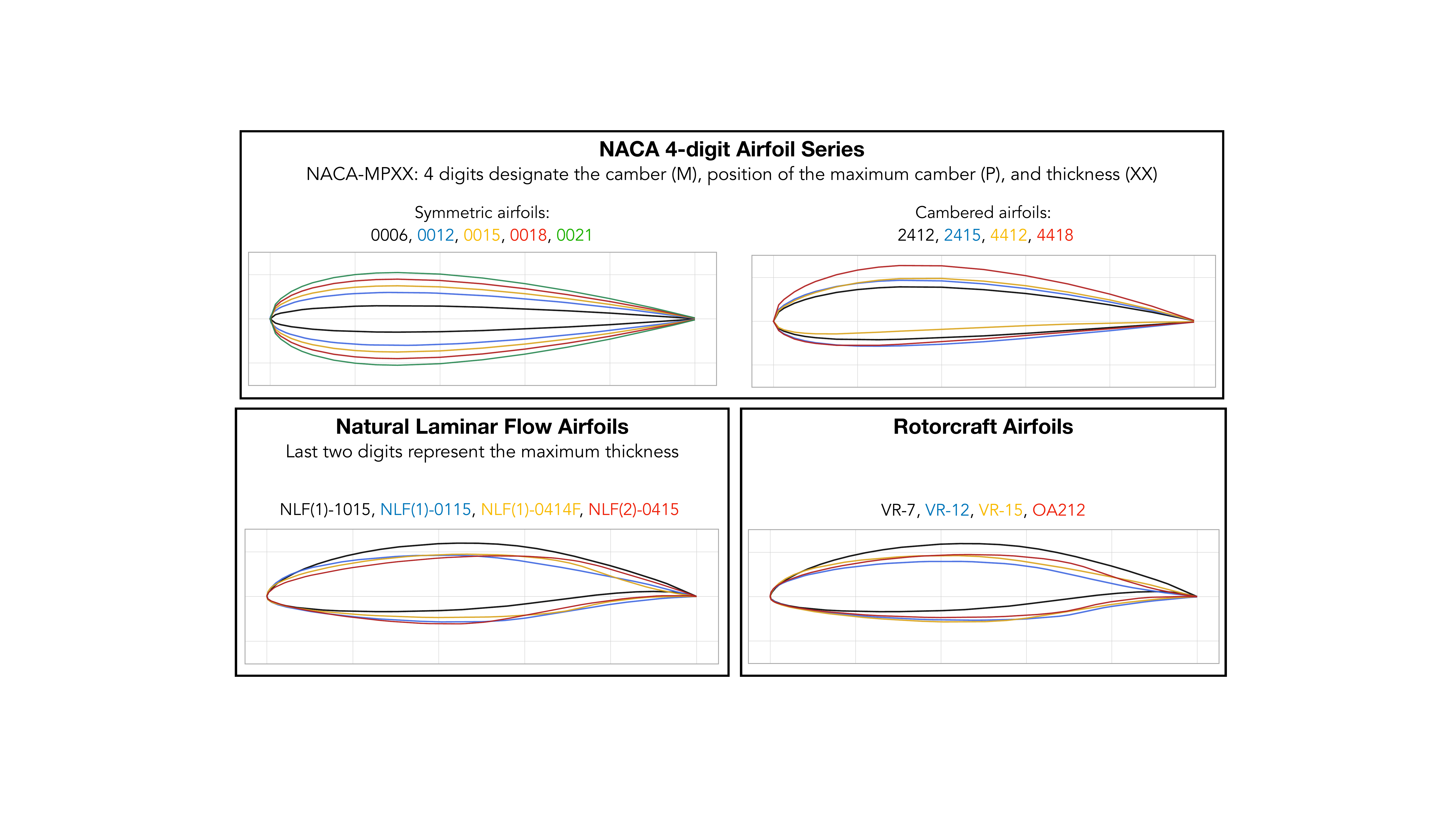}
    \caption{Airfoil sections for three airfoil families in the database. A complete list of airfoils along with their geometries is given in Appendix~\ref{airfoils_list}}
    \label{fig:airfoils}
\end{figure}

Airfoil contours included in the database belong to different categories and were selected randomly to cover a range of practical applications. These categories include different series of NACA airfoils, natural laminar flow airfoils, low Reynolds number airfoils, rotorcraft airfoils, and airfoil contours designed for transonic flows, etc. Selected airfoils from three of these categories have been plotted in Fig.~\ref{fig:airfoils}, which illustrates the markedly different airfoil contours included in the database. Data corresponding to both upper and lower surfaces of the airfoils has been considered, except for the symmetrical airfoil sections for which the lower-surface data have been excluded to avoid duplication, and hence, a bias in the data sampling. For reference, all 53 airfoils from the database are listed as well as plotted in Appendix~\ref{airfoils_list}. The range of chord Reynolds numbers ($Re_c$) included in the database extends over nearly four orders of magnitude ($[5\times 10^4, 2\times 10^8]$) and a broad range of angles of attack (AOA) $[-6^\circ, 8^\circ]$ has been considered for each of these airfoils. However, some of the boundary layers within the above range of Reynolds numbers and angles of attack are either stable or only weakly unstable (i.e., corresponding to a rather small peak N-factor, $N < 3$). Those flows were excluded from the database, still yielding a total of 31,247 flow cases corresponding to the 53 airfoils in this database. Although the computational cost of generating such database using is only few hours, the associated human hours are significantly higher since the process requires manual interventions and expert judgements to avoid spurious modes in linear stability computations. Furthermore, pre-processing of geometrical data to ensure smooth surface curvature also added significant human cost to the database generation.


The database documents the mean flow features and the relevant linear stability characteristics in a sequential manner along 121 streamwise stations, starting from a station close to the onset of instability
and extending up to either the point where the N-factor envelope reaches $N_\text{env}=25$ or to the end of the chordwise domain (which can be upstream of the trailing edge if the boundary-layer solution terminates due to an incipient flow separation). 
We keep the sequence length fixed at 121 for all airfoils and flow conditions, which makes it more efficient in handling the data during the training and testing of the RNN model. The location of each station is defined in terms of the arc length along the airfoil surface ($s$). 
Besides the parameters given in Table~\ref{features}, sequential information for several other relevant parameters has also been included in the database, such as the chordwise location of each station ($x/c$), 
local edge velocity ($U_e$), local boundary-layer edge density ($\rho_e$) and viscosity ($\mu_e$), boundary-layer momentum thickness ($\theta$), etc. 

The present work is aimed at developing an RNN model, which is trained over a subset of the complete dataset, and has the ability to predict the transition location for any boundary-layer flow from the complete dataset with reasonable accuracy. Computational constraints limit the size of the training dataset for RNNs, since they are more expensive to train as compared to simple feedforward neural networks. Reducing the size of the training data via subsampling of the overall database would require the sampling process to avoid any bias toward any specific subset of the database. Such bias can have a dominant effect on the efficacy of the loss function used for training, resulting in a potential overfitting across certaint parts of the training data, and worse predictive performance for other subsets of the database. Hence, a large database of this type requires adequate sampling procedures for the selection of the training data so that the resulting model can provide a balanced representation of the entire database.

\section{Results \label{results}}
The proposed RNN model predicts the sequence of growth rates of the most amplified disturbances, i.e., the slope values of the N-factor envelope as a function of distance along the airfoil contour.  The N-factor envelope can then be determined as the cumulative integral over this sequence, with the lower limit of integration corresponding to the airfoil location where the boundary-layer flow first becomes unstable (or, equivalently, the station across which the slope first changes in sign from a negative value to a positive one).  The transition onset location can then be estimated as the location where the envelope reaches the critical N-factor determined via correlation with a relevant set of measurements. The sequential data are defined at a fixed number of stations for each flow, but the physical domain length can vary from case to case due to the potential onset of flow separation in a laminar boundary-layer computation.  
The loss function used for the training process includes a weighting function corresponding to the cell size ($d\text{s}$) in the vicinity of each station. Specifically, the loss function used for training the neural network is defined in terms of a weighted sum over the square of the local error:
\begin{equation}
\mathcal{L} = \sum_{j=1}^{m}\left(l_j \right) \quad \quad \text{where} \quad \quad l_j= \sum_{i=1}^{n}\left((Y_{i}-\hat{Y}_{i})^{2}\cdot ds_i\right)
\end{equation}
where $m$ denotes the number of sequences in the dataset, $n$ denotes the number of stations in a sequence, and $Y$ and $\hat{Y}$ represent the true and predicted values, respectively, corresponding to the output quantity. The weighting function serves to reduce the bias due to a nonuniform streamwise grid used in the boundary-layer calculation. 

The primary performance indicator of the proposed model is the prediction of the chordwise location of the laminar-turbulent transition. However, besides its obvious dependence on the stability characteristics of the airfoil boundary layer, the transition location also depends on the disturbance environment via the correlating N-factor value, $N_\text{tr}$. The measured transition locations in a broad range of flows typically correlate with a finite band of N-factor values in the vicinity of $N_\text{tr} = 9$ under a benign disturbance environment, such as those encountered by external flows at typical flight altitudes. For this reason, the predictive performance has been assessed considering the flow cases for which N-factor envelope reaches values of upto $N_\text{env} = 13$. To help provide a meaningful assessment of the model accuracy across a broad range of flows and correlating N-factors, three different error metrics have been defined separately from the loss function, as indicated below.
\begin{equation}
E_\text{env} = 100 \times \frac{1}{\tilde{m}} \sum_{j=1}^{\tilde{m}}\left( \frac{||N_\text{env} - \hat{N}_\text{env}||_{ds}}{||N_\text{env}||_{ds}} \right)_j
\label{eqn:env_error}
\end{equation}

\begin{equation}
E_\text{tr} = 100 \times \frac{1}{\tilde{m}} \sum_{j=1}^{\tilde{m}}\left\vert \frac{x/c_\text{tr} - \tilde{x}/c_\text{tr}}{x/c_\text{tr}} \right\vert_j
\label{eqn:rel_tr_error}
\end{equation}

\begin{equation}
E_{x/c} = 100 \times \frac{1}{\tilde{m}} \sum_{j=1}^{\tilde{m}}\left\vert x/c_\text{tr} - \tilde{x}/c_\text{tr} \right\vert_j
\label{eqn:abs_tr_error}
\end{equation}
where $\tilde{m}$ denotes the number of sequences in the dataset for which the N-factor envelope reaches values of upto $N_\text{env} = 13$. The first error metric ($E_\text{env}$) is based on the $L2$ norm to evaluate the accuracy of the predicted N-factor envelope ($N_\text{env}$), determined by integrating the predicted slope values $dN/ds$. To emphasize a finite band of N-factor values in the vicinity of $N_\text{tr} = 9$, only the range of $5<N_\text{env}<13$ has been considered for each flow case.
The second error metric ($E_\text{tr}$) corresponds to the relative discrepancy between the true and predicted chordwise locations of transition onset for the case of $N_\text{tr}=9$. 
The third error metric ($E_{x/c}$) relates to the absolute error in the predicted transition location, scaled by the airfoil chord length.

\subsection{Demonstration of predictive performance}
Selection of training data for the development of a general purpose model for the airfoil universe requires a balance between multiple requirements that may conflict with each other.  It is clearly desirable for the size of the training data to be moderate enough to minimize the training cost.  However, the training data must also be large enough in scope to represent the broad application space and must be designed to avoid an unfair bias toward specific subregions from the parameter space of latent features. Translating these requirements into a practical procedure is not a straightforward task. Given the availability of the large database of stability characteristics as described in the previous section, we have evaluated several different strategies for the selection of an appropriate subset of that database for the training process.

We begin by using a smaller portion of the available database for training purposes.  This baseline case is representative of less ambitious efforts at database generation, as well as being better suited for the case involving a broader application space that includes additional flow parameters such as, for instance, nonzero Mach numbers, nonzero surface transpiration, and surface heating/cooling, etc. The baseline training set consists of five out of the total 53 airfoils, with each of these five airfoils representing a different subgroup of airfoils from Table~\ref{tab:list}, namely, the
$$ \text{NACA 0012, \ ONERA M6, \ NACA 2412, \ NACA 63-415, \ NLF(1)-0416} $$ airfoils.  These airfoils correspond to airfoil indices of 2, 44, 6, 15, and 25, respectively.  Here, the first two airfoils have symmetric contours whereas the latter three correspond to asymmetrical airfoil sections. Table~\ref{aoa_re} lists the various flow conditions at which boundary-layer solutions are available for each airfoil in the database. To assess the RNN model for interpolation and extrapolation with respect to the angle of attack and chord Reynolds number, respectively, flow conditions marked in red have been included in the testing dataset, while the remaining flow cases constitute the training dataset. Such an arrangement of the total available cases from these five airfoils results in a 60\%--40\% split between the training and testing data.

\begin{table}[tbp]
\centering
\caption{Flow conditions for all the cases in the airfoil database. For evaluation of the RNN model, flow conditions used for model testing are marked in red color whereas the flow conditions used for training are indicated in black color. \label{aoa_re}}
\begin{tabular}{|>{\centering\arraybackslash}m{0.34\textwidth} |  >{\centering\arraybackslash}p{0.5\textwidth}|} \hline
\textbf{Angles of Attack} (deg) & \textbf{Reynolds Numbers} \\ \hline
{\small \textcolor{red}{$-6^\circ$}, $-5^\circ$, $-4.5^\circ$, $-4^\circ$, $-3.5^\circ$, \textcolor{red}{$-3^\circ$},} & {\small \textcolor{red}{$3.5\times10^4$}, $5.0\times10^4$, $7.0\times10^4$, $1.0\times10^5$, $1.4\times10^5$,} \\
{\small $-2.5^\circ$, $-2^\circ$, $-1.5^\circ$, $-1^\circ$, $-0.5^\circ$, \textcolor{red}{$0^\circ$},} & {\small \textcolor{red}{$2.8\times10^5$}, $4.0\times10^5$, $5.6\times10^5$, $8.0\times10^5$, $1.1\times10^6$,} \\
{\small $0.5^\circ$, $1^\circ$, $1.5^\circ$, $2^\circ$, $2.5^\circ$,  \textcolor{red}{$3^\circ$}, $3.5^\circ$, $4^\circ$,} & {\small $1.6\times10^6$, $2.3\times10^6$, $3.2\times10^6$, \textcolor{red}{$4.5\times10^6$}, $6.4\times10^6$,} \\
{\small $5^\circ$, $6^\circ$, $7^\circ$, \textcolor{red}{$8^\circ$}} & {\small $9.0\times10^6$, $1.3\times10^7$, $1.8\times10^7$, \textcolor{red}{$3.6\times10^7$}, $5.1\times10^7$,} \\
{\small } & {\small $7.2\times10^7$, $1.0\times10^8$, \textcolor{red}{$1.4\times10^8$}} \\ \hline
\end{tabular}
\end{table}

\begin{figure}
    \centering
    \includegraphics[width=0.77\textwidth]{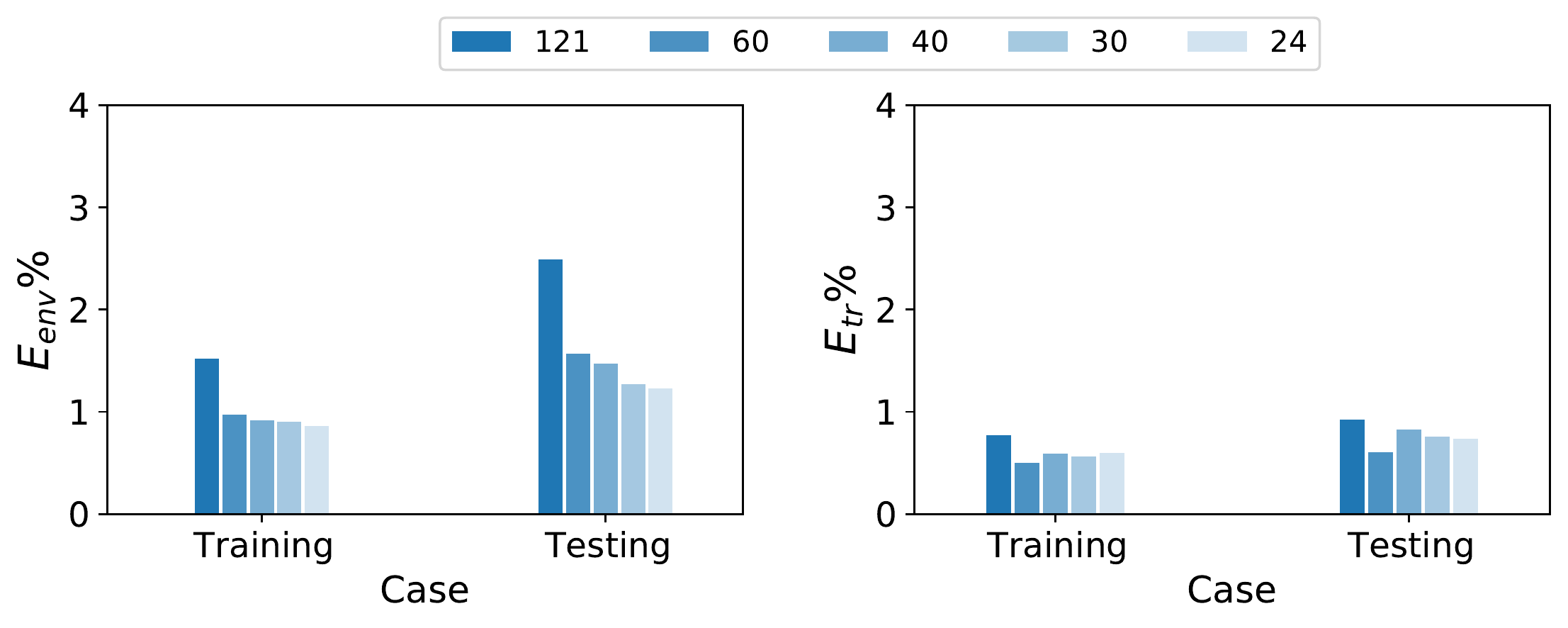}
    \caption{Comparison of prediction error percentage for training and testing datasets with different sequence lengths. Training and testing datasets have been defined based on flow conditions as given in Table~\ref{aoa_re}}
    \label{fig:seq_len}
\end{figure}

The sequential information corresponding to the evolution of mean boundary-layer profiles along the airfoil surface has been documented in the database, with a uniform sequence length of 121 stations for all flow cases. An initial assessment was conducted to ascertain the effect of different sequence lengths. The results of this assessment are shown in Fig.~\ref{fig:seq_len}, wherein the error metrics defined in Eqs.~\ref{eqn:env_error} and \ref{eqn:rel_tr_error} have been plotted for different sequence lengths. Significant improvements can be observed by reducing the sequence length from 121 to 60, the reason for which is not entirely clear. It could be one of the areas in future investigations of such models.  In general, however, this trend may be related to the fact that more information sometimes lead to diluting of information and, consequently, to worse results. 
Further shortening of the sequence length yields relatively little benefit in terms of model accuracy for either training or test data. In fact, the accuracy of predicting the transition location worsens when the sequence length is reduced beyond 60. This observation is likely to be related to a poorer resolution of the shape of the N-factor envelope achieved when fewer stations are used across the same physical domain length along the airfoil surface. Looking at the similar trends for both error metrics, it was decided that a sequence length of 60 stations would provide an optimal choice for all of the results to be presented in this paper. Fig.~\ref{fig:seq_len} shows that the predictive performance for the testing dataset is comparable to that for the training dataset, demonstrating the interpolating and extrapolating capability of the proposed RNN model with respect to both AOA and $\text{Re}_c$.

\begin{figure}
    \centering
    \includegraphics[width=0.42\textwidth]{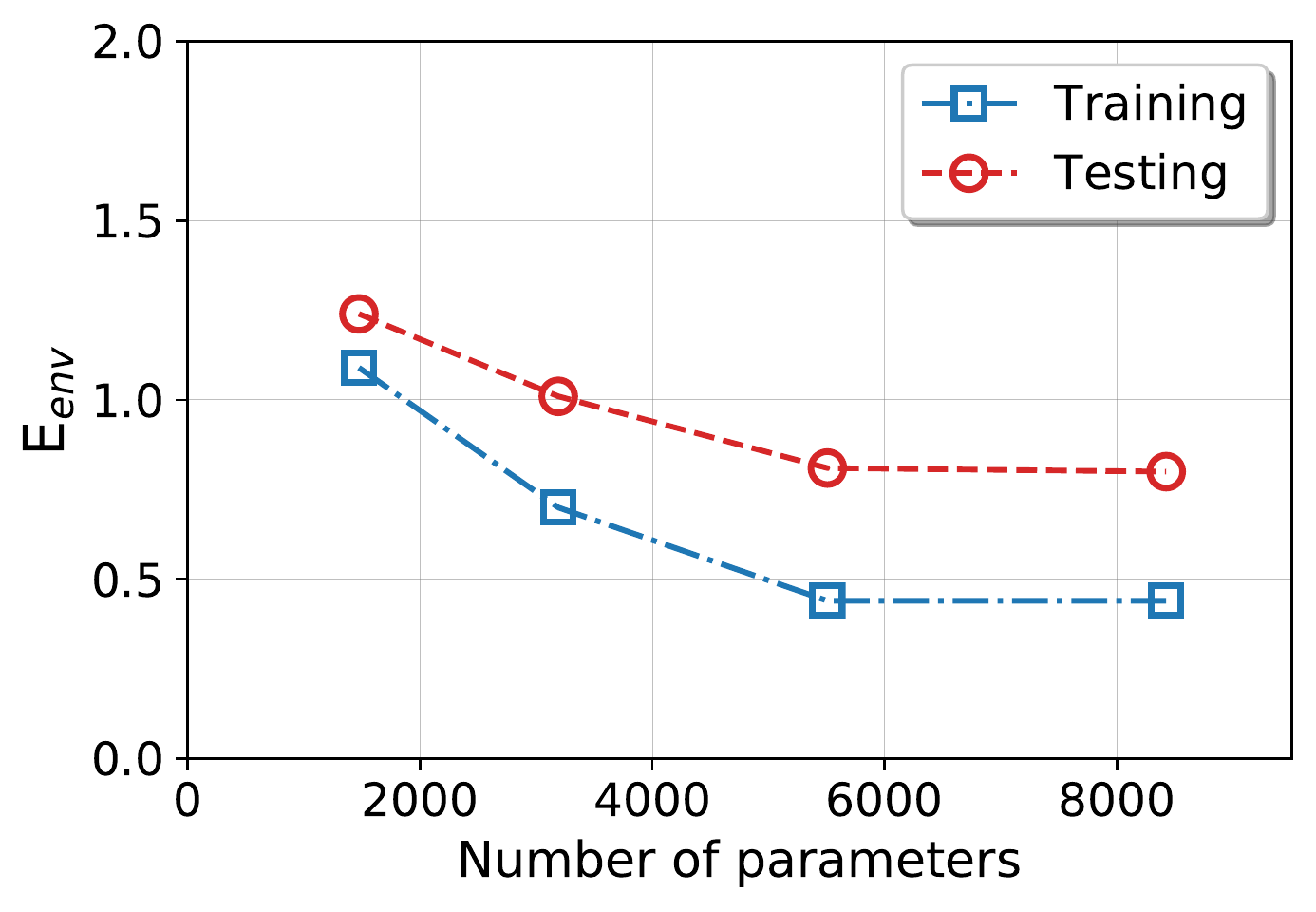} 
    \caption{Training and testing errors for a range of sizes of the RNN model indicated by the number of learnable parameters. The number of layers is kept the same while the parameters are varied proportionately in all three mappings, $\boldsymbol{W}_{\text{hh}}$, $\boldsymbol{W}_{\text{qh}}$, and $\boldsymbol{W}_{\text{hy}}$}
    \label{fig:parameters}
\end{figure}

Next, we assess the effect of the size of the RNN model on the prediction error. Figure~\ref{fig:parameters} displays the variation in error percentage as a function of the number of learnable parameters in the RNN model. While the error metric $E_\text{env}$ decreases as the number of learnable parameters is increased up to 5500, the error remains nearly constant with a further increase in the number of parameters. One may deduce from these results that an RNN model with 5500 learnable parameters provides near-optimal learning capability without causing overfitting. Consequently, this model size will be maintained for all the results presented for the current dataset. We note that the training dataset for this baseline case is of much smaller size in comparison to the complete airfoil database and that the use of a larger training dataset will most likely require a larger number of learnable parameters to enhance the learning capacity of the RNN model. Thus, the selection of model size will be discussed again when we work with somewhat larger training datasets in the following subsections.

The architecture of the proposed neural network model has direct correlation with the underlying physics of flow transition. Previously proposed fully-connected neural network based transition models~\citep{fuller1997, crouch2002} didn't distinguish the evolution of flow in wall-normal and streamwise directions. At any station along airfoil surface, propagation of boundary layer flow in wall-normal direction is instantaneous (analogous to elliptic behavior of diffusion equation). Hence, \cite{zafar2020} used CNN to encode the information from the boundary layer profiles in wall-normal direction in a vector of latent features. Such treatment of boundary layer profiles provides a stronger correlation with the underlying physics of the flow and also allows the application of CNN based transition model to various instability mechanisms~\citep{paredes2020_journal}. These characteristics were clearly lacking in previously proposed neural network based transition models. 

\begin{figure}
    \centering
    \includegraphics[width=0.8\textwidth]{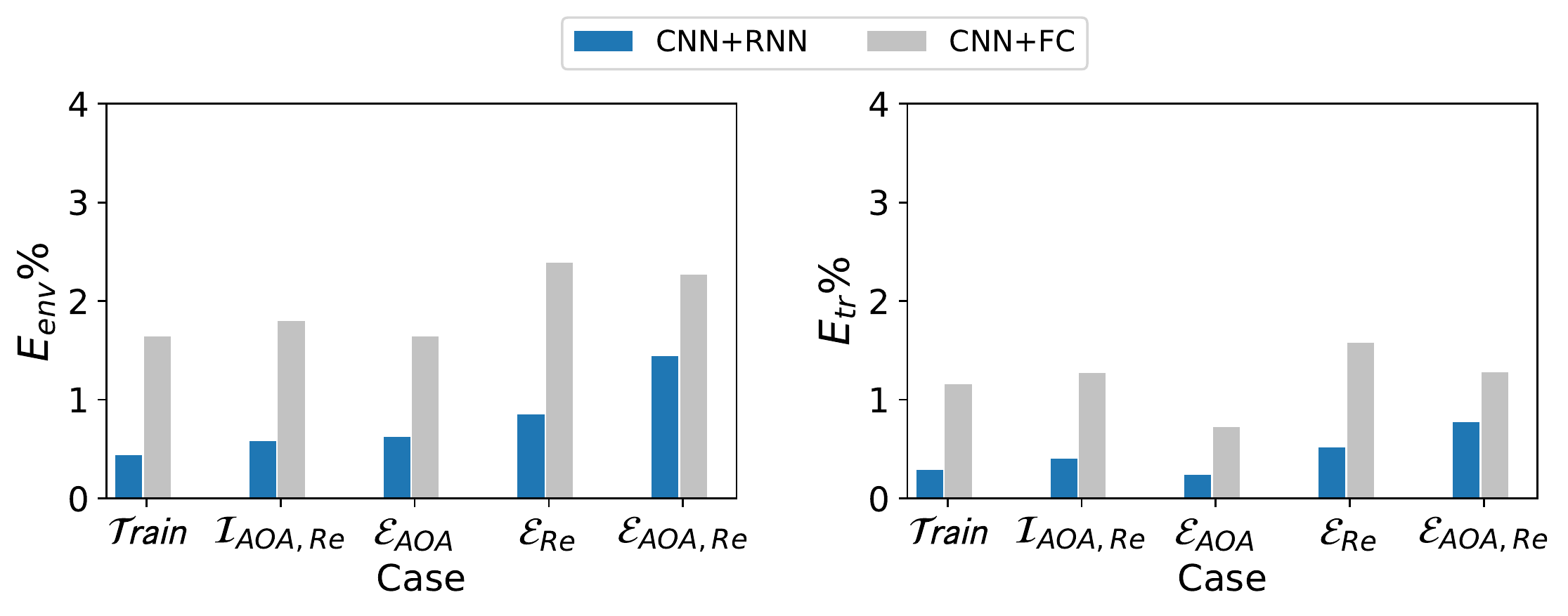} 
    \caption{Comparison of results for the recurrent neural network and fully connected network. Testing cases have been sub-categorised as: Interpolation with respect to both AOA and Re ($\mathcal{I}_\text{AOA,Re}$), Extrapolation with respect to AOA ($\mathcal{E}_\text{AOA}$), Extrapolation with respect to Re ($\mathcal{E}_\text{Re}$), Extrapolation with respect to both AOA and Re ($\mathcal{E}_\text{AOA,Re}$)}
    \label{fig:rnn_vs_fc}
\end{figure}

The current work uses both CNN and RNN in tandem where the RNN has been used to encapsulate the underlying physics of streamwise evolution of flow instability along with CNN which process the boundary layer profiles in wall-normal direction. Along the streamwise direction, elliptic behavior has already been taken care of in the stability theory however hyperbolic nature of the instability amplification from upstream to downstream requires sequence-to-sequence modeling, for which RNN has been used. To assess the benefit of sequence transduction in the RNN model, its predictive performance has compared with that of a previously proposed model which which does not account for the sequence information among the inputs at different stations~\citep{zafar2020}.
For a fair comparison, almost equal number of learnable parameters were used for both networks. The comparison is presented in Fig.~\ref{fig:rnn_vs_fc}, wherein we include the error percentages for both training and testing datasets. The testing cases have been further categorised in terms of interpolation and extrapolation with respect to the flow conditions. The results corresponding to each dataset show a clearly superior predictive performance for the RNN model vis-a-vis the fully connected neural network.   

We also note that, in comparison to transition model based on fully-connected neural network, the RNN model has a moderately higher training cost associated to it. For the given training data (comprising of five airfoil dataset with sequence length of 60) and a given model size ($\sim$5500 learnable parameters), the RNN model took 29 GPU hours to train as compared to 19 GPU hours for the fully connected neural network. A single NVIDIA V100 GPU was used for training purposes.

In comparison to direct stability computations, the RNN model can estimate the transition location three to four orders of magnitude faster. Actual times for direct computations of linear stability can vary depending on various factors, including the specific flow and the level of resolution (in terms of the number of stations in a sequence and the number of frequencies used to determine the N-factor envelope). In that regard, the above estimate is believed to provide a reasonable, if somewhat conservative, estimate of the speed-up due to RNN.

\subsection{Evaluation of predictive performance for complete database}
We now perform a comparative assessment of the accuracy of the RNN models based on different selections of training datasets. These training datasets have been summarised in the Table~\ref{tab:cases}. 
The rationale behind the selection of each training dataset from Table~\ref{tab:cases} will be outlined in the course of the discussion of the results, especially as the results for Case I provide the baseline for the selection of training data for the subsequent cases. Since the analysis is focused on the performance of the RNN model in predicting the transition location over an arbitrary airfoil, no distinction has been made between the airfoils and flow conditions used for training and testing, and all the flow cases included in the training dataset are considered alongside the cases that were not used during training. Because the sizes of the training sets from Table~\ref{tab:cases} are comparable to each other and always less than 24 percent of the total available data, the error metric based on the entire database was deemed to be a meaningful measure of the model's predictive accuracy.

\begin{table}[tbp]
\centering
\caption{Summary of training dataset cases. Flow cases corresponding to the mentioned airfoils are included in the training dataset. \label{tab:cases}}
\begin{tabular}{|>{\centering\arraybackslash}m{1cm} | p{2.1cm} | p{7.5cm} | >{\centering\arraybackslash}m{1.7cm}|} \hline
\textbf{Index} & \textbf{\quad \ Label} & \textbf{\qquad \qquad \qquad Training Dataset} & \textbf{Number of flow cases} \\ \hline
I & Five airfoils & {\small NACA-0012, NACA-2412, NACA-63-415, NLF(1)-0416, ONERA-M6} & 2624 \\ \hline
II & Random augmentation & Case I + 100 random flow cases from each of the other airfoils & 7026 \\ \hline
III & Augmented airfoils set & Case I + Five more airfoils with largest mean error ({\small LRN(1)-1007, NACA-6712, NLF(1)-1015, NLF(2)-0415, CLARK-Y}) & 4455 \\ \hline
IV & Error-based augmentation & Case I + Specific flow cases of other airfoils with $E_{\text{env}} \% > 3$  in Case I & 5024 \\ \hline
V-A & Random selection (\%) & Randomly selected 20\% flow cases of each airfoil & 6233 \\ \hline
V-B & Random selection (\#) & Randomly selected 100 flow cases of each airfoil & 5300 \\ \hline
\end{tabular}
\end{table}

\begin{figure}
    \centering
    \subfloat[Mean error ($E_\text{env}$) percentage for N-factor envelope ]{\includegraphics[width=0.9\textwidth]{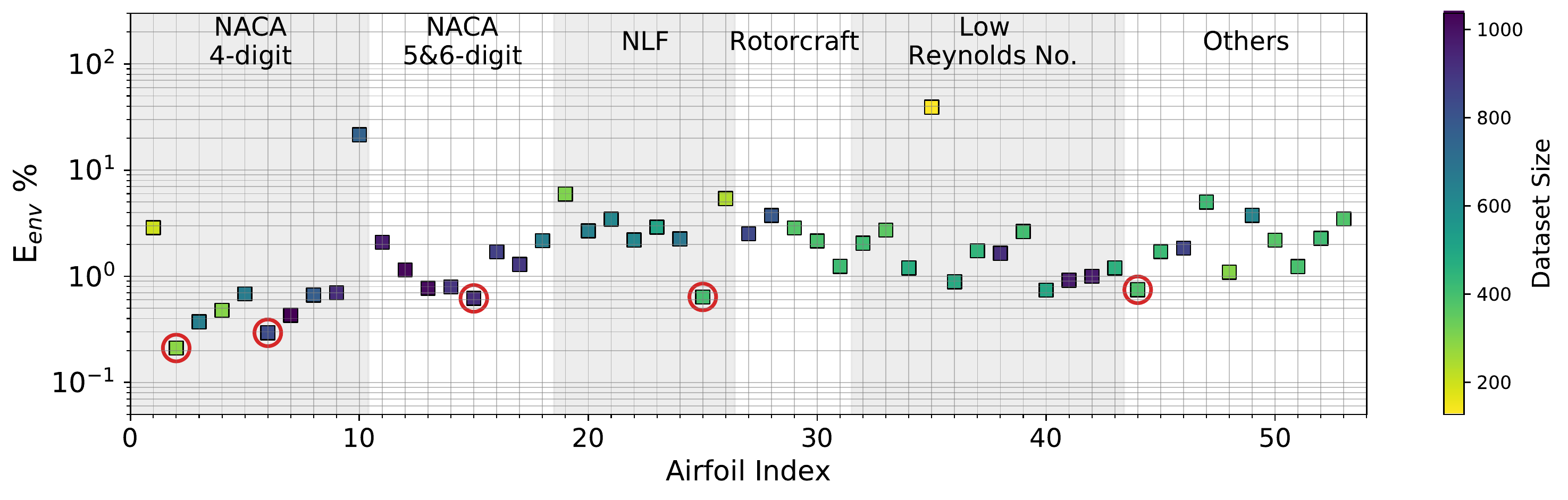}} \\
    \subfloat[Mean relative error ($E_\text{tr}$) percentage for transition location prediction ]{\includegraphics[width=0.9\textwidth]{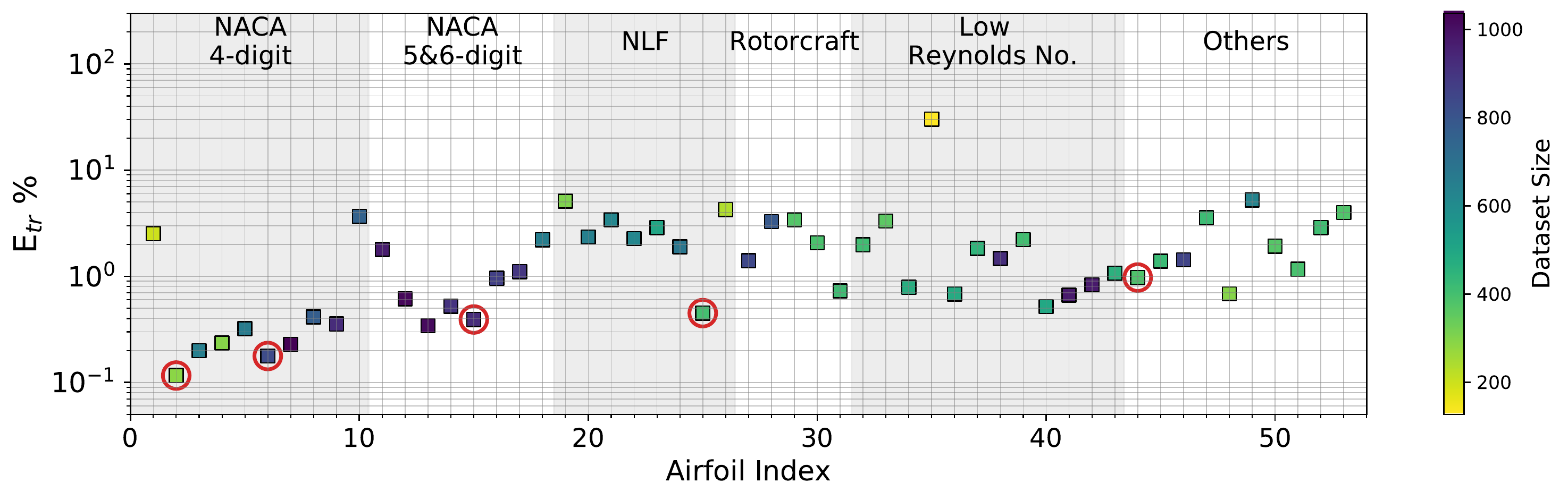}} \\
    \subfloat[Mean absolute error ($E_{x/c}$) percentage for transition location prediction ]{\includegraphics[width=0.9\textwidth]{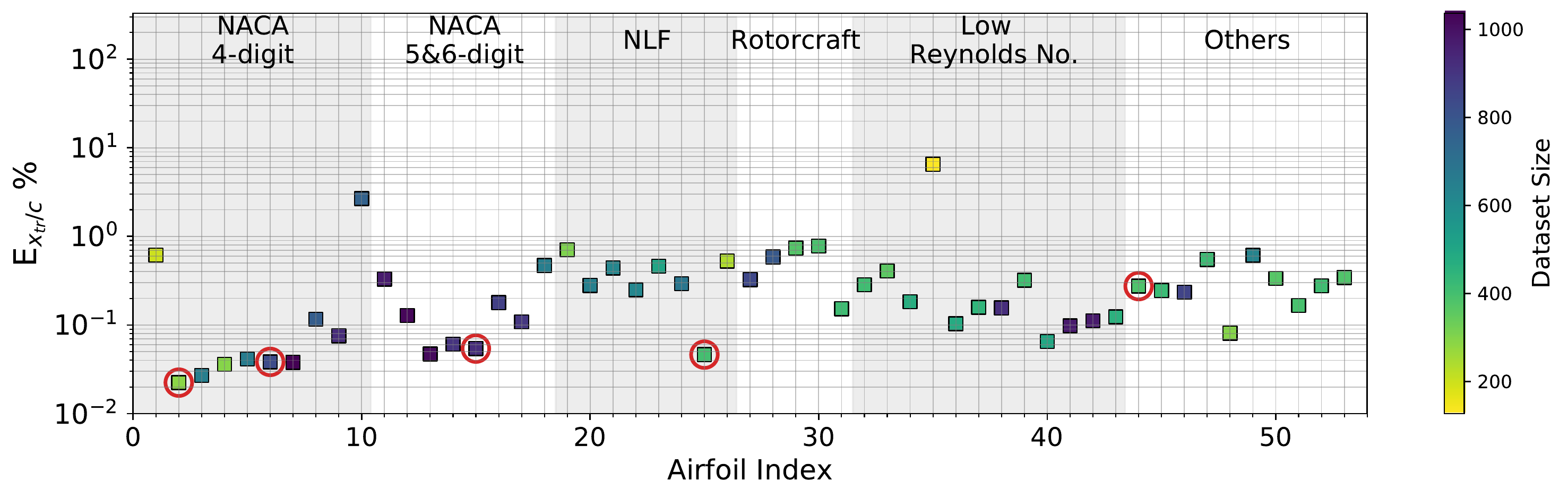}}
    \caption{Mean error percentage for each airfoil in the database, corresponding to training dataset of Case I (five airfoils) given in Table~\ref{tab:cases}. Airfoils corresponding to training dataset have been encircled in red color. Markers' color represent the dataset size (number of flow cases) of each airfoil in the database}
    \label{fig:error_db_1}
\end{figure}

\begin{figure}
    \centering
    \subfloat[Relative error ($E_\text{tr}$) percentage of flow cases at different angles of attack]{\includegraphics[width=0.77\textwidth]{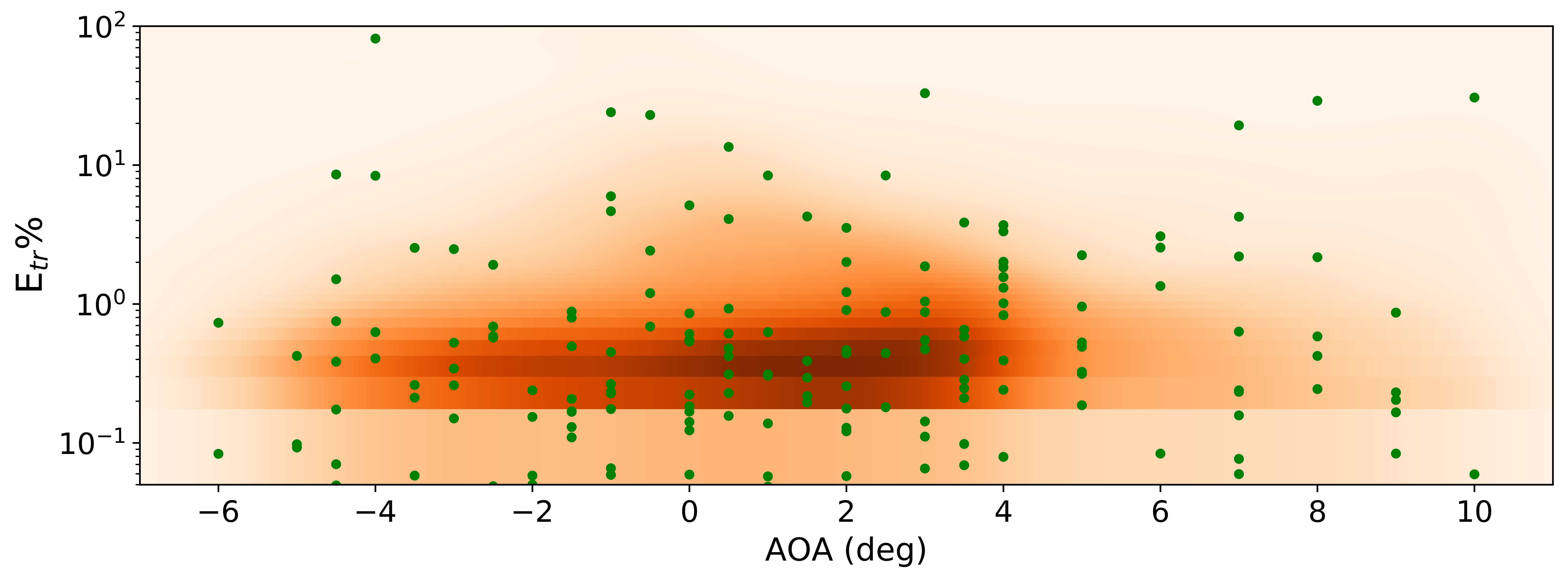}} \\
    \subfloat[Relative error ($E_\text{tr}$) percentage of flow cases at different Reynolds number]{\includegraphics[width=0.77\textwidth]{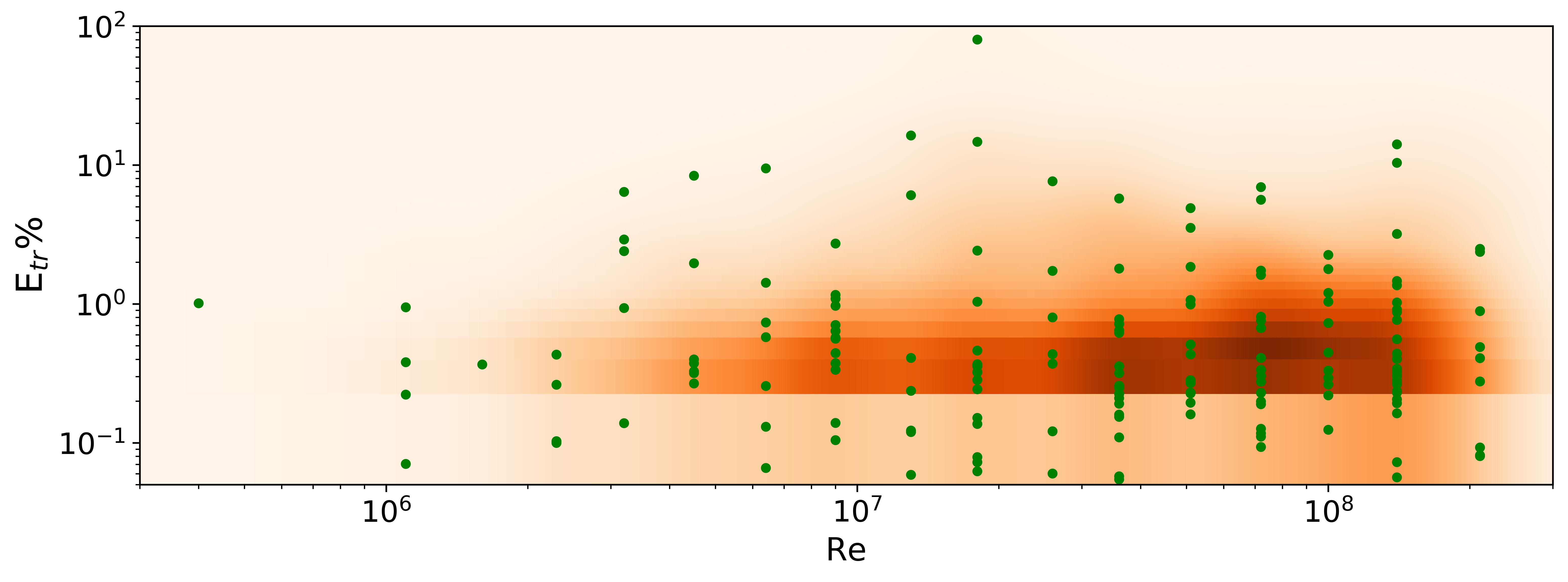}}
    \caption{Relative error ($E_\text{tr}$) percentage for all flow cases, corresponding to training dataset of Case I (five airfoils) given in Table~\ref{tab:cases}. Green markers (filled circles) show only 1\% of the randomly sampled flow cases. The contour shows the kernel density estimated from all the flow cases. Darker region indicates higher probability density. The horizontal lines appearing in the contour plots such as that near an error of 0.2\% are due to technical reason (bins have been defined in linear--scale while the vertical axis of the plot is in log--scale) and don't depict any real discontinuity}
    \label{fig:aoa_re_db_1}
\end{figure}

Case I (denoted as "Five Airfoils") from Table~\ref{tab:cases} involves a training dataset that is comprised of the same five airfoils that were used in the earlier assessment of the RNN model size, comparison with fully connected network, etc. Results for the RNN model trained using the Case I dataset are shown in Figs.~\ref{fig:error_db_1} and \ref{fig:aoa_re_db_1}. Figure~\ref{fig:error_db_1} presents the mean error percentages for the predicted N-factor envelope and transition location corresponding to all airfoils from the overall database. The figures have been shaded to distinguish between the different groups of airfoils belonging to the airfoil families included within the overall database. Airfoil names corresponding to the indices from Fig.~\ref{fig:error_db_1} are given in Table~\ref{tab:list}. In general, the mean error percentage for most of the airfoils is below about 3\%, which demonstrates the general capability of the RNN model based on the Case I training data. Even though the model has been trained with a significantly smaller subsample of airfoils from the overall database, it is still able to predict the $N_\text{env}$ and transition location for the entire set of airfoils including different categories with a reasonable accuracy. Laminar to turbulent transition due to TS amplification within the attached flow region is achieved at varying numbers of flow conditions across the different airfoils, and the markers for each airfoil in the figure have been colored on the basis of the dataset size of that airfoil.  This feature will be used to gain additional insights during the subsequent discussion as we describe the results for the remaining cases from Table~\ref{tab:list}.

Predictive performance as a function of the angle of attack and chord Reynolds number is shown in Fig.~\ref{fig:aoa_re_db_1}, which indicates the distribution of error percentage across the overall database via a color map for the kernel density estimate. In addition, 1\% of randomly sampled data points from the overall database have also been included as green dots within the figure. No bias in predictive errors toward specific flow conditions may be observed within the figure, indicating that the model is able to yield comparable accuracy across the entire range of flow conditions. The transition locations for most of the flow cases is predicted with a relative error percentage of ($E_\text{tr}$) $<$ 2\%, as shown by the higher density region with a darker color in the color map from Fig.~\ref{fig:aoa_re_db_1}. 

Results for Case I, shown in Fig.~\ref{fig:error_db_1}, indicate a few outlying airfoils corresponding to a high error percentage in the predictions of the RNN model. In particular, the average error in the prediction of the N-factor envelope for the LRN(1)-1007 airfoil is significantly higher ($E_\text{env}$>30\%) as compared to that for most of the other airfoils. LRN(1)-1007, designed for high lift and low drag at Re=$[5\times 10^4, 1.5\times 10^5]$, has a peculiar airfoil contour and aerodynamic behavior, which is markedly different with respect to the other airfoils in the training dataset. This may explain why the predictive error percentage for the LRN(1)-1007 airfoil is higher by almost an order of magnitude. Similarly, the NACA 6712 airfoil with a highly aft-cambered airfoil section also has a significantly higher error percentage ($E_\text{env}$>20\%). In comparison, the predictive performance for the other NACA airfoils is reasonably good, with the average absolute error in predicting the transition location below 1\% for most of those airfoils. 

\begin{figure}
    \centering
    \subfloat[Mean error ($E_\text{env}$) percentage for N-factor envelope ]{\includegraphics[width=0.9\textwidth]{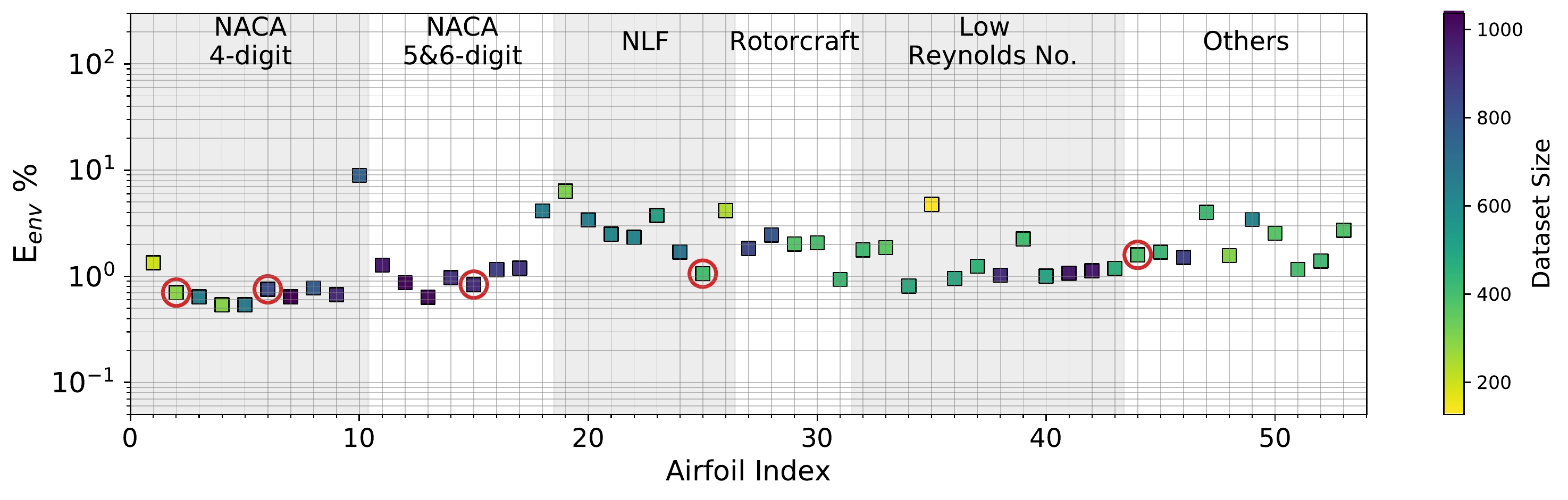}} \\
    \subfloat[Mean relative error ($E_\text{tr}$) percentage for transition location prediction ]{\includegraphics[width=0.9\textwidth]{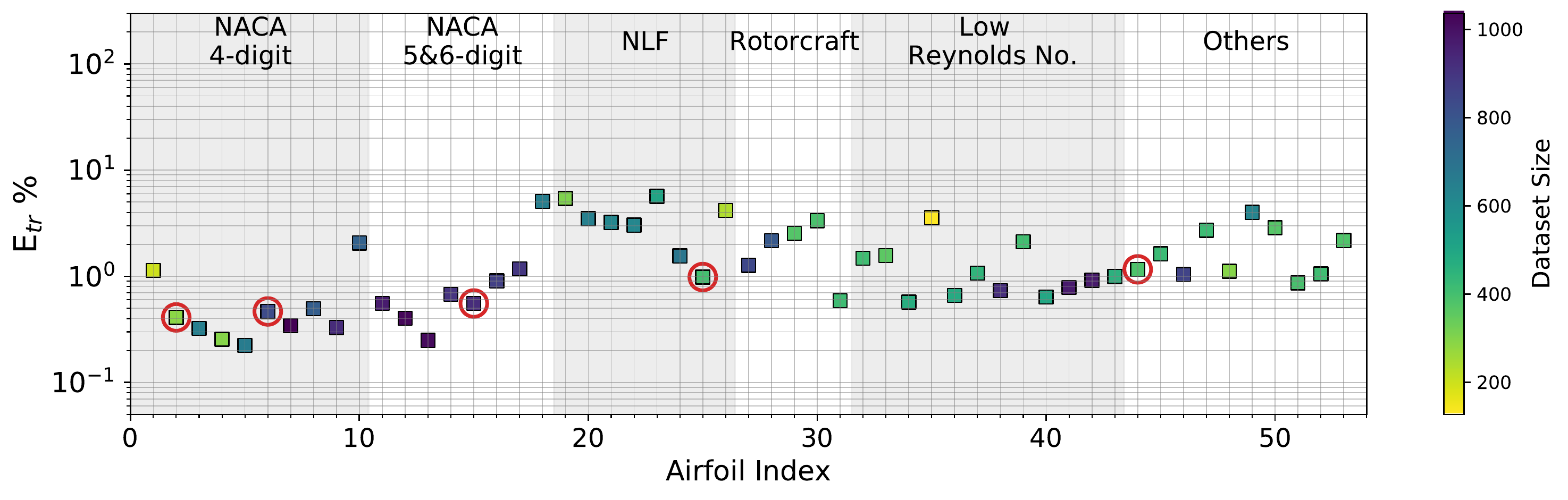}} \\
    \subfloat[Mean absolute error ($E_{x/c}$) percentage for transition location prediction ]{\includegraphics[width=0.9\textwidth]{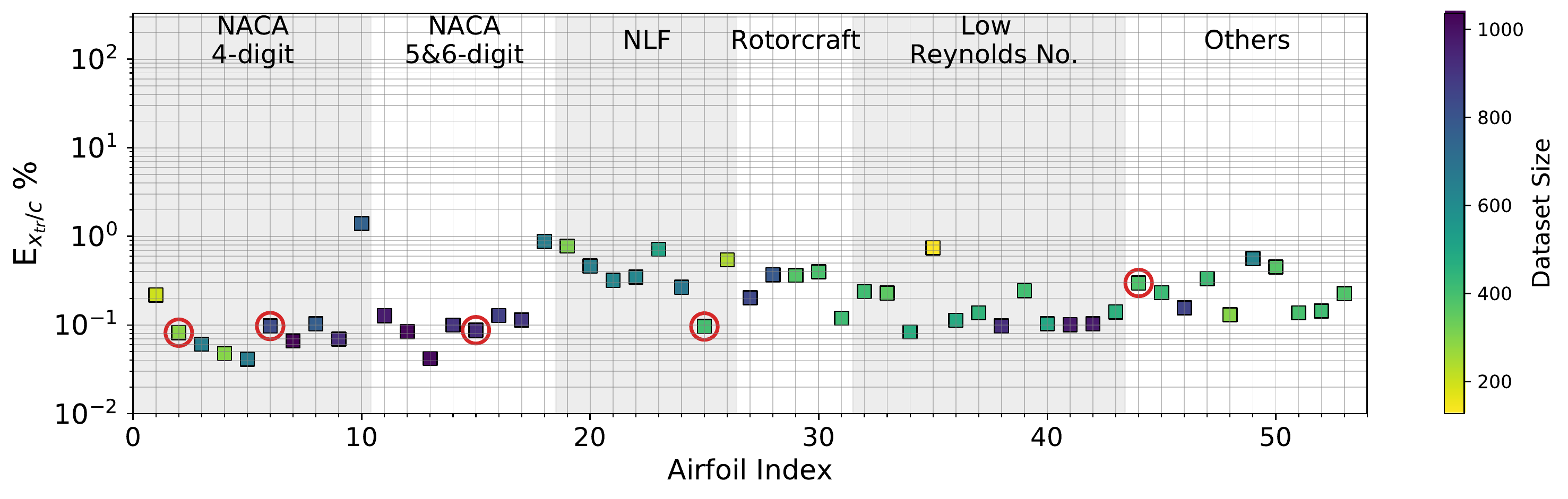}}
    \caption{Mean error percentage for each airfoil in the database, corresponding to training dataset of Case II (random augmentation) given in Table~\ref{tab:cases}. Airfoils corresponding to training dataset have been encircled in red color. Markers' color represent the dataset size (number of flow cases) of each airfoil in the database}
    \label{fig:error_db_2}
\end{figure}

Augmenting the training dataset from Case I with additional data provides the most obvious way of improving on the predictive performance of the above RNN model. Several different strategies for data augmentation were evaluated in the course of this work, and they are denoted as Cases II, III, and IV in Table~\ref{tab:list}. For Case II, 100 randomly selected flow cases from every other airfoil have been added to the training dataset from Case I.
Even though this data augmentation causes the size of the training dataset to increase almost threefold with respect to that in Case I, the inclusion of flow cases for every airfoil within the training dataset leads to significantly improved predictive performance of the RNN model. The results for Case II are shown in Fig.~\ref{fig:error_db_2}. The overall prediction error percentages have decreased significantly in comparison with Case I, and the maximum absolute error in predicting the transition location ($E_{x/c}$) for any airfoil has reduced from $6.5\%$ in Case I to approximately 1\% in Case II.

\begin{figure}
    \centering
    \subfloat[Mean error ($E_\text{env}$) percentage for N-factor envelope ]{\includegraphics[width=0.9\textwidth]{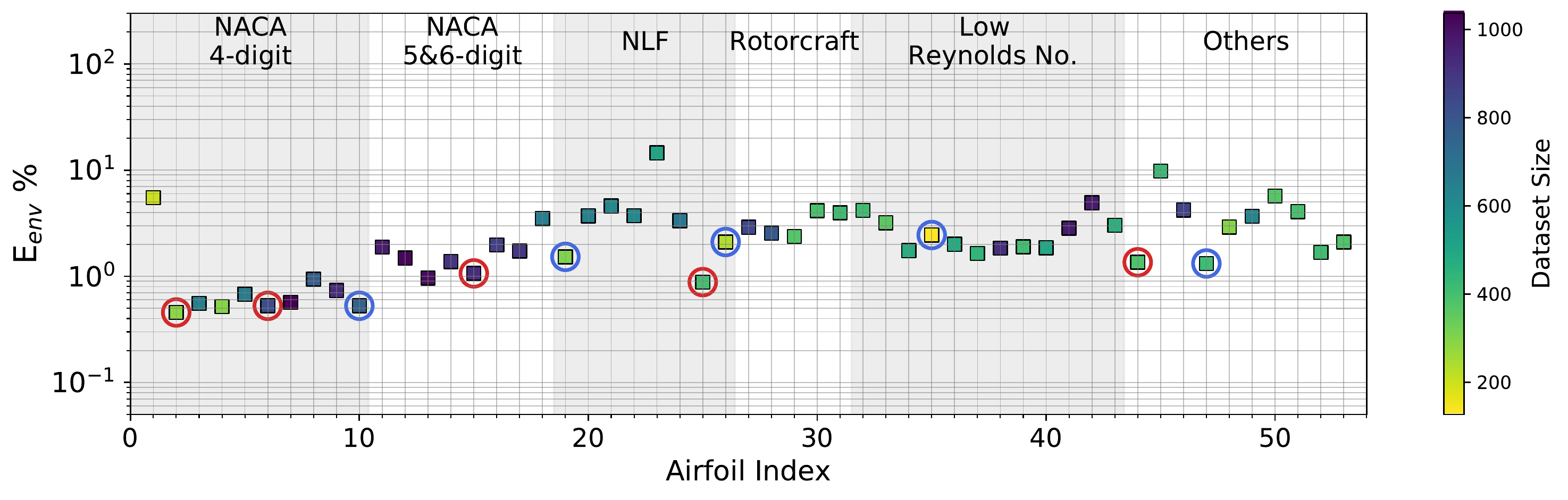}} \\
    \subfloat[Mean relative error ($E_\text{tr}$) percentage for transition location prediction ]{\includegraphics[width=0.9\textwidth]{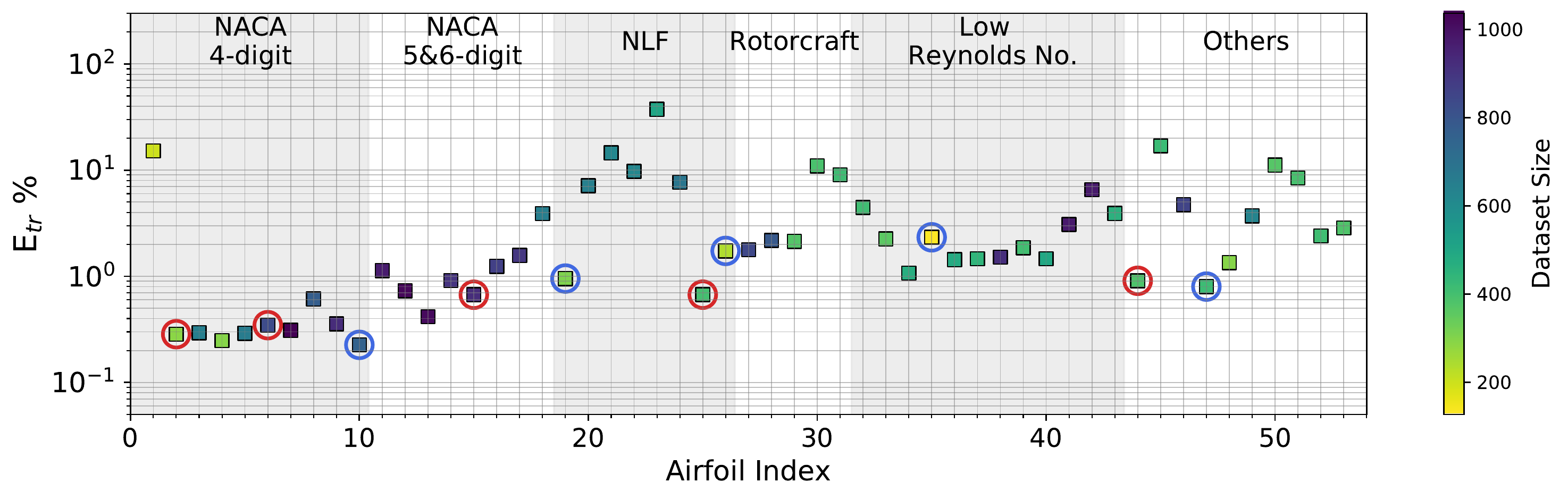}} \\
    \subfloat[Mean absolute error ($E_{x/c}$) percentage for transition location prediction ]{\includegraphics[width=0.9\textwidth]{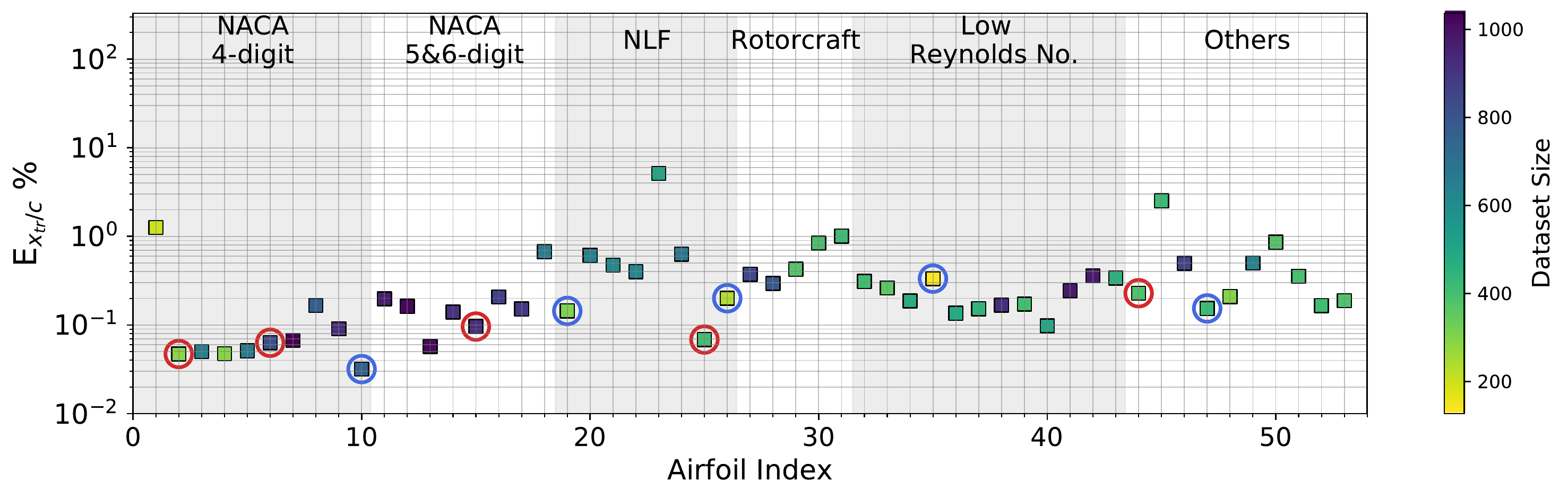}}
    \caption{Mean error percentage for each airfoil in the database, corresponding to training dataset of Case III (augmented airfoils set) given in Table~\ref{tab:cases}. Airfoils corresponding to training dataset have been encircled, where airfoils already in the training dataset from Case I have been encircled in red color, while the augmented set of airfoils have been encircled in blue color. Markers' color represent the dataset size (number of flow cases) of each airfoil in the database}
    \label{fig:error_db_3}
\end{figure}

For Case III, the training dataset has been augmented by including an additional set of airfoils for which the average error ($E_\text{env}$) is greater than 3\%. Five such airfoils, mentioned in Table~\ref{tab:cases}, along with the original five airfoils from Case I constitute the training dataset for Case III. Results for this training case are plotted in Fig.~\ref{fig:error_db_3}. The figure shows that, despite a larger size of the training dataset with respect to that in Case I, the predictive performance for Case III has worsened. The overall trend can be summarised by looking at the group of natural laminar flow airfoils in Fig.~\ref{fig:error_db_3}, for which the model predictions are now significantly worse (~7\%<$E_\text{tr}$<~13\%), except for those airfoils that have been included within the training dataset (~0.7\%<$E_\text{tr}$<~2\%).  This observation points to a possible overfitting of the data by the RNN model under consideration. Hence, one may conclude that the augmentation of the original set by five additional airfoils does not provide a good representation of the overall database.

\begin{figure}
    \centering
    \subfloat[Relative error ($E_\text{tr}$) percentage of flow cases at different angles of attack]{\includegraphics[width=0.77\textwidth]{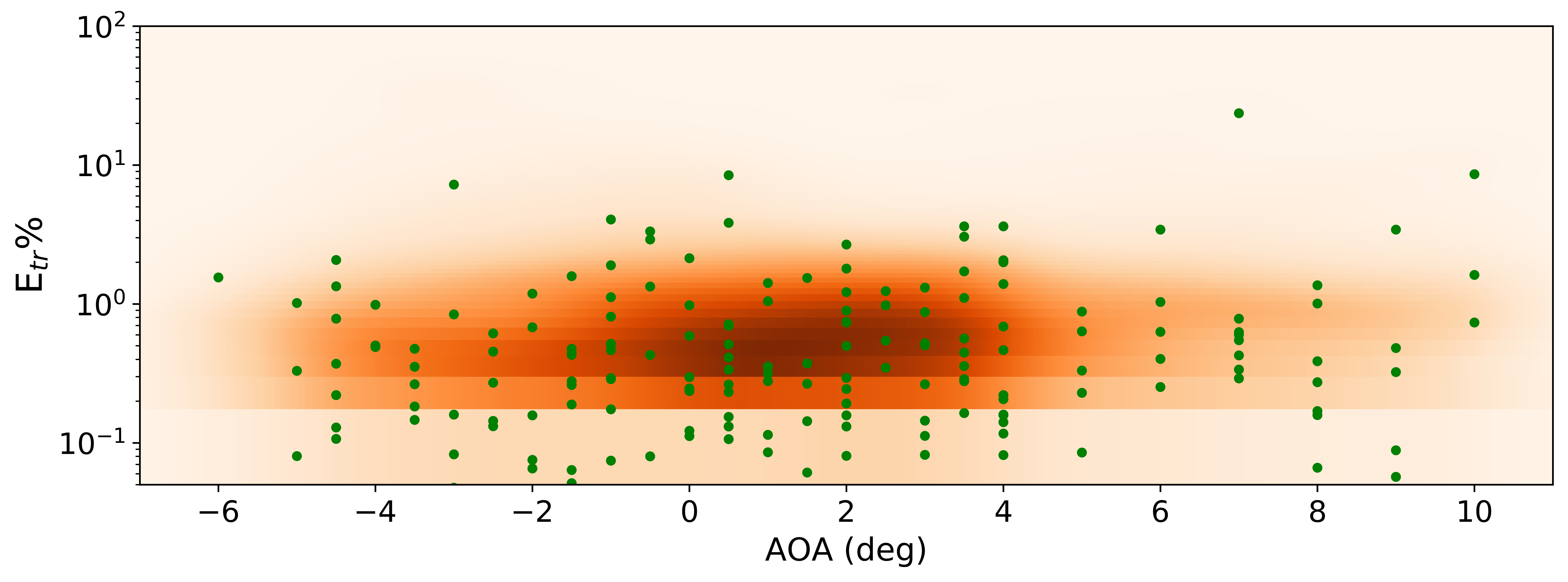}} \\
    \subfloat[Relative error ($E_\text{tr}$) percentage of flow cases at different Reynolds number]{\includegraphics[width=0.77\textwidth]{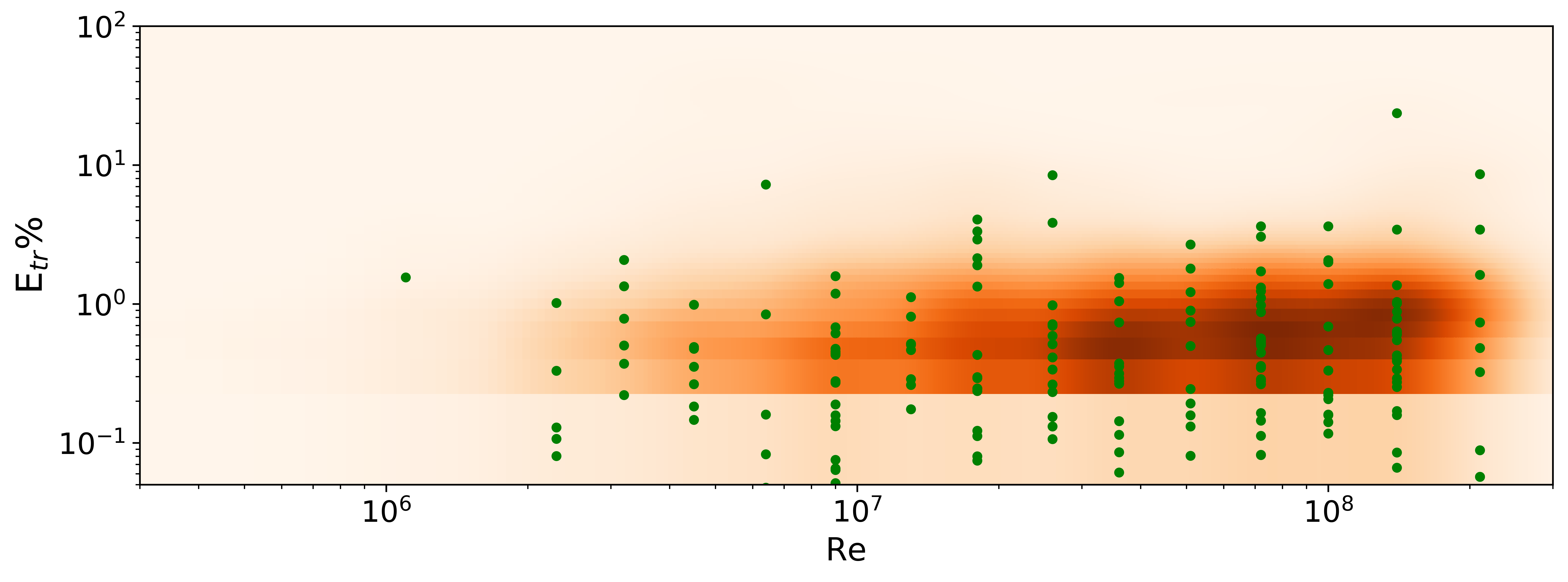}}
    \caption{Relative error ($E_\text{tr}$) percentage for all flow cases, corresponding to training dataset of Case IV given in Table~\ref{tab:cases}. Green markers (filled circles) show only 1\% of the randomly sampled flow cases. The contour shows the kernel density estimated from all the flow cases. Darker region indicates higher probability density. The horizontal lines appearing in the contour plots such as that near an error of 0.2\% are due to technical reason (bins have been defined in linear--scale while the vertical axis of the plot is in log--scale) and don't depict any real discontinuity}
    \label{fig:aoa_re_db_4}
\end{figure}

For Case IV, the training dataset from Case I has been augmented with the flow cases from overall database that correspond to the highest predictive error percentage ($E_\text{env}$ > 3\%) associated with the RNN model from Case I. A significant improvement can be observed in the overall predictive performance of the RNN model as compared to all of the previous cases. Based on this error-based data augmentation, the absolute error percentage in the predicted transition location ($E_{x/c}$) over any airfoil from the database is less than approximately 0.7\%. Fig.~\ref{fig:aoa_re_db_4} shows the distribution of the error percentage as a function of the flow conditions. Due to the kind of data augmentation used for this case, one finds that most flow cases tend towards smaller error values in comparison to those in Case I. 
\begin{figure}
    \centering
    \subfloat[Mean error ($E_\text{env}$) percentage for N-factor envelope ]{\includegraphics[width=0.9\textwidth]{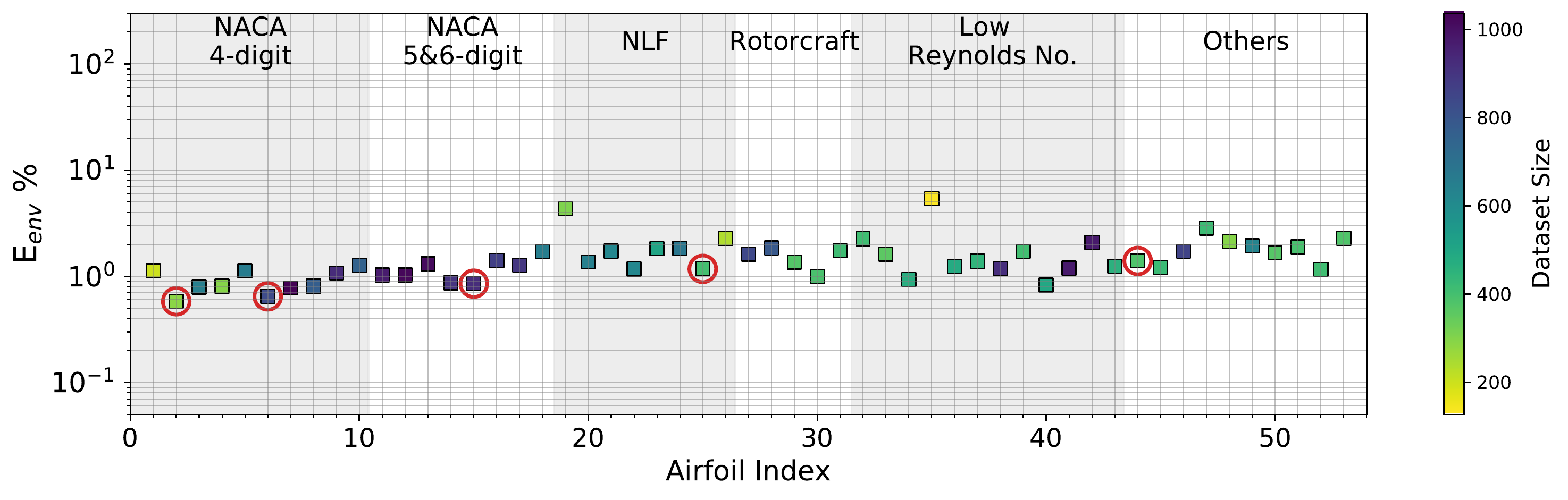}} \\
    \subfloat[Mean relative error ($E_\text{tr}$) percentage for transition location prediction ]{\includegraphics[width=0.9\textwidth]{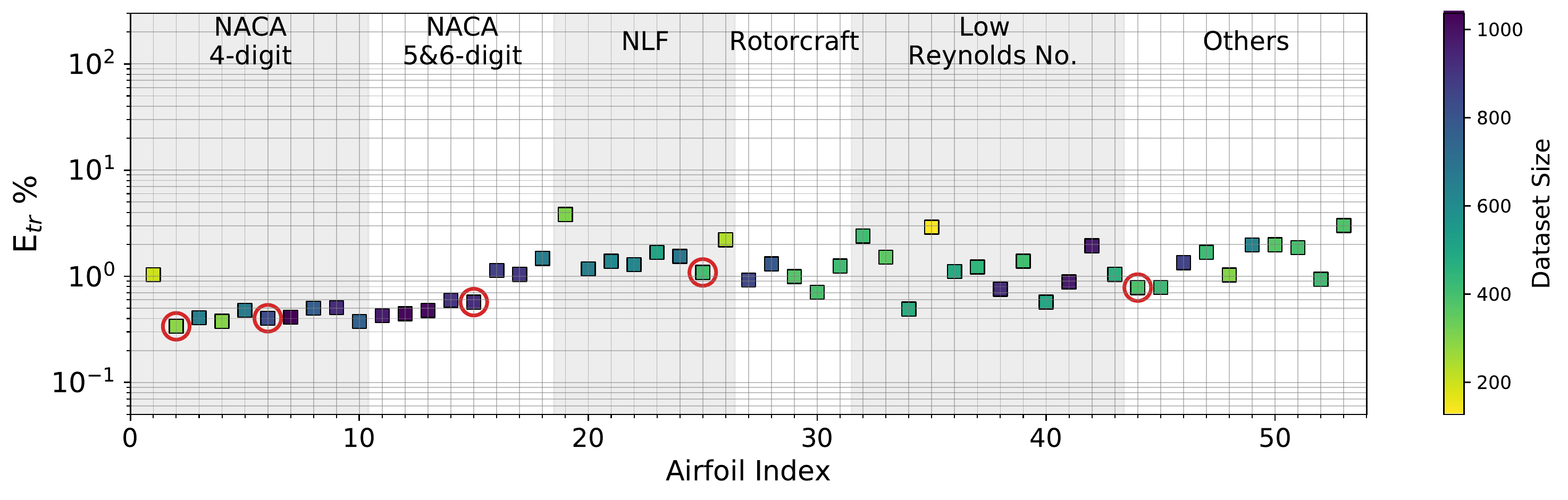}} \\
    \subfloat[Mean absolute error ($E_{x/c}$) percentage for transition location prediction ]{\includegraphics[width=0.9\textwidth]{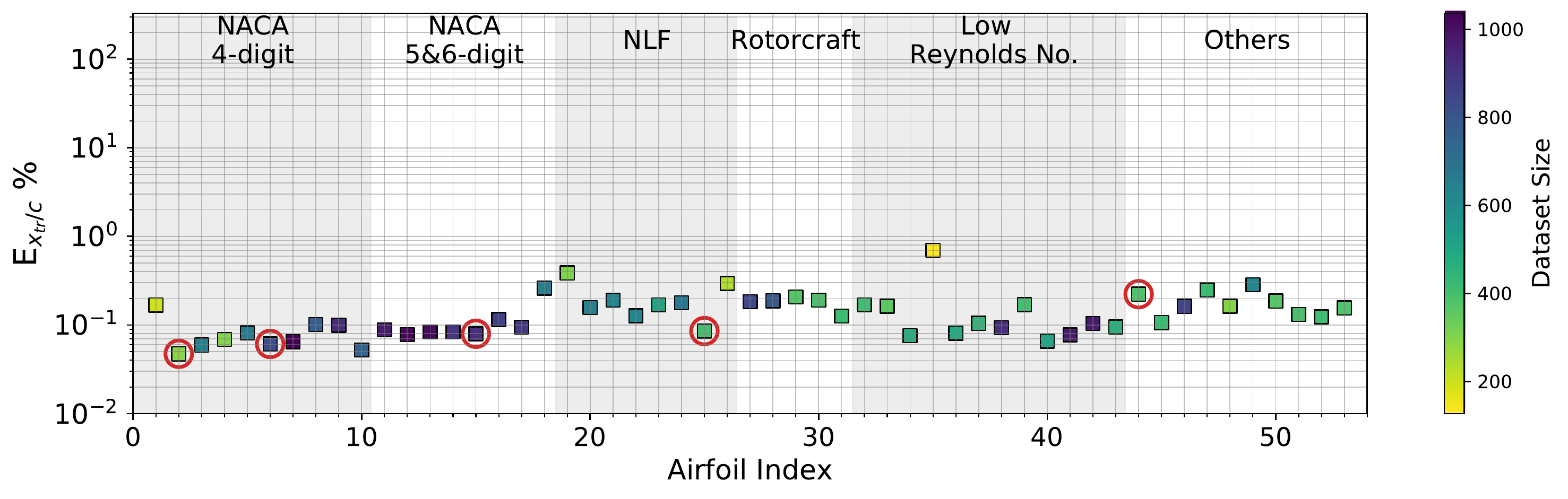}}
    \caption{Mean error percentage for each airfoil in the database, corresponding to training dataset of Case IV (error based augmentation) given in Table~\ref{tab:cases}. Airfoils corresponding to training dataset have been encircled in red color. Markers' color represent the dataset size (number of flow cases) of each airfoil in the database} 
    \label{fig:error_db_4}
\end{figure}

The selection of training data for the cases II, III, and IV was based on the results of Case I, which consisted of five airfoils chosen somewhat arbitrarily (except for the attempt to include some representation from five different groups of airfoils). Although Case IV provides quite good results, such that the absolute error percentage associated with the prediction of the transition location ($E_{x/c}$) over any airfoil is 0.7\% or less, the selection of the training dataset has been made in an indirect manner on the basis of the results obtained in Case I. 

\begin{figure}
    \centering
    \subfloat[Maximum error percentage for any airfoil dataset]{\includegraphics[width=0.4\textwidth]{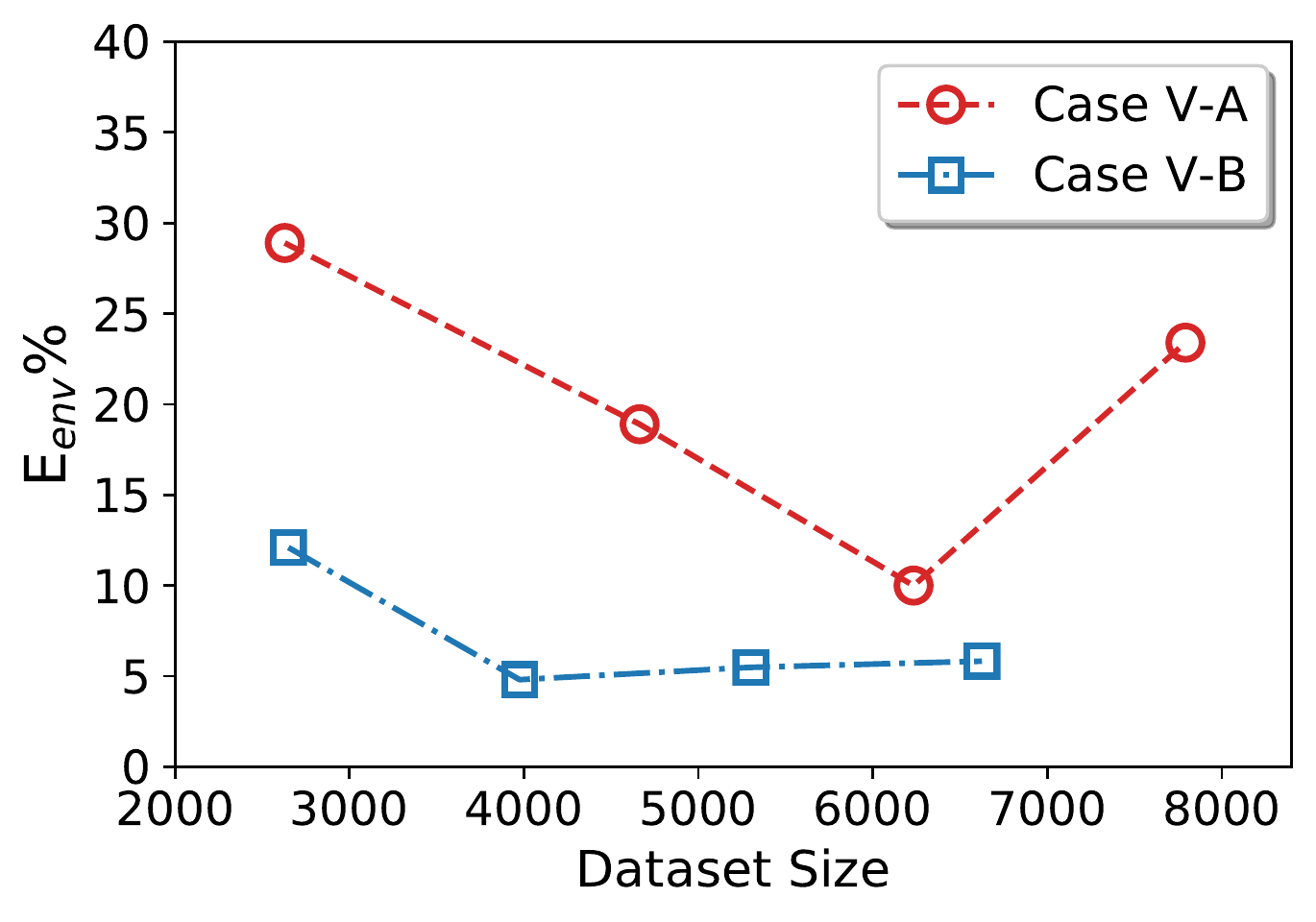} \quad \quad \includegraphics[width=0.4\textwidth]{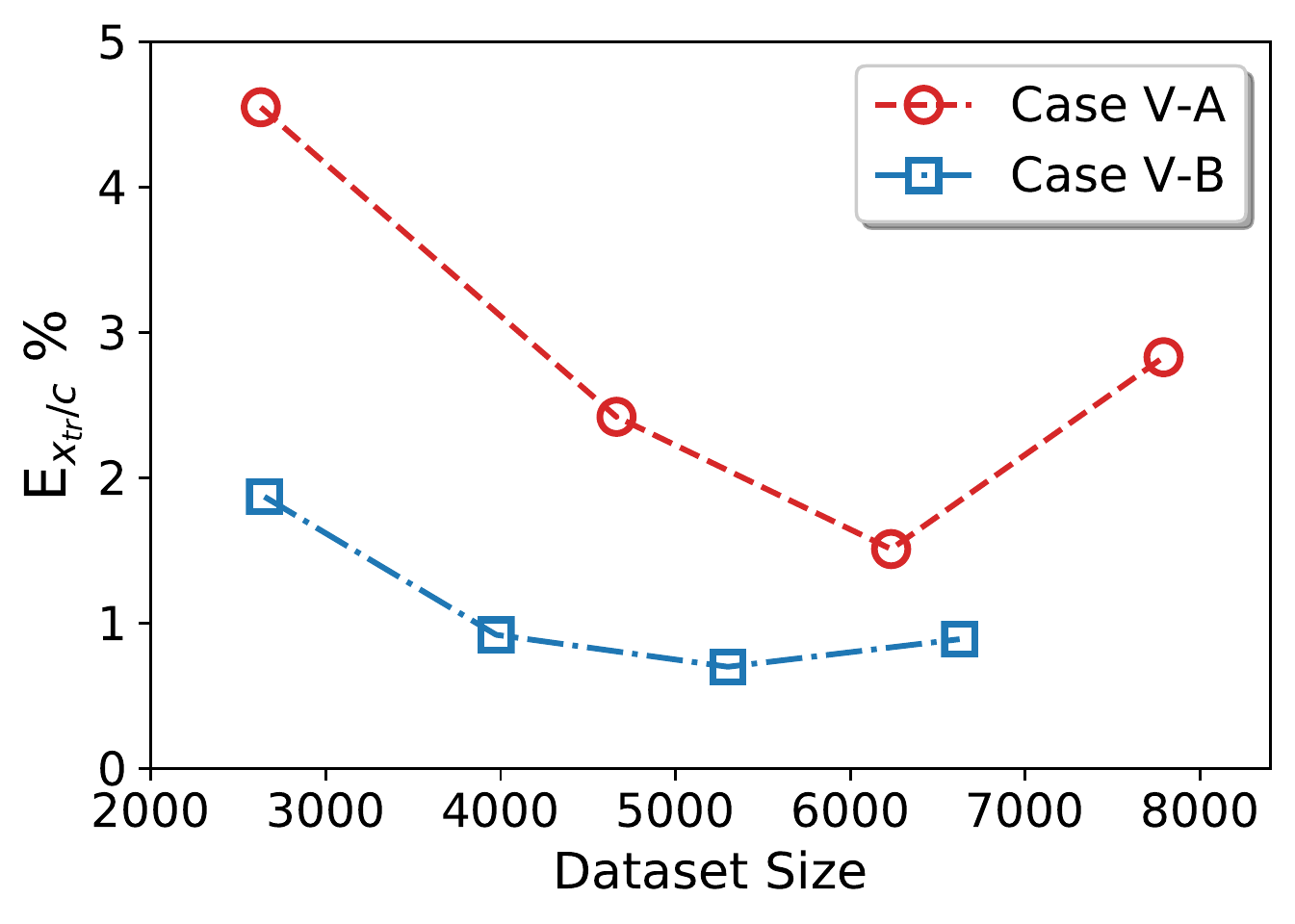}} \\
    \subfloat[Average error percentage for whole airfoil database ]{\includegraphics[width=0.4\textwidth]{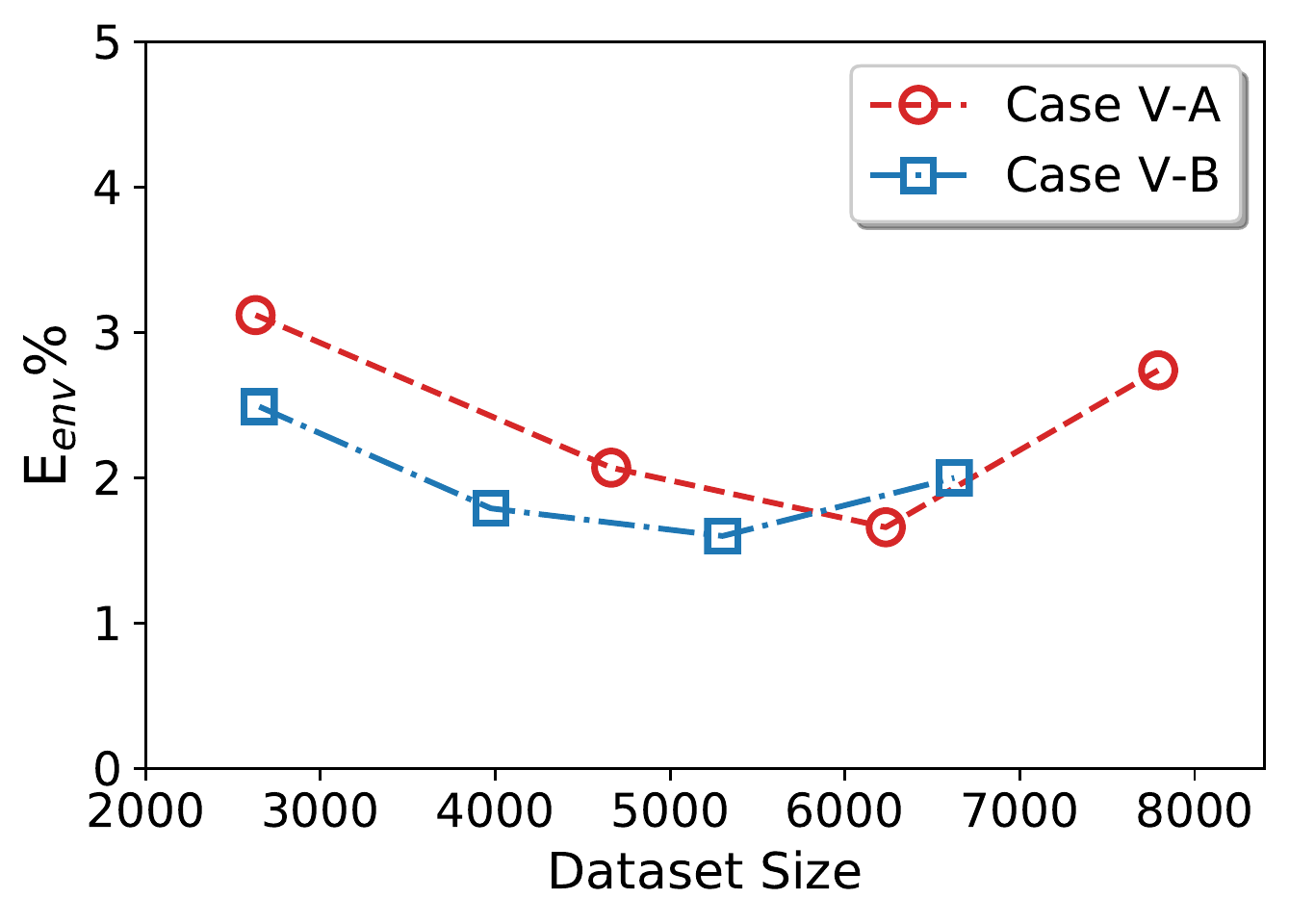} \quad \quad \includegraphics[width=0.4\textwidth]{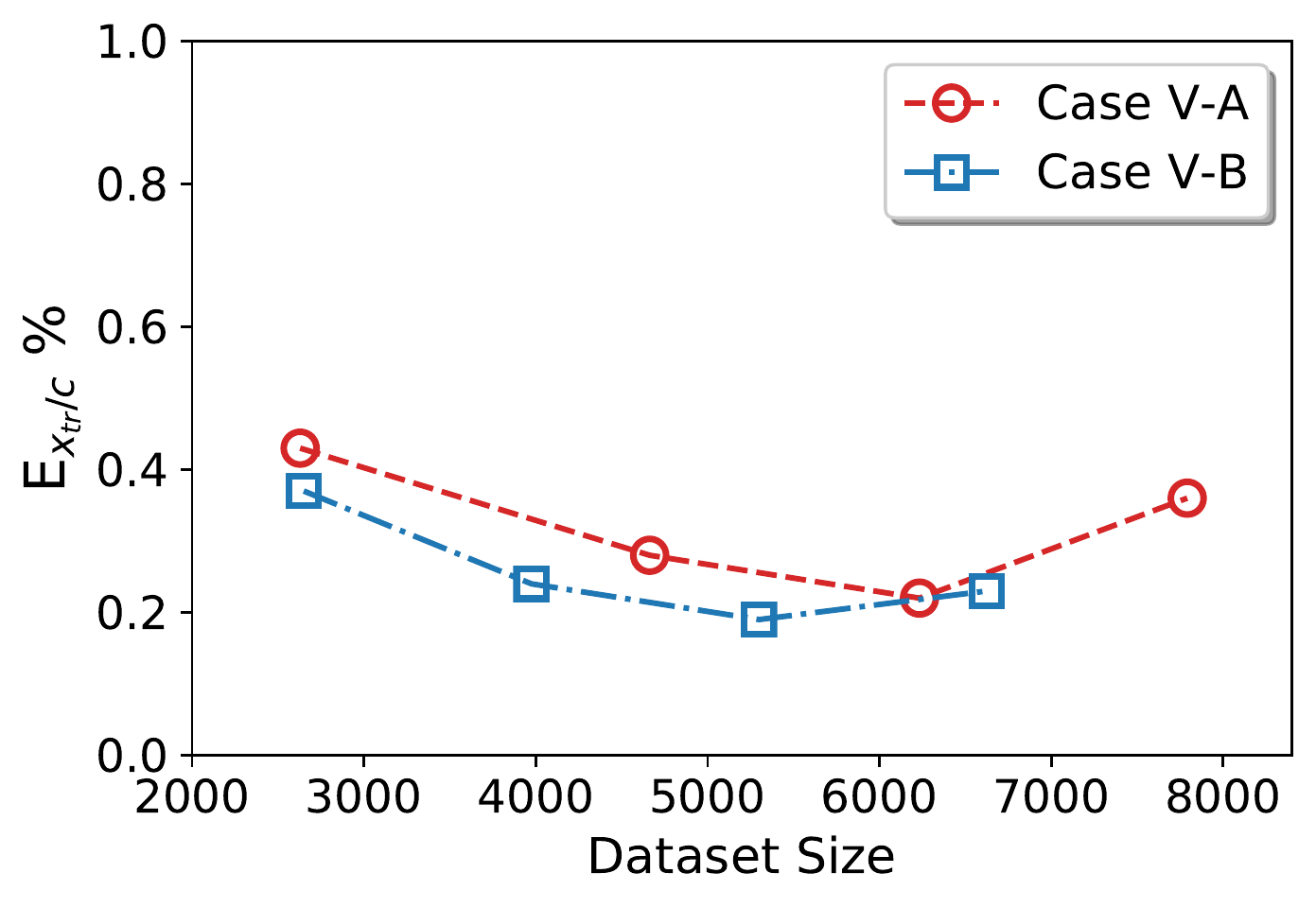}} \\
    \caption{Case V: Variation of error percentage with respect to training dataset size (number of flow cases)}
    \label{fig:case_5_comp}
\end{figure}

A more direct strategy for subsampling a training dataset from the entire database has been assessed in Case V, where a completely random subset of varying magnitudes has been selected from the overall database. Two subcases (V-A and V-B) have been assessed in this regard, as summarised in Table~\ref{tab:cases}. Case V-A involves the selection of a fixed percentage of flow cases corresponding to each airfoil, which results in a different number of flow cases for each airfoil in the training dataset. As mentioned earlier, this uneven number of flow cases in the database results naturally from the fact that different airfoils achieve laminar to turbulent transition at different flow conditions and the fact that only upper surface boundary layers are retained in the case of airfoils with symmetric contours. For Case V-B, a fixed number of flow cases corresponding to each airfoil have been selected as training data in order to provide a uniform weighting to each of the airfoils. With this arrangement to define the two sub-cases, different sizes of training datasets have been used to analyze the variation of error percentage with respect to the size of the training dataset. The results of this study are shown in Fig.\ref{fig:case_5_comp}. The figure shows that, in both cases (V-A and V-B), there is an optimal size of the training dataset that leads to a minimum prediction error. For Case V-A, a training dataset size of $\sim$6200 (20\% flow cases from each airfoil) provides the best predictive performance. Similarly, for Case V-B, a training dataset size of $\sim$5300 (100 flow cases of each airfoil) provides the best predictive performance. 

\begin{figure}
    \centering
    \subfloat[Case V-A with randomly selected 20\% flow cases from each airfoil ]{\includegraphics[width=0.9\textwidth]{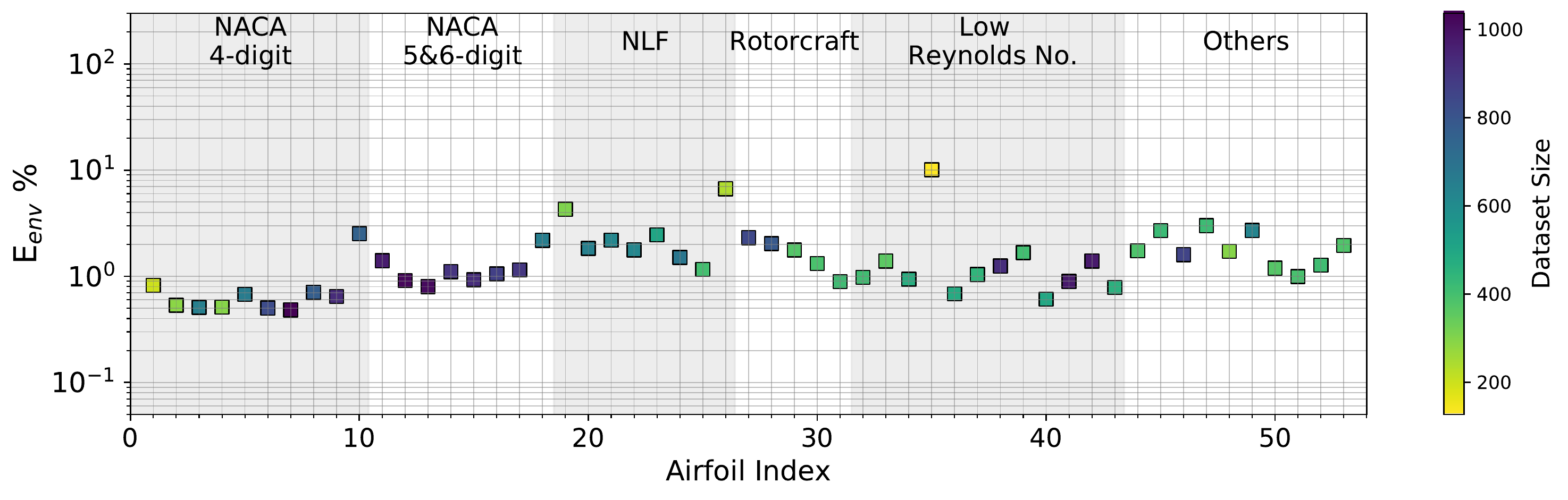}} \\
    \subfloat[Case V-B with randomly selected 100 flow cases from each airfoil ]{\includegraphics[width=0.9\textwidth]{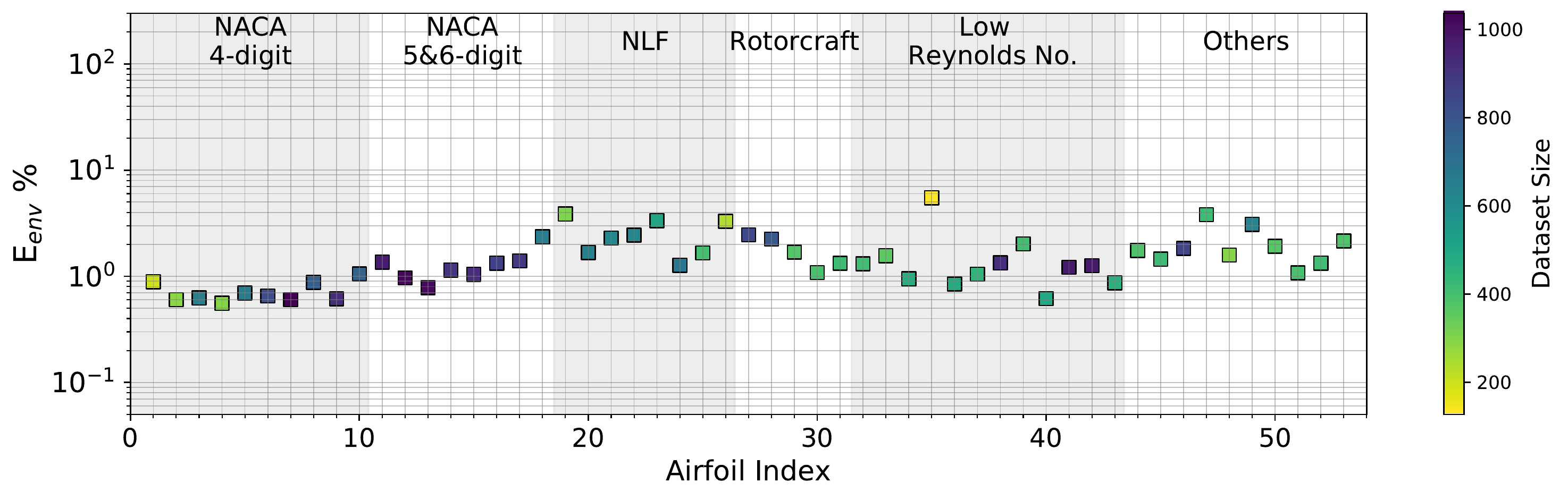}} \\
    \caption{Comparison of mean error ($E_\text{env}$) percentage for N-factor envelopes in Case V-A and V-B. Markers' color represent the dataset size (number of flow cases) of each airfoil in the database\label{fig:error_db_5}}
\end{figure}

Comparing the results for both subcases in Fig.\ref{fig:case_5_comp} shows that Case V-B provides a better prediction accuracy, which can be explained based on the error percentages of three airfoils, NLF(1)-1015, NLF(2)-0415, and LRN(1)-1007. 
These three airfoils with the highest percentage of error in Fig.~\ref{fig:error_db_5}(a) correspond to a relatively smaller number of flow cases in the database, as seen from the colors of their respective markers. Because the training set in Case V-A includes a fixed percentage of flow cases for each airfoil, the above three airfoils remain relatively underweight with respect to the other airfoils with a larger number of flow cases.  
On the other hand, using a fixed number of flow cases for each airfoil provides a more balanced representation of the various airfoils within the training dataset, which results in a better overall predictive performance.  

\begin{figure}
    \centering
    \subfloat[NACA-4418 ($-0.5^\circ$, $1\times10^8$)]{\includegraphics[width=0.31\textwidth]{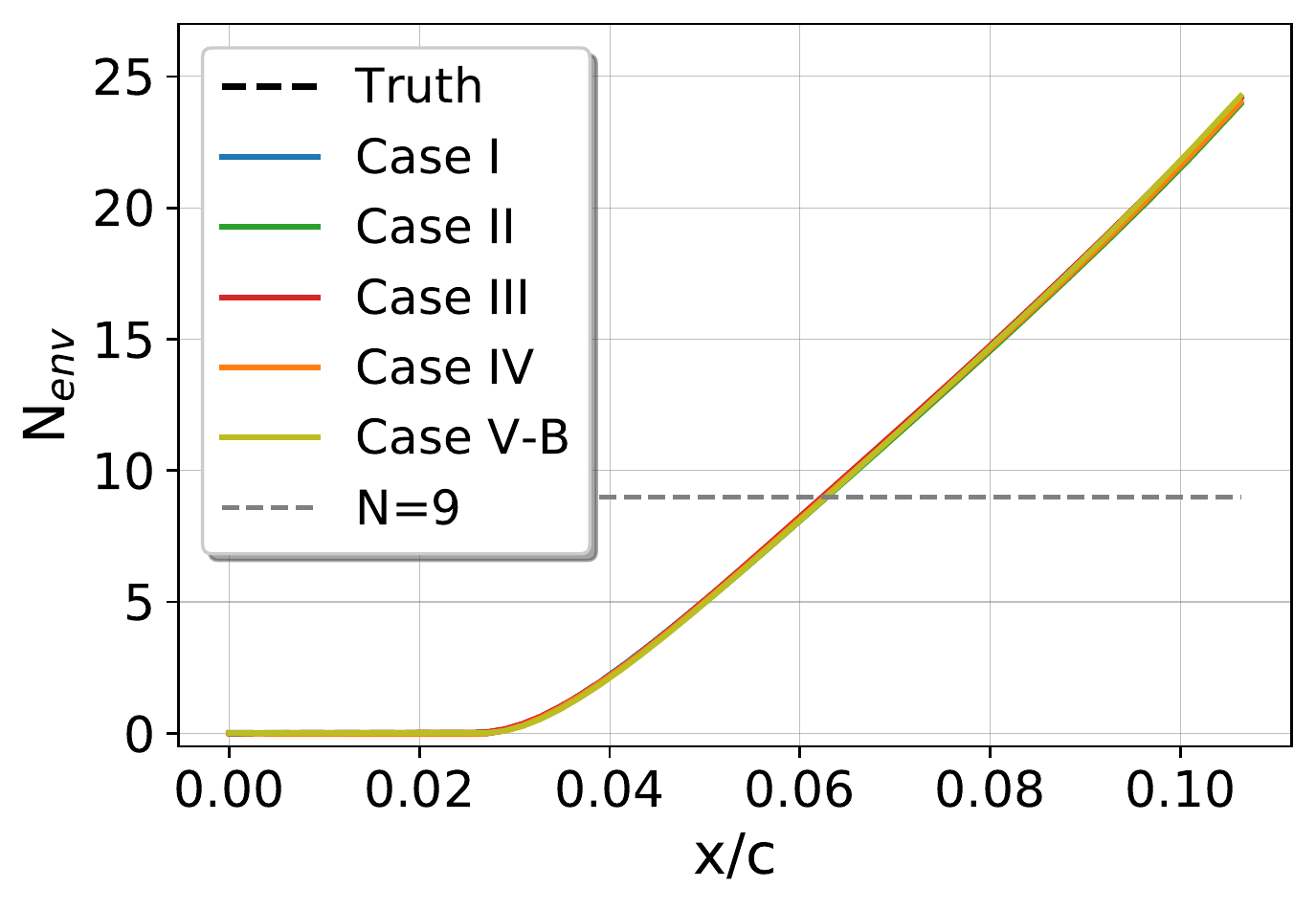}} \quad
    \subfloat[NACA-6712 ($-2.5^\circ$, $2\times10^8$)]{\includegraphics[width=0.31\textwidth]{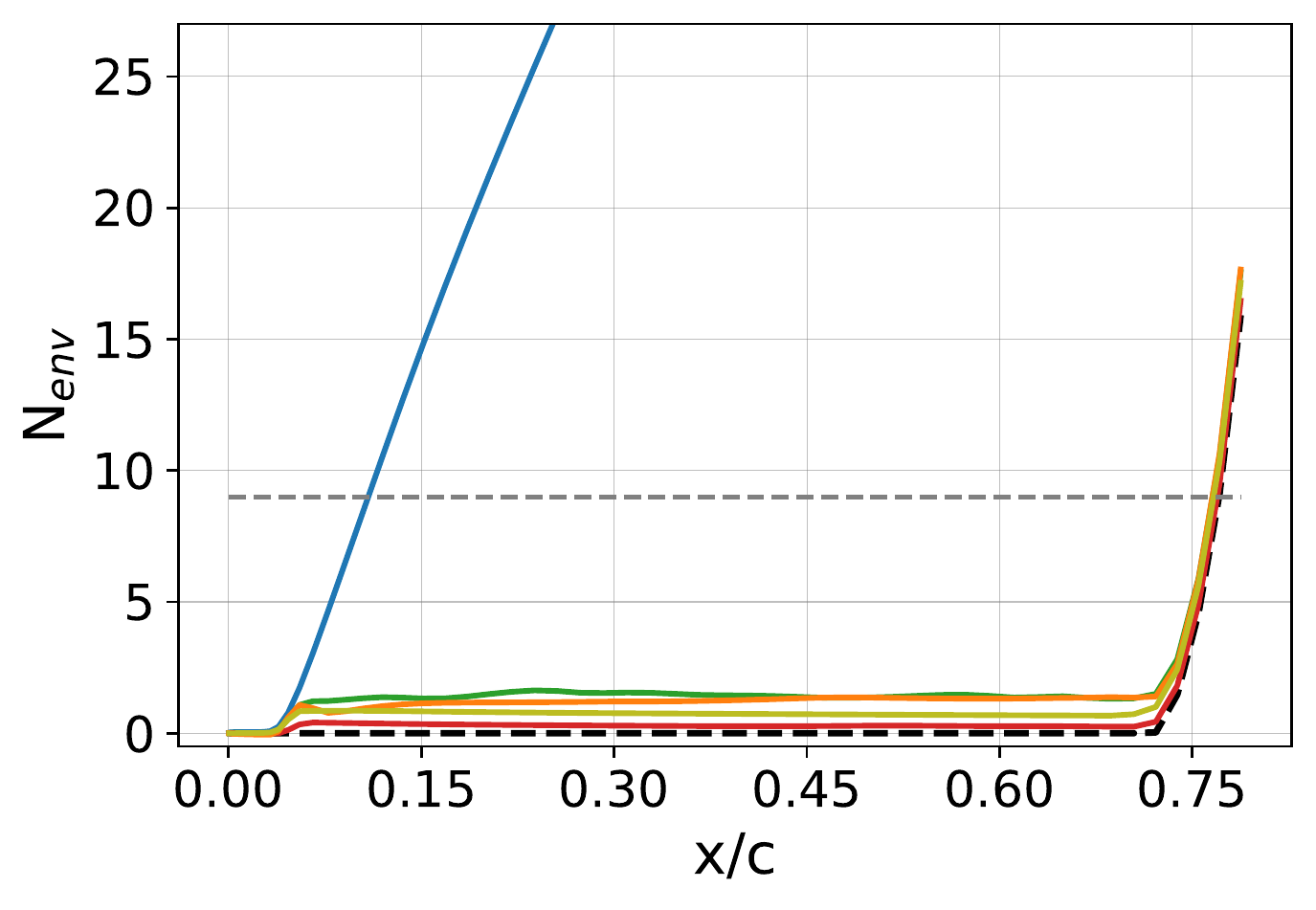}} \quad
    \subfloat[HSNLF(1)-0213 ($-2.0^\circ$, $1\times10^8$)]{\includegraphics[width=0.31\textwidth]{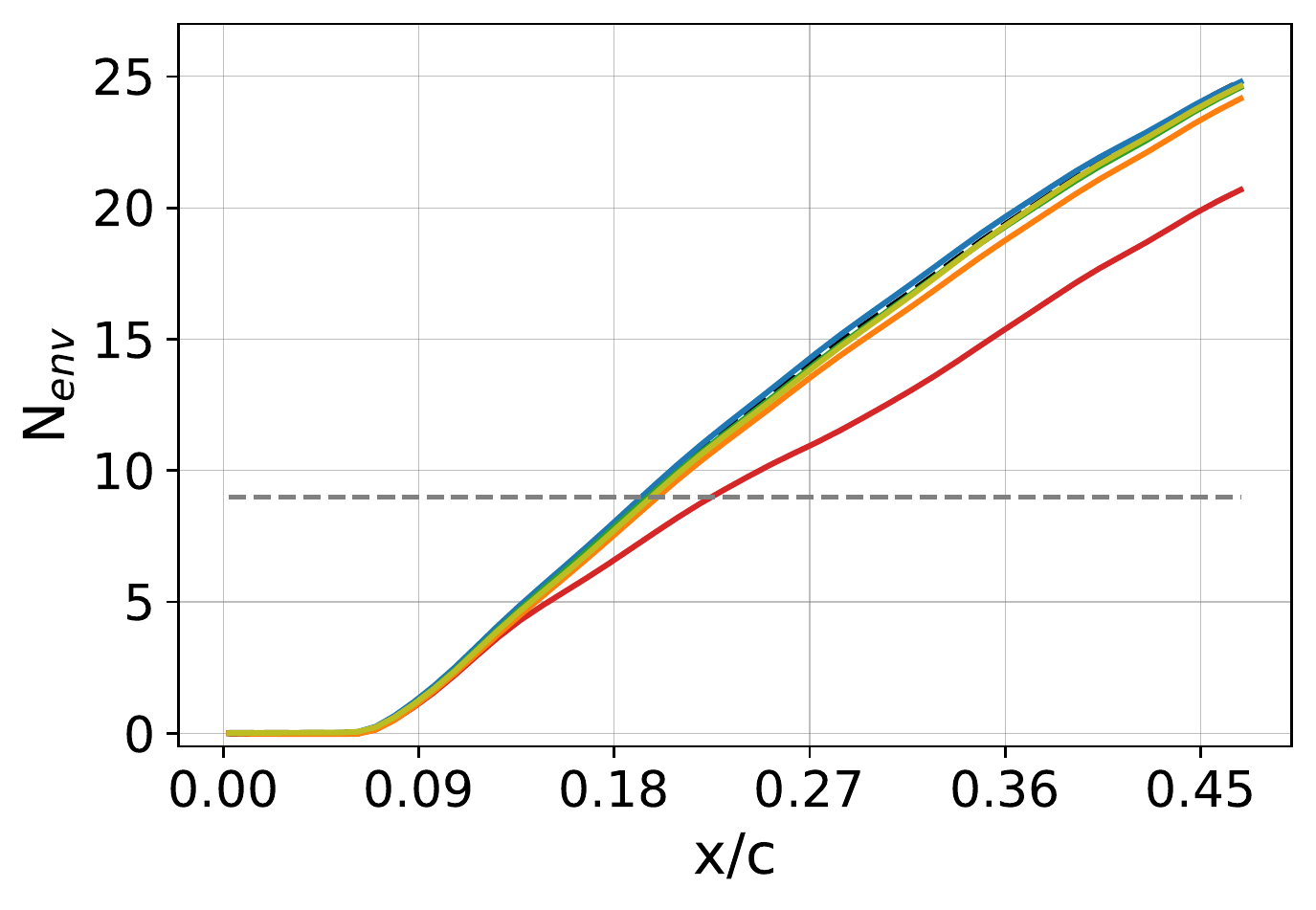}} \\
    \subfloat[LRN(1)-1007 ($-1^\circ$, $1.4\times10^8$)]{\includegraphics[width=0.31\textwidth]{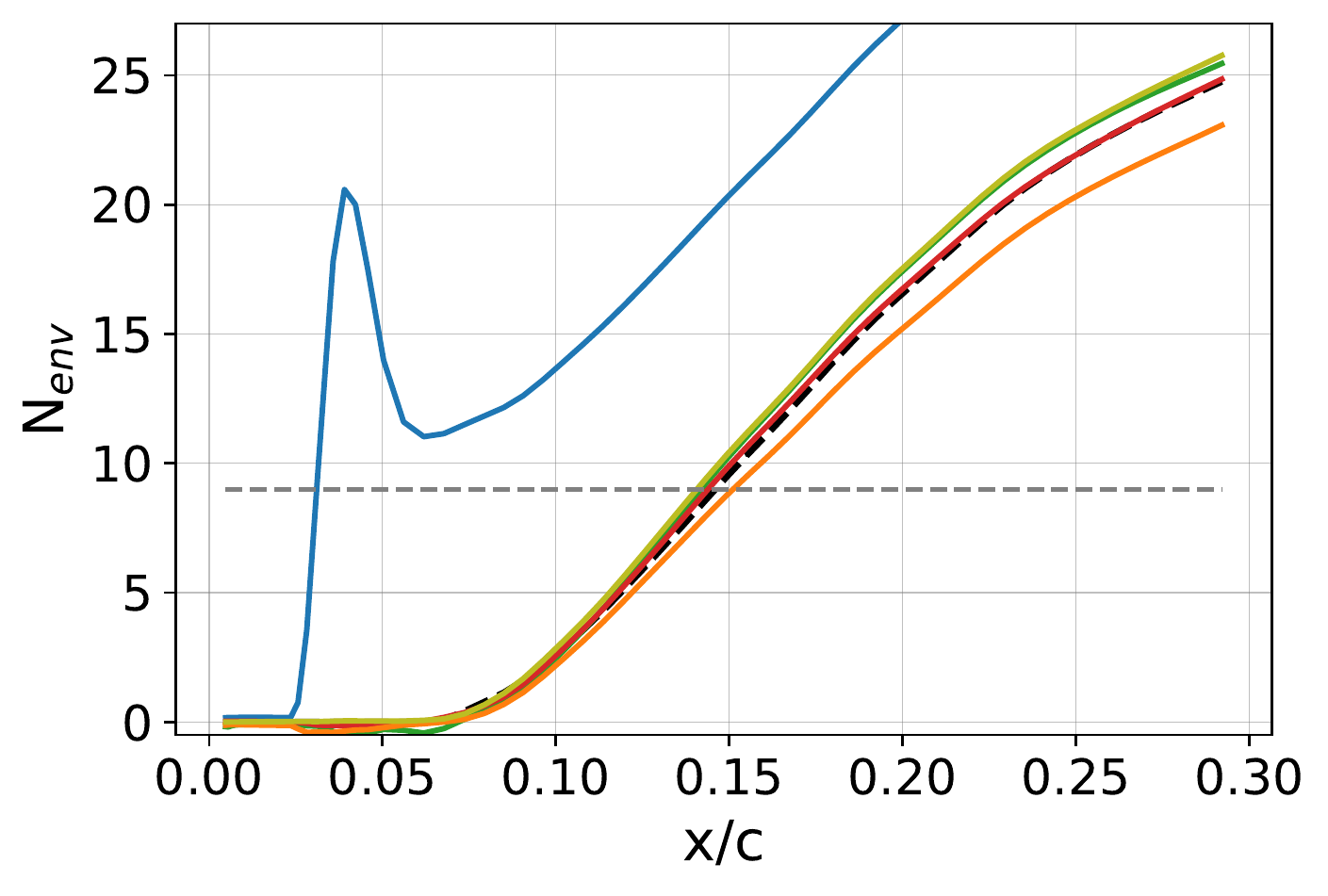}} \quad
    \subfloat[NACA 63(2)-615 ($2^\circ$, $1.4\times10^8$)]{\includegraphics[width=0.31\textwidth]{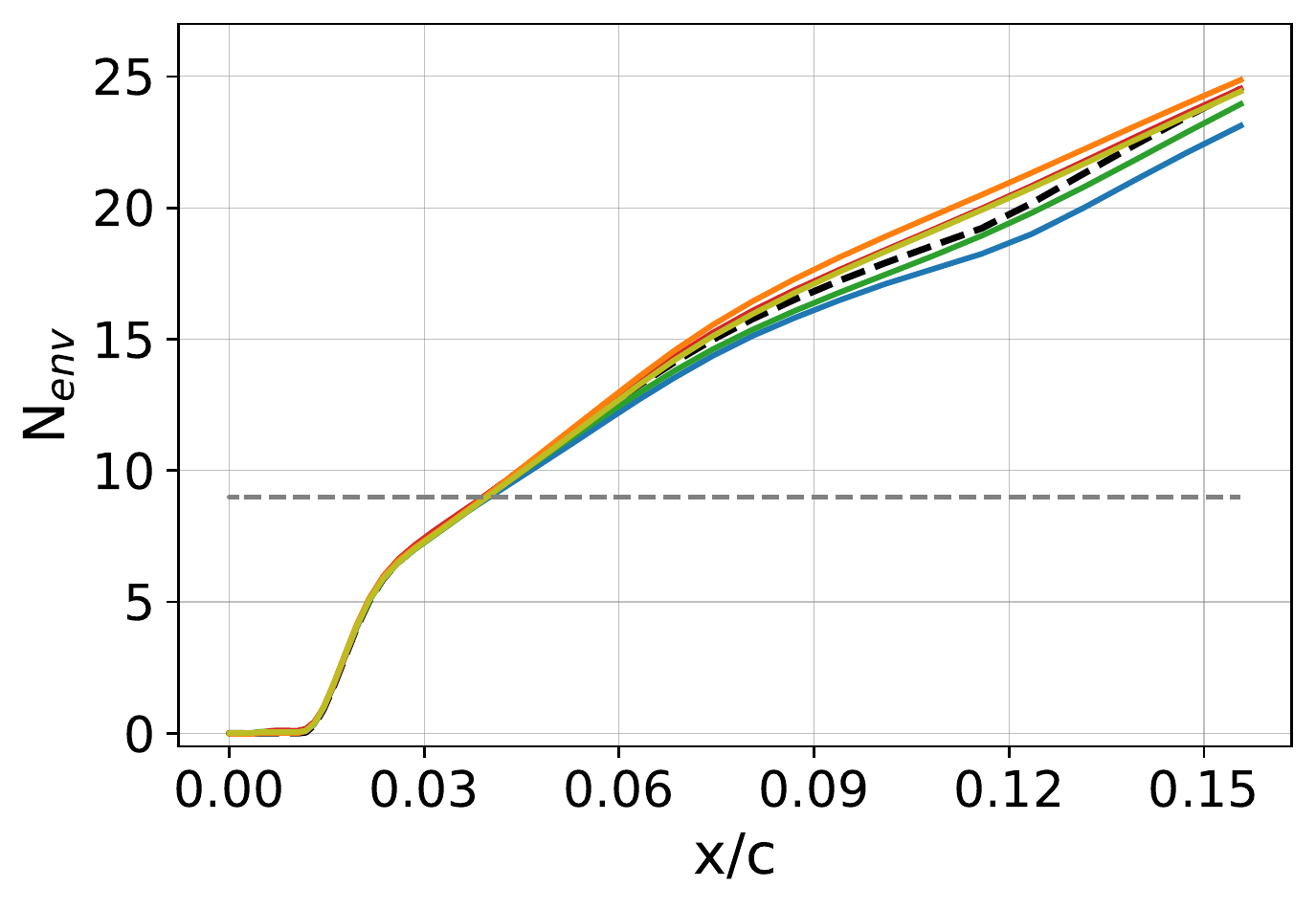}} \quad
    \subfloat[GAW-1 ($3.5^\circ$, $2.6\times10^7$)]{\includegraphics[width=0.31\textwidth]{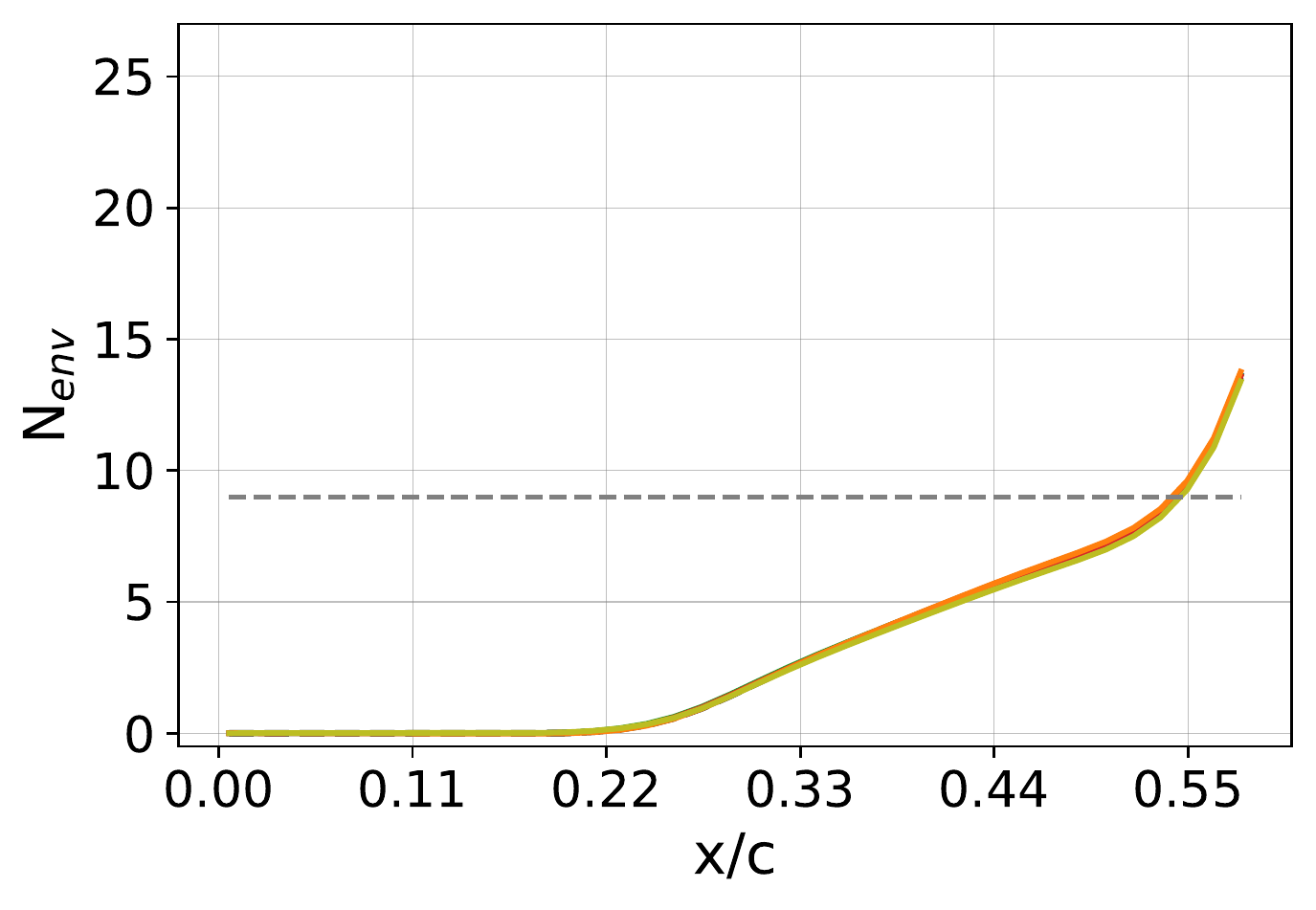}} \\
    \caption{N-factor envelope plots for arbitrarily chosen flow cases to illustrate the comparison of prediction by different training cases. Corresponding airfoil name and flow conditions (AOA, $\text{Re}_c$) have been mentioned for each plot\label{fig:N_plots}}
\end{figure}

Results for all of the cases discussed in this section are summarised in Table~\ref{tab:cases_results}. It is interesting to note that the results of Case V-B, which provides a more direct approach for selecting the training dataset, are very comparable in terms of prediction accuracy with the results from Case IV, which uses an indirect approach to select the training dataset and provides the best results among all of the cases discussed herein. Moreover, the training dataset for both of these cases is of almost equal size. Hence, Case V-B provides a direct and convenient approach for selecting a subsample from a large database as the training data, while also yielding a good predictive performance over the entire database. Sample plots of the N-factor envelope for arbitrary combinations of airfoil contours and flow conditions are shown in Fig.~\ref{fig:N_plots}.  These plots illustrate a qualitative comparison of the N-factor predictions based on the different training cases. One may clearly see that the predictions for certain flows in cases IV and V-B are accurate even if the corresponding predictions for cases I-III include significantly larger error. 
\begin{table}[tbp]
\centering
\caption{Results for different training dataset cases. \label{tab:cases_results}}
\begin{tabular}{|>{\centering\arraybackslash}m{0.9cm} | p{2.1cm} | >{\centering\arraybackslash}m{1.7cm} | >{\centering\arraybackslash}m{0.9cm} | >{\centering\arraybackslash}m{0.9cm} | >{\centering\arraybackslash}m{0.9cm} | >{\centering\arraybackslash}m{1cm} | >{\centering\arraybackslash}m{1cm} | >{\centering\arraybackslash}m{1cm} |} \hline
 &  &  & \multicolumn{3}{c|}{
 \begin{tabular}{>{\centering\arraybackslash}m{3cm}}
\textbf{Maximum error} \\ \hline
\end{tabular}
 } & \multicolumn{3}{c|}{
\begin{tabular}{>{\centering\arraybackslash}m{3.3cm}}
\textbf{Average error} \\ \hline
\end{tabular}
} \\ 
\textbf{Index} & \textbf{Label} & \textbf{Number of flow cases} & $\mathbf{E_\text{env}}$ & $\mathbf{E_\text{tr}}$ & $\mathbf{E_{x/c}}$ & $\mathbf{E_\text{env}}$ & $\mathbf{E_\text{tr}}$ & $\mathbf{E_{x/c}}$ \\ \hline
I & Five airfoils & 2624 & 39.2\% & 30.2\% & 6.52\% & 2.95\% & 2.24\% & 0.42\% \\ \hline
II & Random augmentation & 7026 & 8.95\% & 5.66\% & 1.40\% & 1.92\% & 1.64\% & 0.26\% \\ \hline
III & Augmented airfoils set & 4455 & 14.5\% & 37.4\% & 5.15\% & 2.71\% & 4.14\% & 0.43\% \\ \hline
IV & Error-based augmentation & 5024 & 5.38\% & 3.82\% & 0.70\% & 1.53\% & 1.17\% & 0.15\% \\ \hline
V-A & Random selection (\%) & 6233 & 10.1\% & 35.0\% & 1.51\% & 1.66\% & 2.04\% & 0.22\% \\ \hline
V-B & Random selection (\#) & 5300 & 5.49\% & 4.22\% & 0.70\% & 1.60\% & 1.38\% & 0.19\% \\ \hline
\end{tabular}
\end{table}

\subsection{Working with limited database}
The database of airfoil flows generated during the present effort is relatively extensive in comparison to what may be generally available in a majority of practical situations. For this reason, assessments have been made to understand the predictive performance of the RNN model in selected possible scenarios. Such assessments are based on interpolation and extrapolation with respect to the airfoil contours, under the assumption that a relatively smaller training dataset based on just a few NACA 4-digit series airfoils is available. Table~\ref{tab:small_cases} outlines a summary of these cases and the corresponding results. Case VI and VII provide an assessment with respect to the interpolation of airfoil contours within a family of selected symmetric and asymmetric airfoils, respectively. The resulting predictions are seen to be reasonably accurate for the testing dataset with average error of 0.12\% and 0.04\% of the chord length for both symmetric and asymmetric airfoils, respectively. 
Similarly, Case VIII targets the evaluation of model performance with respect to the extrapolation of the airfoil thickness, and again, the predictions for the testing dataset are found to be reasonably accurate with average error of 0.06\% of the chord length for the given airfoil.
Hence, it appears that the RNN model is able to predict well for previously unknown airfoil sections within the same family, regardless of whether the test data involves an interpolation within the distribution of the training data or an extrapolation beyond its boundaries. These findings support the selection strategy underlying case I, which included five airfoils representing multiple groups from the overall database.

\begin{table}[tbp]
\centering
\caption{Assessment cases for different training cases based on NACA 4-digit series airfoils. These cases have been studied to understand the model performance when a limited training dataset is available. \label{tab:small_cases}}
\begin{tabular}{|>{\centering\arraybackslash}m{0.9cm} | p{5cm} | >{\centering\arraybackslash}m{2.4cm} | >{\centering\arraybackslash}m{0.9cm} | >{\centering\arraybackslash}m{0.9cm} | >{\centering\arraybackslash}m{0.9cm} |} \hline
 &  &  & \multicolumn{3}{c|}{
 \begin{tabular}{>{\centering\arraybackslash}m{3cm}}
\textbf{Testing Error \%} \\ \hline
\end{tabular}
 } \\ 
\textbf{Index} & \textbf{\qquad \quad Training Dataset} & \textbf{Testing Dataset} & $\mathbf{E_\text{env}}$ & $\mathbf{E_\text{tr}}$ & $\mathbf{E_{x/c}}$  \\ \hline
VI & {\small NACA-0006, \ NACA-0018} & {\small NACA-0012} & 0.97\% & 0.55\% & 0.12\% \\ \hline
VII & {\small NACA-2412, \ NACA-4412} & {\small NACA 2415} & 0.34\% & 0.17\% & 0.04\% \\ \hline
VIII & {\small NACA-0006, \ NACA-0012} & {\small NACA-0018} & 0.56\% & 0.24\% & 0.06\% \\ \hline
\end{tabular}
\end{table}

\begin{table}[tbp]
\centering
\caption{Results for a training dataset comprised of a single family of airfoils and a testing dataset comprised of the rest of the airfoils in the database. \label{tab:airfoil_extra}}
\begin{tabular}{|>{\centering\arraybackslash}m{0.75cm} | p{3.5cm} | >{\centering\arraybackslash}m{1.7cm} | >{\centering\arraybackslash}m{0.8cm} | >{\centering\arraybackslash}m{0.8cm} | >{\centering\arraybackslash}m{0.8cm} | >{\centering\arraybackslash}m{0.8cm} | >{\centering\arraybackslash}m{0.8cm} | >{\centering\arraybackslash}m{0.8cm} |} \hline
 &  &  & \multicolumn{3}{c|}{
 \begin{tabular}{>{\centering\arraybackslash}m{2.5cm}}
\textbf{Maximum error} \\ \hline
\end{tabular}
 } & \multicolumn{3}{c|}{
\begin{tabular}{>{\centering\arraybackslash}m{2.4cm}}
\textbf{Average error} \\ \hline
\end{tabular}
} \\ 
\textbf{Index} & \textbf{\quad Training dataset} & \textbf{Number of flow cases} & $\mathbf{E_\text{env}}$ & $\mathbf{E_\text{tr}}$ & $\mathbf{E_{x/c}}$ & $\mathbf{E_\text{env}}$ & $\mathbf{E_\text{tr}}$ & $\mathbf{E_{x/c}}$ \\ \hline
IX & {\small NACA-0006, \ NACA-0018, NACA-2412, \ NACA-4412, NACA-6712} & 2841 & 67.5\% & 44.8\% & 9.51\% & 9.55\% & 9.43\% & 1.52\% \\ \hline
\end{tabular}
\end{table}

\begin{figure}
    \centering
    \subfloat[Mean error ($E_\text{env}$) percentage for N-factor envelope ]{\includegraphics[width=0.9\textwidth]{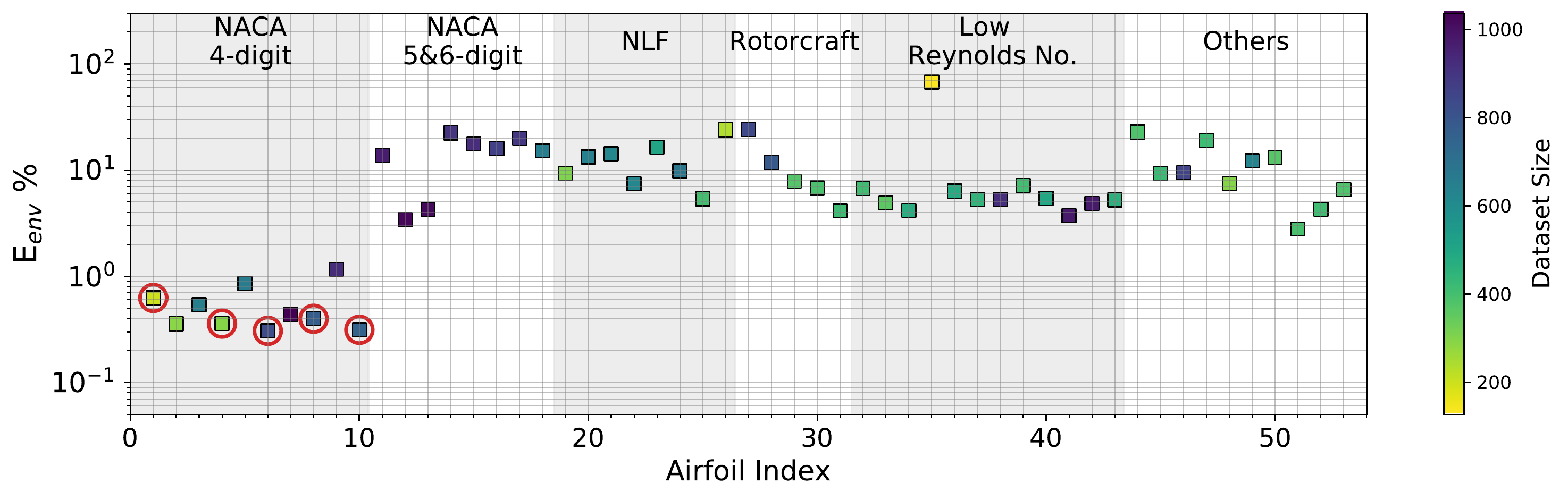}} \\
    \subfloat[Mean relative error ($E_\text{tr}$) percentage for transition location prediction ]{\includegraphics[width=0.9\textwidth]{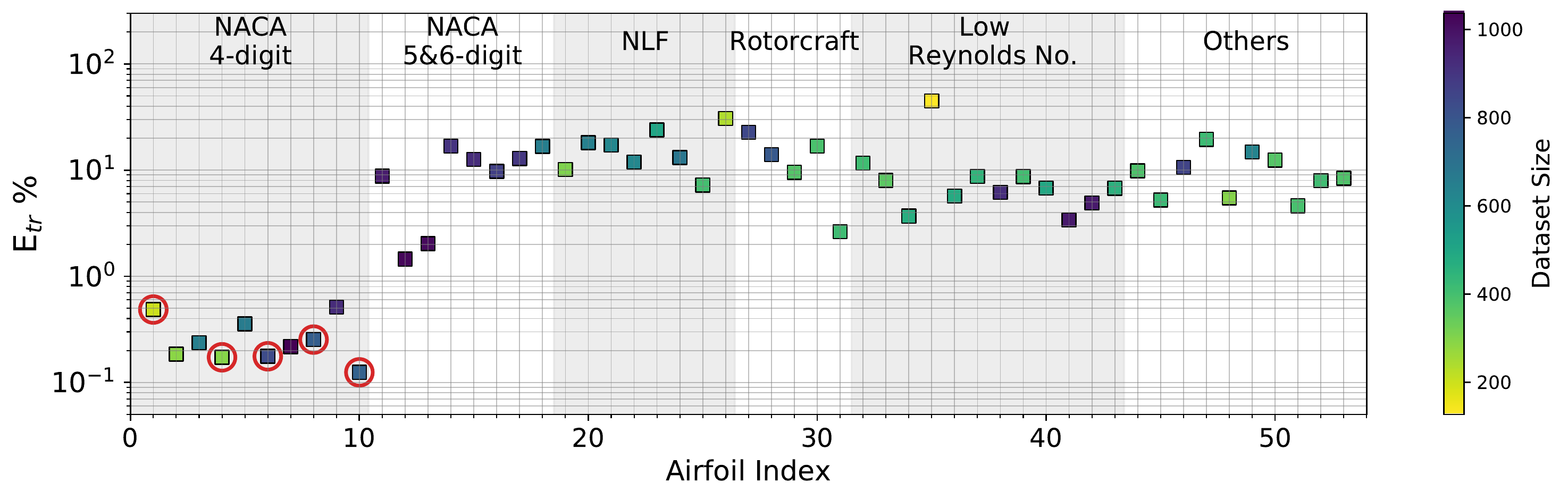}} \\
    \subfloat[Mean absolute error ($E_{x/c}$) percentage for transition location prediction ]{\includegraphics[width=0.9\textwidth]{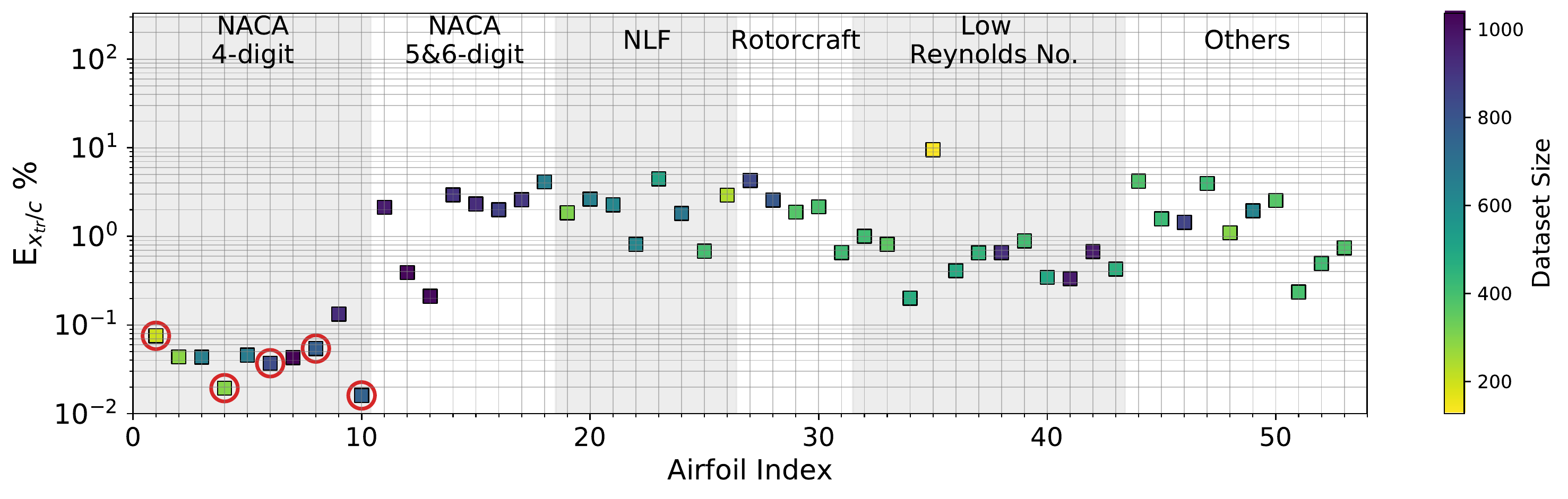}}
    \caption{Mean error percentage for each airfoil in the database, corresponding to training dataset of Case IX given in Table~\ref{tab:airfoil_extra}. Airfoils corresponding to training dataset have been encircled in red color. Markers' color represent the dataset size (number of flow cases) of each airfoil in the database\label{fig:error_db_extrap}}
\end{figure}

Assessment in Case IX involves a training dataset of five NACA 4-digit series airfoils, and the predictive performance is evaluated using a testing dataset based on the rest of the airfoils. Results for this case have been shown in Table~\ref{tab:airfoil_extra}, where it can be observed that the predictive performance in this case is far worse in comparison to Case I, where five airfoils were taken from a different family of airfoils. Hence, a model trained using just a single family of airfoils does not extrapolate well to the other families of airfoils. Results for Case IX have also been shown in Fig.~\ref{fig:error_db_extrap}, wherein the mean error percentages for the remaining families of airfoils are seen to be an order of magnitude higher than the error magnitudes associated with airfoils included in the training dataset. This finding reinforces the method adopted in Case I, namely, that a balanced training dataset should contain representation from different families of airfoils to achieve reasonably accurate predictive performance for the overall database. 

\section{Conclusion \label{conclusion}}
A sequence-to-sequence modeling approach based on a recurrent neural network has been proposed to predict the location of laminar-turbulent transition via linear amplification characteristics of hydrodynamic instabilities in boundary-layer flows. This approach provides an end-to-end transition model, which maps the sequence of mean boundary-layer profiles to corresponding growth rates along the N-factor envelope, and then, to the estimated transition location. In this regard, a large database comprised of the linear growth characteristics of over 33,000 boundary-layer flows over 53 airfoils from a disparate range of applications has been used to train and test the proposed model. The results demonstrate that the RNN model is able to predict the transition location at various test flow conditions and for the entire range of airfoil contours with good accuracy (average error of less than 0.70 percent of the chord length for any given airfoil) despite being trained by a small subsample (about 16\%) of the complete database. To our knowledge, the database used herein is one of the largest of its kind, presumably representing a significant cross-section of the airfoil universe.  To assess the techniques to facilitate the selection of representative yet computationally efficient training data, several alternate strategies have been investigated to provide insights into working with large amounts of data. The more easily realizable training set based on a small group of five airfoils (one each from five different groups of airfoil contours) forms the baseline for the selection of training data.  A limited database of this type is found to result in substantial errors in transition prediction for a number of other airfoils, with average errors in transition location prediction across multiple test conditions for a single airfoil approaching as much as 6.52\% of the chord length for the given airfoil. Data augmentation with additional cases from other airfoils that correspond to worst prediction errors from the baseline model is found to provide the best choice for improving the predictive performance of the RNN model, reducing the average error across all flow conditions in predicting transition location to 0.7\% of the chord length for any airfoil. An alternate strategy of using a training dataset consisting of an equally weighted representation of each airfoil was also evaluated and was found to provide equally good predictive performance. Further assessments also showed that the RNN-based model is able to extrapolate/interpolate well within a family of similar airfoils and the predictive performance worsened while extrapolating the predictions to airfoil from other families. 

Transition estimates based on the RNN model are easily three to four orders of magnitude faster than those based on direct stability computations. However, the main benefit of the deep learning models is an improved robustness of the prediction process, making it easier for non-experts in laminar-turbulent transition to perform such computations. On the other hand, the deep learning models are restricted in their generalizability and this paper has addressed some of the issues related to the development of models that cover a broad space of flows. We believe the two types of models to be complementary in nature.

A significant advantage of the proposed RNN model over the previously proposed neural network-based transition models is that by using the sequential information of the underlying mean flow, the RNN model is able to directly predict the required information of the N-factor envelope and the transition location, without requiring the user to define a range of critical frequencies and predicting the instability growth rates at a number of frequencies in this range. On the other hand, since the RNN model predicts the growth rates of the N-factor envelope in a sequential manner, it cannot be applied in a parallel manner, unlike conventional methods or previously proposed neural networks that can predict the local growth rates at each station in a parallel manner.

The proposed architecture processes the boundary-layer profiles at each station in a physically consistent manner using the convolutional neural network. This attribute enables its generalization to other instability mechanisms involving three-dimensional boundary-layer profiles involving crossflow velocity components or second-mode instabilities in high speed flows involving the profiles of thermodynamic quantities such as density and/or temperature. Future work could involve the application of the proposed architecture to one of the other instability mechanisms. Furthermore, since the RNN model uses input data for the boundary-layer profiles, which depends on the airfoil contour and flow conditions, future explorations could involve airfoil contours along with angle of attack and Reynolds number as global inputs to predict the N-factor envelope. Use of vector-cloud neural network can also be explored, as it would allow the user to employ boundary layer profiles defined at any arbitrary and variable number of grid points~\citep{zhou2021frame}.

\appendix
\section{List of Airfoils \label{airfoils_list}}
A listing of the 53 airfoils included in the database of stability characteristics is given in Table~\ref{tab:list}. Furthermore, a graphical catalog of all airfoil contours is also included, since that may be of interest to the readers. 
\begin{table}[htbp]
    \centering
    \caption{List of airfoils in the database. \label{tab:list}}
    \begin{tabular}{|m{0.05\textwidth} | m{0.14\textwidth} || m{0.05\textwidth} | m{0.16\textwidth}|| m{0.05\textwidth} | m{0.14\textwidth}|| m{0.05\textwidth} | m{0.14\textwidth}|} \hline
    \textbf{Index} & \textbf{Airfoil} & \textbf{Index} & \textbf{Airfoil} & \textbf{Index} & \textbf{Airfoil} & \textbf{Index} & \textbf{Airfoil} \\ \hline \hline
    1 & {\small NACA 0006} & 15 & {\small NACA 63-415} & 29 & {\small VR-15} & 43 & {\small S8052 } \\ \hline
    2 & {\small NACA 0012} & 16 & {\small NACA 64-215} & 30 & {\small OA209} & 44 & {\small ONERA M6} \\ \hline
    3 & {\small NACA 0015} & 17 & {\small NACA 63(2)-615} & 31 & {\small OA212} & 45 & {\small RAE 2822} \\ \hline
    4 & {\small NACA 0018} & 18 & {\small NACA 66(1)-212} & 32 & {\small E374} & 46 & {\small GA(W)-1} \\ \hline
    5 & {\small NACA 0021} & 19 & {\small NLF(1)-1015} & 33 & {\small E387} & 47 & {\small CLARK Y} \\ \hline
    6 & {\small NACA 2412} & 20 & {\small HSNLF(1)-0213} & 34 & {\small E472} & 48 & {\small LNV109A} \\ \hline
    7 & {\small NACA 2415} & 21 & {\small NLF(1)-0115} & 35 & {\small LRN(1)-1007} & 49 & {\small S8055} \\ \hline
    8 & {\small NACA 4412} & 22 & {\small NLF(1)-0215F} & 36 & {\small SD7003} & 50 & {\small S805a} \\ \hline
    9 & {\small NACA 4418} & 23 & {\small NLF(1)-0414F} & 37 & {\small SD7032} & 51 & {\small PSU 94-097} \\ \hline
    10 & {\small NACA 6712} & 24 & {\small NLF(1)-0414D} & 38 & {\small SD7062} & 52 & {\small SA7036} \\ \hline
    11 & {\small NACA 23015} & 25 & {\small NLF(1)-0416} & 39 & {\small SD7080} & 53 & {\small SA7038} \\ \hline
    12 & {\small NACA 23017} & 26 & {\small NLF(2)-0415} & 40 & {\small SD8020} &  &  \\ \hline
    13 & {\small NACA 23024} & 27 & {\small VR-7} & 41 & {\small S8036} &  &  \\ \hline
    14 & {\small NACA 63-215} & 28 & {\small VR-12} & 42 & {\small S8037} &  &  \\ \hline
    \end{tabular}
\end{table}
\begin{figure}
    \centering
    \includegraphics[width=0.925\textwidth]{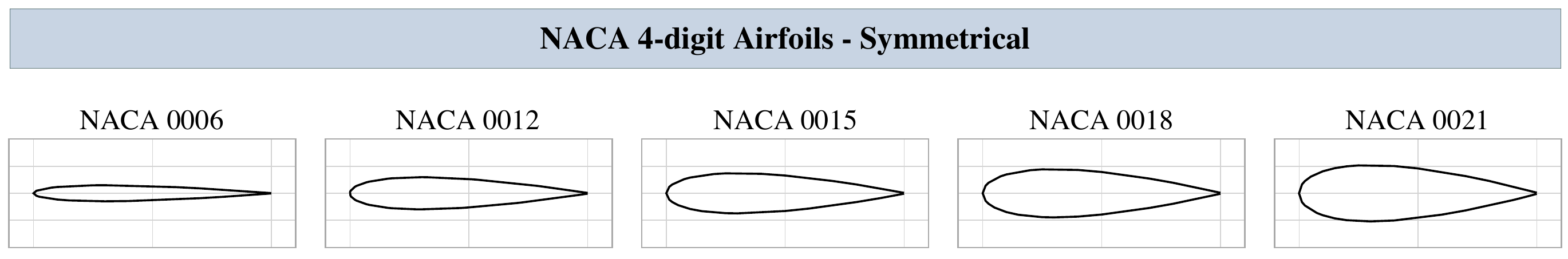}
    \includegraphics[width=0.925\textwidth]{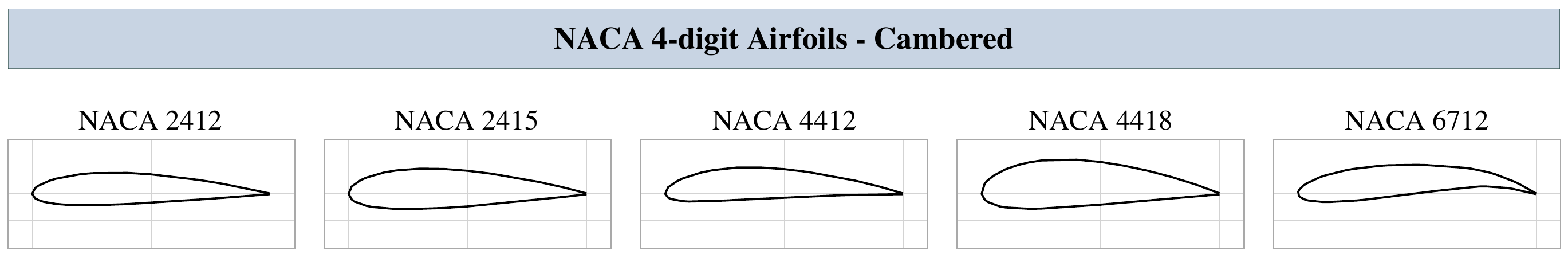}
    \includegraphics[width=0.925\textwidth]{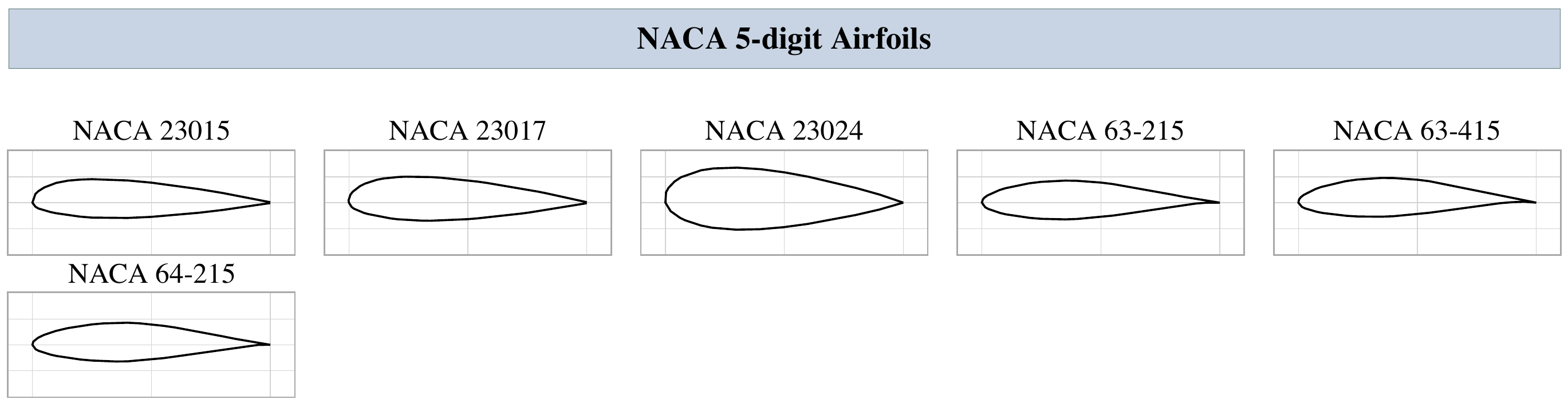}
    \includegraphics[width=0.925\textwidth]{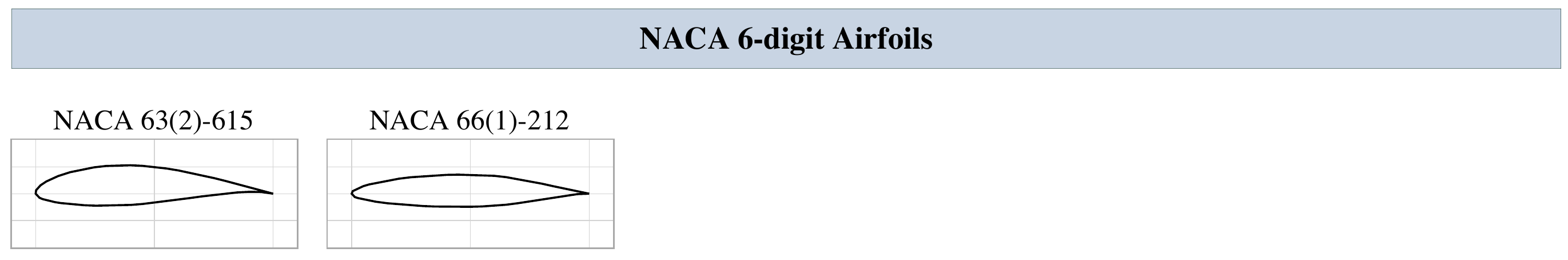}
    \includegraphics[width=0.925\textwidth]{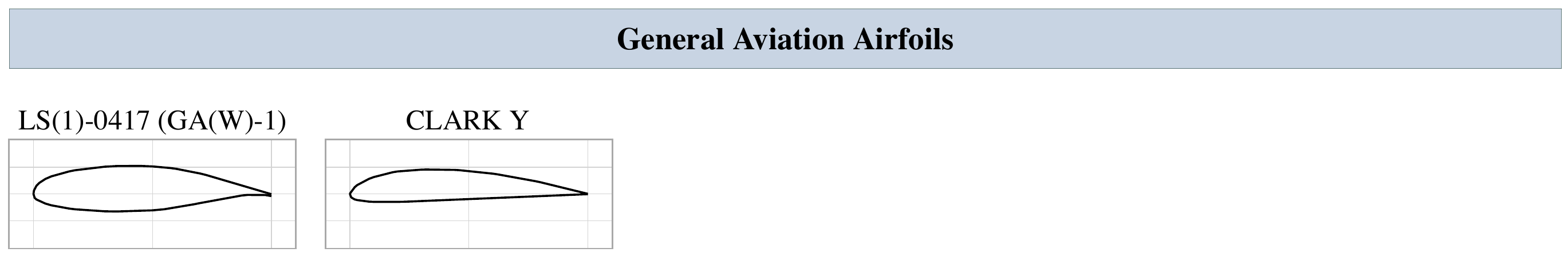}
    \includegraphics[width=0.925\textwidth]{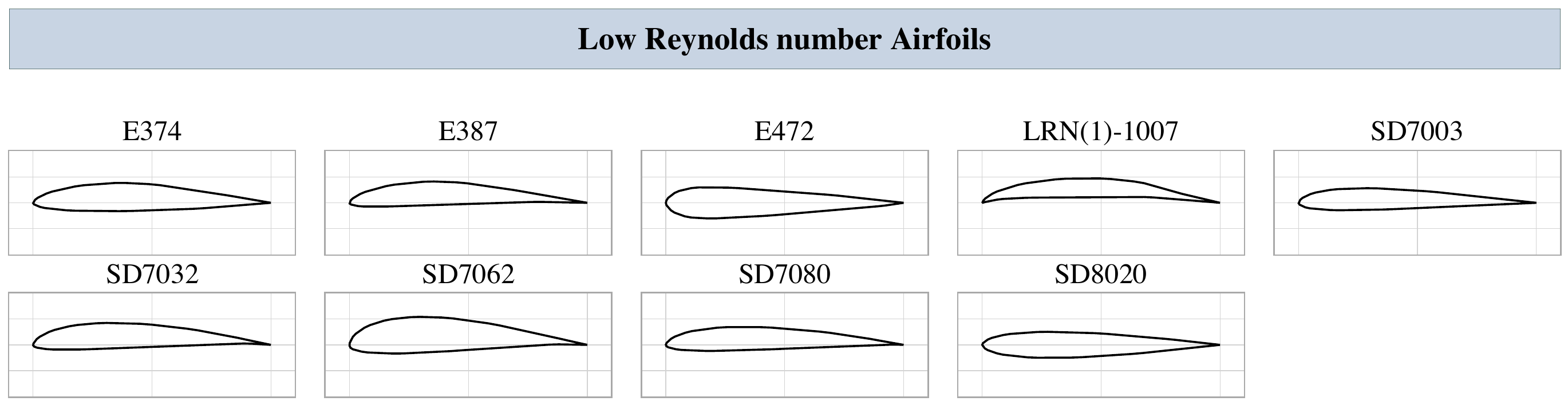}
    \includegraphics[width=0.925\textwidth]{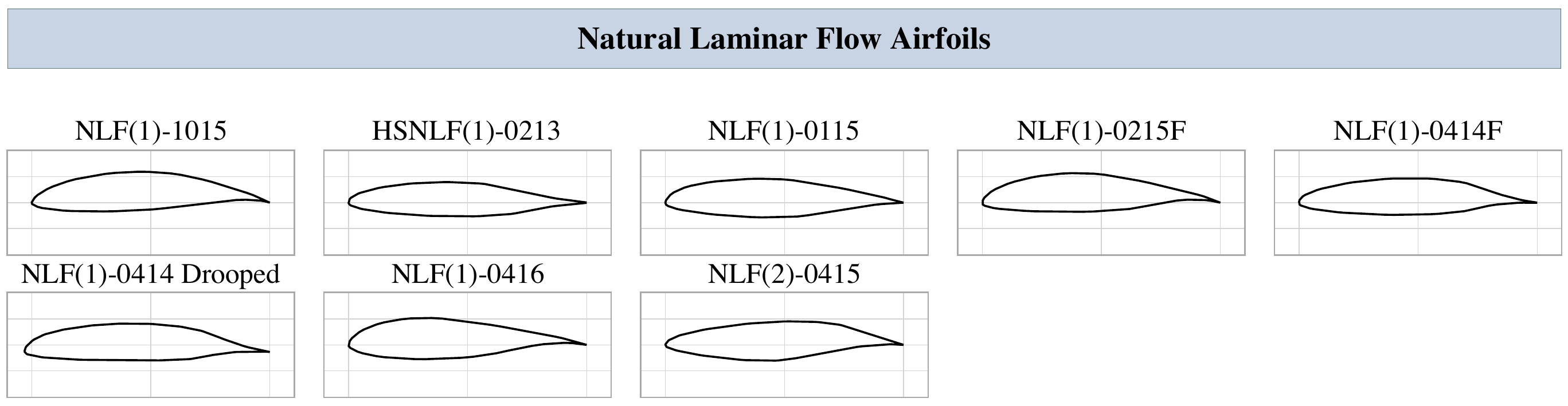}
    \label{fig:airfoil_db_1}
\end{figure}
\begin{figure}
    \centering
    \includegraphics[width=0.925\textwidth]{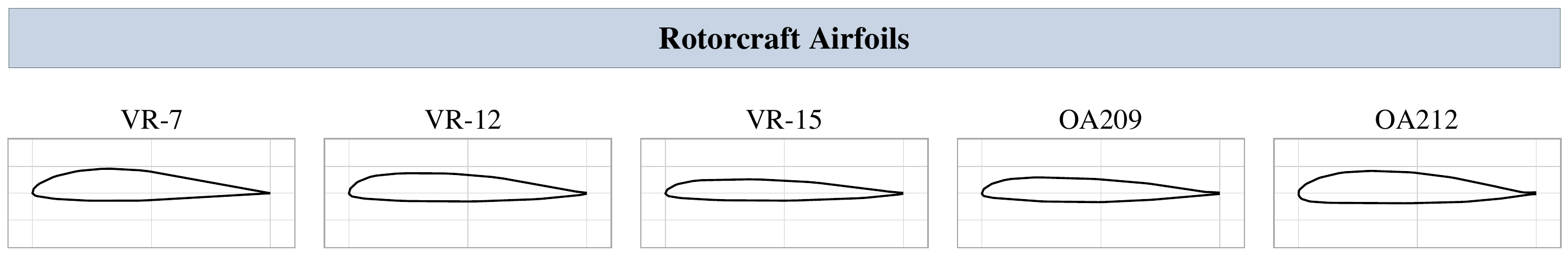}
    \includegraphics[width=0.925\textwidth]{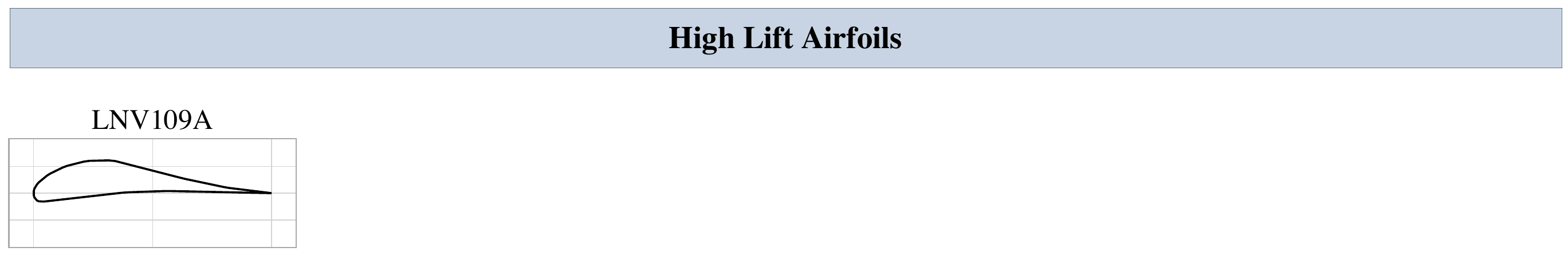}
    \includegraphics[width=0.925\textwidth]{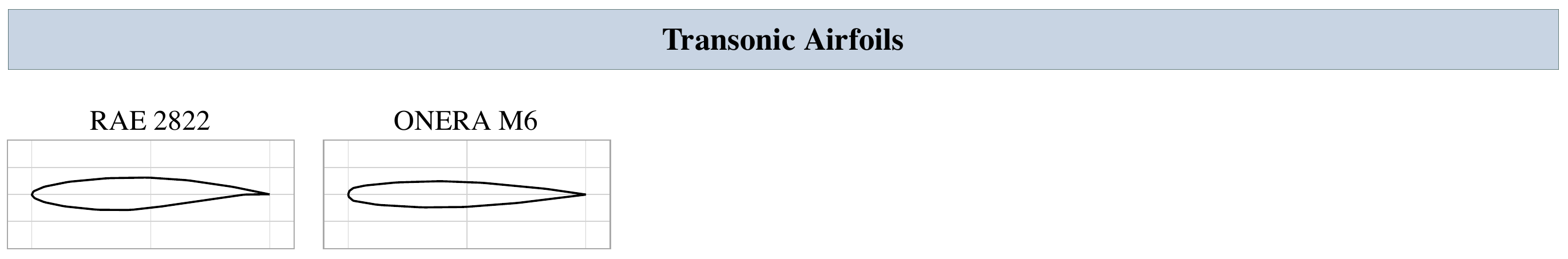}
    \includegraphics[width=0.925\textwidth]{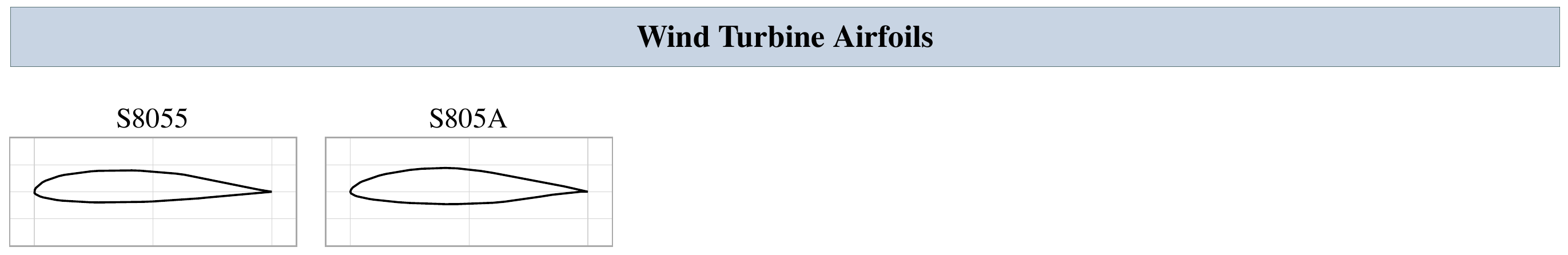}
    \includegraphics[width=0.925\textwidth]{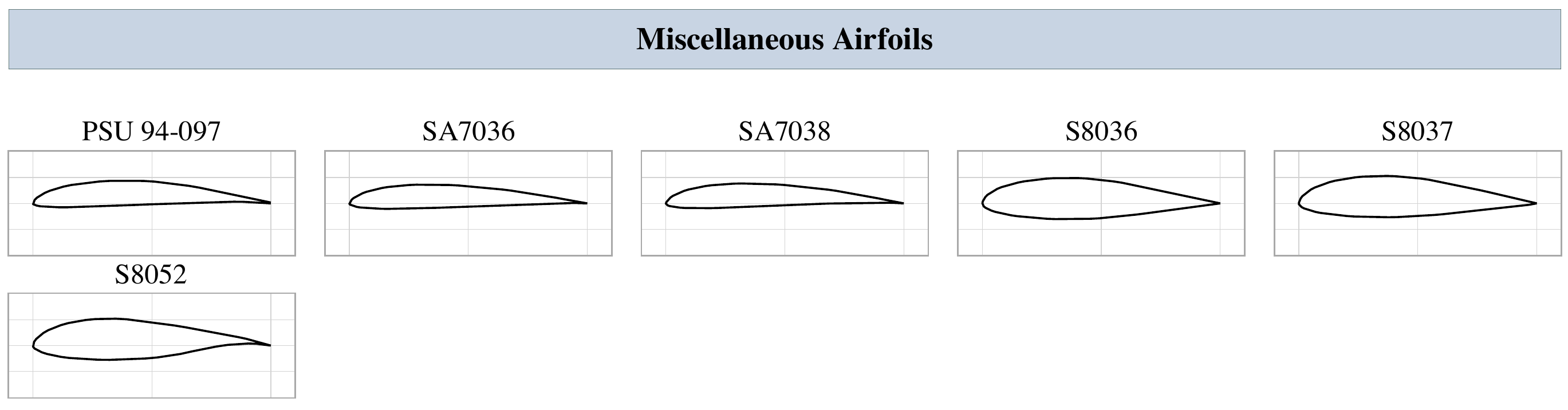}
    \label{fig:airfoil_db_2}
\end{figure}

\newpage
\begin{Backmatter}

\paragraph{Acknowledgments}
The computational resources used for this project were provided by the Advanced Research Computing (ARC) of Virginia Tech, which is gratefully acknowledged. 

\paragraph{Funding Statement}
This research was supported by the Revolutionary Computational AeroSciences discipline of NASA's Transformational Tools and Technologies Project.

\paragraph{Competing Interests}
The authors declare no competing interests.

\paragraph{Data Availability Statement}
The data and code used in this work is currently proprietary and confidential; it will be considered for public release in the future.

\paragraph{Author Contributions}
Conceptualization: M.-M.C.; H.X.; Methodology: M.-M.C.; H.X.; M.-I.Z.; Data Curation: P.P.; M.-M.C.; M.-I.Z.; Formal Analysis: M.-I.Z. ; Writing original draft: M.-I.Z.; M.-M.C.; Writing-review editing: all. All authors approved the final submitted draft.

\bibliographystyle{apalike}

\end{Backmatter}

\end{document}